\documentclass[aps,onecolumn,11pt,secnumarabic,nobalancelastpage,amsmath,amssymb,
nofootinbib,numbers]{revtex4-2}

\usepackage{amsmath, amssymb, amsthm} 
\usepackage{mathtools}
\usepackage{thmtools}
\usepackage{bm}
\usepackage{dsfont}
\usepackage{braket}

\numberwithin{equation}{section}

\usepackage{enumitem}
\setlist[itemize]{noitemsep}
\setlist[description]{noitemsep}

\usepackage[dvipsnames]{xcolor}
\usepackage{tikz}
\usepackage{pgfplots}
\usetikzlibrary{intersections,backgrounds}
\usepgfplotslibrary{fillbetween}
\pgfplotsset{compat = newest}

\usepackage{natbib}
\usepackage{hyperref}
\hypersetup{colorlinks=true, linktoc=page, linkcolor=purple, citecolor=blue}

\newcommand{\floor}[1]{\cramped{\left\lfloor #1 \right\rfloor}}

\newcommand{\negphantom}[1]{
    \ifmmode\settowidth{\dimen0}{$#1$}
    \else\settowidth{\dimen0}{#1}
    \fi
    \hspace*{-\dimen0}}
    
\makeatletter
\newcommand{\mask}[2]{{\mathpalette\mask@{{#1}{#2}}}}
\newcommand{\mask@}[2]{\mask@@{#1}#2}
\newcommand{\mask@@}[3]{%
  \settowidth{\dimen@}{$\m@th#1#2$}%
  \makebox[\dimen@]{$\m@th#1#3$}%
}
\makeatother
    
\newcommand{\sm}{\smallskip}

\newcommand{\fsl}[1]{\ensuremath{\mathrlap{\!\not{\phantom{#1}}}#1}}

\begin{document}

\title{Covariant actions and propagators for all spins, masses, and dimensions}
\author{Lukas W. Lindwasser}
\email[email:]{lukaslindwasser@physics.ucla.edu}
\affiliation{\textit{Mani L. Bhaumik Institute for Theoretical Physics,} \\ \textit{Department of Physics and Astronomy,\\ University of California, Los Angeles, CA 90095, USA}}
\date{\today}
\begin{abstract}
The explicit covariant actions and propagators are given for fields describing particles of all spins and masses, in any spacetime dimension. Massive particles are realized as ``dimensionally reduced" massless particles. To obtain compact expressions for the propagators, it was useful to introduce an auxiliary vector coordinate $s^{\mu}$ and consider ``hyperfields" that are functions of space $X^{\mu}$ and $s^{\mu}$. The actions and propagators serve as a basic starting point for concrete high spin computations amenable to dimensional regularization, provided that gauge invariant interactions are introduced.
\end{abstract}
\maketitle
\pagebreak

\tableofcontents
\pagebreak


\maketitle

 \newpage

\section{Introduction}
\label{sec:intro}
The modelling of particles with spin higher than 3/2 has a long history, dating back to at least the work of Dirac in 1936 \cite{Dirac:1936tg}. Equations of motion for free massive particles of arbitrary spins were written down in \cite{Fierz:1939ix}, and it was noted then that introducing interactions in this framework will generically not preserve the physical degrees of freedom of the high spin particle. It was suggested that a way to fix this was to formulate an action principle for the higher spin field. Subsequently, the first actions for free massive particles of arbitrary spins were written down in \cite{Singh:1974qz, Singh:1974rc}, and then their massless counterparts in \cite{Fronsdal:1978rb, Fang:1978wz}. 

\sm

The problem of introducing interactions persisted. For massless higher spin particles, general arguments \cite{Weinberg:1964ew,Weinberg:1980kq,Porrati:2008rm} were made which essentially invalidate the existence of spins higher than 2 in asymptotically flat space, assuming they interact with gravity in a way which obeys the equivalence principle\footnote{There is of course, a notable theory \cite{Vasiliev:1990en} of massless higher spins in anti-de Sitter space.}. We know on the other hand that massive higher spins exist in nature \cite{ParticleDataGroup:2022pth}, and so we expect there to be a way to model them and their interactions. 

\sm

In spite of this, the task of finding consistent interactions for massive higher spins remains a challenging one. Along with the aforementioned change in degrees of freedom issue, it was realized that minimally coupling spins higher than 3/2 to an electromagnetic background will generically allow superluminal propagation \cite{PhysRev.186.1337}. Coupling massive higher spins to gravity has also been shown to violate causality \cite{Camanho:2014apa, Afkhami-Jeddi:2018apj}. For spin 3/2, both of these problems can be solved when it is realized within the framework of supergravity \cite{Deser:1977uq}. For higher spins, the only known solution to both is to realize them within string theory \cite{Porrati:2010hm}. It is clear based on the explicit construction in \cite{Porrati:2010hm} that a consistent set of interactions with electromagnetism is by no means unique. A construction of the space of consistent interactions from first principles is still warranted.

\sm 

To facilitate this, one must formulate massive higher spins in a way which makes the introduction of interactions straightforward. The actions \cite{Singh:1974qz, Singh:1974rc} are not suited for this task for at least two reasons. First, these actions are only consistent in four spacetime dimensions, and based on experience with string theory, the consistency of interactions appears to depend sensitively on the spacetime dimension. Thus we should find a way to formulate massive higher spins in arbitrary spacetime dimensions $d$. Finally, these actions do not have an organizing principle such as gauge invariance which helps decide what kinds of interactions will preserve the particle's degrees of freedom. In this paper, we will write down covariant actions for all massive spins which addresses both of these points.

\sm

Another important aspect of massive higher spins is the complexity of their propagators. Explicit propagators for the physical spin $s$ field resulting from the actions \cite{Singh:1974qz, Singh:1974rc} were written in \cite{Singh:1981aw}. A different set of propagators were written in \cite{Weinberg:1964cn,Sagnotti:2010at}, without any reference to an action. They both find spin $s$ propagators which grow like $\sim p^{-2+2s}$ in momentum space. Although these works find propagators for the physical spin $s$ field, they do not find propagators for the various auxiliary fields needed in any action formulation of massive particles with spin $s\geq 3/2$. These must be included in your Feynman rules, and are therefore important for doing any concrete computation of Feynman diagrams. Propagators for spins 1 and higher are not unique, and we demonstrate this non-uniqueness as a manifestation of gauge invariance within this formalism. In this paper, we instead find propagators for all fields needed in the description which grow like $\sim p^{-2}$ and $\sim p^{-1}$ for all integer and half integer spins, respectively. This discrepancy is understood as a natural extension of the well known fact that massive spin 1 propagators can have different high energy behavior depending on the gauge one picks, e.g. unitarity or $R_{\xi}$ gauges. The improved high energy behavior is achieved at the cost of introducing unphysical polarizations into the propagator, but as in Yang-Mills theory, as long as gauge invariant interactions are present, these unphysical polarizations should not contribute to observables. Furthermore, because the propagators are valid in any spacetime dimension $d$, dimensional regularization can be used readily. 

\sm

The expressions for the propagators are compactly written in terms of Gegenbauer polynomials. This is made possible by introducing an auxiliary vector coordinate $s^{\mu}$, and writing propagators for the fields $\frac{1}{n!}\phi_{\mu_1\cdots\mu_n}(X)s^{\mu_1}\cdots s^{\mu_n}$. In particular, this allows us to write the propagators so that the spin $s=n, n+1/2$ is a freely tunable parameter, enabling analytic control over the large spin asymptotic limit $n\to\infty$. Indeed, using known properties of Gegenbauer polynomials \cite{nla.cat-vn2358422}, we find simple asymptotic $n\to\infty$ expressions for the propagators. This may be particularly helpful in the current research program of using effective field theory techniques to model black hole binary dynamics with classical values of spin  \cite{Guevara:2017csg,Chung:2018kqs, Guevara:2018wpp, Arkani-Hamed:2019ymq,Guevara:2019fsj,Bern:2020buy, Bern:2020uwk, Bern:2022kto, Chiodaroli:2021eug, FebresCordero:2022jts, Aoude:2022thd, Cangemi:2022bew}. Indeed, a main motivation for this work is to develop a formalism which allows one to take an $n\to\infty$ limit without computing a few low spin examples and having to extrapolate.

\sm

Finally, by virtue of the gauge symmetry inherent in this formulation, finding interactions which preserve the higher spin particle's degrees of freedom from first principles is equivalent to maintaining some form of gauge invariance. Formulating massive higher spins in a way that can be used as a starting point for straightforwardly introducing interactions is a main motivation for the current work. The formulation we arrive at is similar to previous work \cite{Klishevich:1998ng,Klishevich:1998yt,Zinoviev:2001dt,Metsaev:2006zy,Asano:2019smc}. All such formulations, including ours, are presumably related via field redefinition, yet vary in their complexity.

\sm

Previous attempts at introducing consistent interactions to these actions have restricted to constrained background fields, e.g. \cite{Klishevich:1998ng,Klishevich:1998yt,Buchbinder:2000fy,Zinoviev:2006im,Zinoviev:2008ck}. The difficulty in constructing interactions with general dynamical fields is in part due to the inherent complexity of high spin actions. One virtue of the current work which alleviates this difficulty is the elucidation of natural objects $\mathcal{F}_{\mu_1\cdots\mu_n}$, $\mathcal{F}_{\mu_1\cdots\mu_{n-1}}$, $\mathcal{F}_{\mu_1\cdots\mu_{n-2}}$, $\mathcal{F}_{\mu_1\cdots\mu_{n-3}}$ for the integer spin case, and $\mathcal{S}_{\mu_1\cdots\mu_n}$, $\mathcal{S}_{\mu_1\cdots\mu_{n-1}}$, $\mathcal{S}_{\mu_1\cdots\mu_{n-2}}$ for the half integer spin case, which are gauge invariant and satisfy Bianchi-like identities, described in \autoref{sec5} and \autoref{sec6}. This is possible because we formulate $d$ dimensional massive particles as ``dimensionally reduced" $d+1$ dimensional massless particles, which themselves have gauge invariant field strengths satisfying Bianchi-like identities. Dimensional reduction of massless particles was first discussed in detail for integer spins in \cite{Aragone:1987dtt}, elaborated in \cite{Rindani:1988gb,Rindani:1989ym}, and further studied in \cite{Klishevich:1998ng,Klishevich:1998yt,Bekaert:2003uc,Buchbinder:2008ss}. We differentiate our work from theirs by organizing the explicit actions in terms of the natural objects $\mathcal{F}_{\mu_1\cdots\mu_{n-i}}$ and $\mathcal{S}_{\mu_1\cdots\mu_{n-i}}$, and also finding the explicit propagators, including that of the auxiliary fields. Furthermore, the manifest gauge invariance of the massive higher spin actions we arrive at is guaranteed by the properties of $\mathcal{F}_{\mu_1\cdots\mu_{n-i}}$ and $\mathcal{S}_{\mu_1\cdots\mu_{n-i}}$ discussed above, whereas in previous formulations it is less obvious. This allows a simpler and more organized formulation for constructing gauge invariant interactions. We delay a detailed discussion of interactions within this formalism for future work.

\subsection{Conventions}
Throughout this paper, we use the metric signature $\eta = \text{diag}(-1,+1,\cdots,+1)$. Subsequently, the standard massless and massive Feynman propagators for spins 0 and 1/2 are respectively,
\begin{align}
    &G_0(X-Y) = \int \frac{d^dp}{(2\pi)^d}\frac{-i}{p^2-i\epsilon}e^{ip\cdot(X-Y)},&&G_m(X-Y) = \int \frac{d^dp}{(2\pi)^d}\frac{-i}{p^2+m^2-i\epsilon}e^{ip\cdot(X-Y)} 
\\
    &\Delta_0(X-Y) = \int \frac{d^dp}{(2\pi)^d}\frac{-\fsl{p}}{p^2-i\epsilon}e^{ip\cdot(X-Y)},&&\Delta_m(X-Y) = \int \frac{d^dp}{(2\pi)^d}\frac{-\fsl{p}-im}{p^2+m^2-i\epsilon}e^{ip\cdot(X-Y)}
\end{align}
These will be basic building blocks for their higher spin generalizations. We will define all correlation functions via analytic continuation of their Euclidean space versions. Because of this, when writing correlation functions such as $\langle\mathcal{O}_1(X_1)\cdots\mathcal{O}_n(X_n)\rangle$, we leave the time ordered product $T$ as implicit.

\sm

Frequently used in this paper is the notation $\mathcal{O}_{(\mu_1\cdots\mu_n)}$, which denotes the total symmetrization of the indices enclosed by the parentheses. The normalization we use for instance for the symmetrization of two indices is
\begin{align}
    \mathcal{O}_{(\mu_1\mu_2)} \equiv \frac{1}{2}(\mathcal{O}_{\mu_1\mu_2}+\mathcal{O}_{\mu_2\mu_1})
\end{align}

\subsection{Outline}
In \autoref{sec2}, we introduce the ``hyperfield" formalism, which will enable compact expressions for the covariant actions, propagators, and interactions in subsequent sections. Then in \autoref{sec3} and \autoref{sec4}, we review the formulation of free massless higher spins \cite{Fronsdal:1978rb,Fang:1978wz}, and we compute their propagators in a special gauge, as well as characterize their gauge ambiguities. This review importantly sets the stage for \autoref{sec5} and \autoref{sec6}, where we do the same for massive higher spins, defining them via ``dimensional reduction" of their massless counterparts as in \cite{Aragone:1987dtt}. Finally in \autoref{sec:disc}, we discuss our results and present some directions for future research.

\section{Hyperfields}
\label{sec2}

In this section, we introduce the notion of a hyperfield, which greatly simplifies the analysis of fields with arbitrary spin. Similar techniques have been used in the high spin literature, including but not limited to \cite{Bekaert:2005in, Ponomarev:2016jqk, Roiban:2017iqg,Najafizadeh:2018cpu}. We will find in later sections that covariant actions and propagators admit particularly simple closed form expressions within this formalism.

\sm

Particles with spin are covariantly described by totally symmetric tensors $\phi_{\mu_1\cdots\mu_n}(x)$ or spinor tensors $\psi_{\mu_1\cdots\mu_n}(x)$. Totally symmetric tensors have ${d+n-1\choose n}$ independent components, and $2^{\floor{d/2}}{d+n-1\choose n}$ if there is an additional Dirac index, where $\floor{x}$ is the floor function\footnote{In this paper we consider for simplicity only parity invariant representations.}. This is far in excess of the physical degrees of freedom needed to describe a particle with spin $n$ or $n+1/2$. To maintain covariance, one must impose additional constraints on the fields $\phi_{\mu_1\cdots\mu_n}(x)$, $\psi_{\mu_1\cdots\mu_n}(x)$, outlined in later sections, which remove unnecessary degrees of freedom. This procedure is quite unwieldy for arbitrary spins, as evidenced by their covariant actions \cite{Singh:1974qz,Singh:1974rc,Fronsdal:1978rb,Fang:1978wz}, so much so that many physicists opt to study the effects of high spin particles using alternative methods \cite{Veneziano:1968yb, Virasoro:1969me,Coon:1969yw,Arkani-Hamed:2017jhn, Afkhami-Jeddi:2018apj,Caron-Huot:2016icg, Bern:2020buy, Bern:2022kto}. To study the effects of high spin particles in a fully covariant and off-shell manner, a new formalism which simplifies calculations is warranted.

\sm

In this paper the approach will be to introduce an auxiliary vector coordinate $s^{\mu}$, and consider ``hyperfields" $\Phi_n(X,s) = \frac{1}{n!}i^{-n/2}\phi_{\mu_1\cdots\mu_n}(X)s^{\mu_1}\cdots s^{\mu_n}$, where the factor $i^{-n/2}$ is added for later convenience. If $\Phi_n(X,s)$ has no external Lorentz indices, the component field will be a totally symmetric rank $n$ tensor, which will describe an integer spin $n$ particle after appropriate constraints are imposed. If the hyperfield has an external Dirac index, which we will henceforth denote $\Psi_n(X,s)$, the component field will similarly describe a half integer spin $n+1/2$ particle. More generally, one can consider a hyperfield $\Phi(X,s)$ which is a general function of $s^{\mu}$, with its formal Taylor expansion generating all totally symmetric tensors
\begin{align}
\Phi(X,s) = \sum_{n=0}^{\infty}\frac{1}{n!}i^{-n/2}\phi_{\mu_1\cdots\mu_n}(X)s^{\mu_1}\cdots s^{\mu_n}
\end{align}
Such hyperfields are sufficient to describe the leading Regge trajectory of string theory. One can further consider introducing $N$ auxiliary vectors $s_i^{\mu}$, and if $N>d$, the corresponding hyperfield $\Phi(X,\{s_i\})$ would contain all representations present in string theory. This is of course precisely how string field theory packages all of its particle states into a single object \cite{Witten:1985cc}. For the purposes of the present paper, we make no attempt at making contact with string theory, or its generalizations \cite{Cheung:2022mkw, Cheung:2023adk, Geiser:2022exp}. The author hopes to return to this possibility in future work.

\subsection{Hyperfield basics}
\label{subsec:Hyperbasics}

In this section we quickly review how to perform basic operations on $\phi_{\mu_1\cdots\mu_n}(X)$ relevant for the covariant formulation of spinning particles, at the level of a hyperfield $\Phi_n(X,s)$. For instance, one often has to take traces of the fields $\phi^{\lambda}_{\;\;\lambda\mu_3\cdots\mu_n}(X)$. This can be achieved at the level of $\Phi_n(X,s)$ by taking the Laplacian with respect to $s^{\mu}$
\begin{align}
    \label{eq:trace}
   & i\,\partial_s^2\Phi_n(X,s) = \frac{1}{(n-2)!}i^{-(n-2)/2}\phi^{\lambda}_{\;\;\lambda\mu_1\cdots\mu_{n-2}}(X)s^{\mu_1}\cdots s^{\mu_{n-2}} 
   \intertext{The divergence $\partial^{\lambda}\phi_{\lambda\mu_2\cdots\mu_n}(X)$ may also be written in terms of $\Phi_n(X,s)$}
   \label{eq:diverge}
    & i^{1/2}\,\partial_s\cdot\partial_X\Phi_n(X,s) = \frac{1}{(n-1)!}i^{-(n-1)/2}\partial^{\lambda}\phi_{\lambda\mu_1\cdots\mu_{n-1}}(X)s^{\mu_1}\cdots s^{\mu_{n-1}}
   \intertext{The symmetric derivative $\partial_{(\mu_1}\phi_{\mu_2\cdots\mu_{n+1})}(X)$ is written in terms of $\Phi_n(X,s)$ via}
   \label{eq:symmderiv}
    & i^{-1/2}s\cdot\partial_X\Phi_n(X,s) = \frac{1}{n!}i^{-(n+1)/2}\partial_{(\mu_1}\phi_{\mu_2\cdots\mu_{n+1})}(X)s^{\mu_1}\cdots s^{\mu_{n+1}}
   \intertext{One can also contract indices between two equal rank $n$ hyperfields $A_n(X,s)$ and $B_n(X,s)$}
   \label{eq:contract}
    & \int \frac{d^dsd^ds'}{(2\pi)^d}e^{is\cdot s'}A_n(X,s)B_n(X,s') = \frac{1}{n!}a_{\mu_1\cdots\mu_n}b^{\mu_1\cdots\mu_n}
   \intertext{These four operations are sufficient for the formulation of integer spin fields, as we will see in \autoref{sec3} and \autoref{sec5}. For half integer spin fields, contraction with gamma matrices $\gamma^{\mu}$ will also be necessary. For a Dirac hyperfield $\Psi_n(X,s)$, we have}
   \label{eq:gammatrace}
    & i^{1/2}\fsl{\partial}_s\Psi_n(X,s) = \frac{1}{(n-1)!}i^{-(n-1)/2}\gamma^{\lambda}\psi_{\lambda\mu_1\cdots\mu_n}(X) s^{\mu_1}\cdots s^{\mu_n}
\end{align}

\sm

The formula (\ref{eq:contract}) is particularly interesting, as it suggests introducing a (pseudo) inner product on the space of hyperfields
\begin{align}
    &(A_n,B_n) = \int d^dX\frac{d^dsd^ds'}{(2\pi)^d}e^{is\cdot s'}\tilde{A}_n(X,s)B_n(X,s') = \int d^dX \frac{1}{n!}a^*_{\mu_1\cdots\mu_n}b^{\mu_1\cdots\mu_n} \\
    &\text{where } \tilde{A}_n(X,s) = \frac{1}{n!}i^{-n/2}a^*_{\mu_1\cdots\mu_n}(X)s^{\mu_1}\cdots s^{\mu_n}
\end{align}
and we may consider the (pseudo) Hilbert space of hyperfields $\Phi(X,s)$ with finite (pseudo) norm $(\Phi,\Phi)<\infty$\footnote{$(A,B)$ is only positive definite in Euclidean signature.} \cite{Bargmann:1977gy}. Note that $(i^{-1/2}s^{\mu}A,B)=(A,i^{1/2}\partial_s^{\mu}B)$, and so for instance, the divergence and symmetric derivative are anti-Hermitian adjoints of each other~$(i^{1/2}\partial_s\cdot\partial_X)^\dagger = -i^{-1/2}s\cdot\partial_X$ in this space. 

\sm

Crucial operators for the analysis in subsequent sections are the projection operators $P_{n,\eta}^d(s,t)$, $P_{n,\gamma}^d(s,t)$ on to the subspace of $\eta$ traceless rank $n$ hyperfields $i\,\partial_s^2 \Phi_n(X,s) = 0$ and $\gamma$ traceless rank $n$ hyperfields $i^{1/2}\fsl{\partial}_s \Psi_n(X,s) = 0$, respectively. We define operators to act on the hyperfields via 
\begin{align}
    (P\Phi_n)(X,s) = \int\frac{d^dtd^dt'}{(2\pi)^d}e^{it\cdot t'}P(s,t)\Phi_n(X,t')
\end{align}
The projection operators can be straightforwardly found from the equations $\partial_s^2P_{n,\eta}(s,t) = \partial_t^2P_{n,\eta}(s,t) = 0$, $\fsl{\partial}_sP_{n,\gamma}(s,t) = P_{n,\gamma}(s,t)\overleftarrow{\fsl{\partial}}_t = 0$, and they have simple expansions in terms of classical Gegenbauer polynomials $C_n^{\alpha}(x)$ \cite{nla.cat-vn2358422}, they read
\begin{align}
    \label{eq:eproj}
    &P_{n,\eta}^d(s,t) = \frac{1}{(\frac{d-2}{2})_n}\Big(-\frac{i}{2}\sqrt{s^2t^2}\Big)^n C_n^{\frac{d-2}{2}}\Big(\frac{s\cdot t}{\sqrt{s^2t^2}}\Big) \\
    &P_{n,\gamma}^d(s,t) = \nonumber\\ 
    \label{eq:gproj}
    &\frac{1}{(\frac{d}{2})_n}\Big(-\frac{i}{2}\sqrt{s^2t^2}\Big)^n C_n^{\frac{d}{2}}\Big(\frac{s\cdot t}{\sqrt{s^2t^2}}\Big)+i\frac{\fsl{s}\fsl{t}}{2}\frac{1}{(\frac{d}{2})_{n}}\Big(-\frac{i}{2}\sqrt{s^2t^2}\Big)^{n-1} C_{n-1}^{\frac{d}{2}}\Big(\frac{s\cdot t}{\sqrt{s^2t^2}}\Big)
\end{align}
where $(x)_n = \Gamma(x+n)/\Gamma(x)$ is the Pochhammer symbol.

\section{Massless integer spins}
\label{sec3}
In this section, we will use the hyperfield formalism of \autoref{sec2} to review massless integer spin fields. We will find compact, closed form expressions for their covariant actions and propagators. First, we review the standard description of massless integer spin fields $\phi_{\mu_1\cdots\mu_n}(X)$.

\sm

The covariant formulation of massless integer spin fields $\phi_{\mu_1\cdots\mu_n}(X)$ was first worked out in \cite{Fronsdal:1978rb}. In this description, Fronsdal found that it was possible to write down a covariant action for $\phi_{\mu_1\cdots\mu_n}$ which is quadratic in derivatives, provided that $\phi_{\mu_1\cdots\mu_n}$ is a real, symmetric tensor that satisfies the unusual \textit{double traceless} condition $\phi^{\lambda\;\;\omega}_{\;\;\lambda\;\;\omega\mu_5\cdots\mu_n}=0$. The equations of motion are
\begin{align}
    \label{eq:mlesseom}
    \mathcal{F}_{\mu_1\cdots\mu_n} = 
    \partial^2\phi_{\mu_1\cdots\mu_n}-n\partial_{(\mu_1}\partial^{\lambda}\phi_{\lambda\mu_2\cdots\mu_n)}+\frac{1}{2}n(n-1)\partial_{(\mu_1}\partial_{\mu_2}\phi^{\lambda}_{\;\;\lambda\mu_3\cdots\mu_n)} = 0
\end{align}
where we will call $\mathcal{F}_{\mu_1\cdots\mu_n}$ the Fronsdal field strength, which may be thought of as the spin $n$ generalization of the linearized Ricci curvature tensor $R_{\mu\nu} = \partial^2h_{\mu\nu}-\partial_{\mu}\partial^\lambda h_{\lambda\nu}-\partial_{\nu}\partial^\lambda h_{\lambda\mu}+\partial_{\mu}\partial_{\nu}h$ obtained from the metric $g_{\mu\nu} = \eta_{\mu\nu} - 2 h_{\mu\nu}$. $\mathcal{F}_{\mu_1\cdots\mu_n}$ satisfies the spin $n$ generalization of the second contracted Bianchi identity
\begin{align}
\label{eq:mless2ndBianchi}
\partial^{\lambda}\mathcal{F}_{\lambda\mu_1\cdots\mu_{n-1}}-\frac{1}{2}(n-1)\partial_{(\mu_1}\mathcal{F}^{\lambda}_{\;\;\lambda\mu_2\cdots\mu_{n-1})} = 0
\end{align}

\sm

As with Maxwell theory and linearized gravity, the equations of motion (\ref{eq:mlesseom}) have a gauge redundancy. Under the gauge transformation
\begin{align}
    \phi_{\mu_1\cdots\mu_n}(X)' = \phi_{\mu_1\cdots\mu_n}(X) + 
    n\partial_{(\mu_1}\epsilon_{\mu_2\cdots\mu_n)}(X)
\end{align}
the Fronsdal field strength is invariant provided that $\epsilon_{\mu_1\cdots\mu_{n-1}}$ is a symmetric and traceless, but otherwise arbitrary function, i.e.
\begin{align}
\label{eq:mlessN}
    \delta\mathcal{F}_{\mu_1\cdots\mu_n} = \frac{1}{2}n(n-1)(n-2)\partial_{(\mu_1}\partial_{\mu_2}\partial_{\mu_3}\epsilon^{\lambda}_{\;\;\lambda\mu_4\cdots\mu_n)}
    =
    0
\end{align}

This gauge redundancy allows us to impose a gauge fixing condition. The natural choice is the spin $n$ generalization of the Lorenz and de Donder gauge conditions
\begin{align}
    \label{eq:indexdeDonder}
    \partial^{\lambda}\phi_{\lambda\mu_1\cdots\mu_{n-1}}-\frac{1}{2}(n-1)\partial_{(\mu_1}\phi^{\lambda}_{\;\;\lambda\mu_2\cdots\mu_{n-1})} = 0
\end{align}
In this gauge, the equations of motion (\ref{eq:mlesseom}) reduce to
\begin{align}
\mathcal{F}_{\mu_1\cdots\mu_n}=\partial^2\phi_{\mu_1\cdots\mu_n}=0
\end{align}
and so the particles described by $\phi_{\mu_1\cdots\mu_n}$ are indeed massless. In this gauge there remains a residual gauge symmetry under a gauge parameter $\epsilon_{\mu_1\cdots\mu_{n-1}}$ satisfying $\partial^2\epsilon_{\mu_1\cdots\mu_{n-1}}=0$. In total then, the number of degrees of freedom removed from $\phi_{\mu_1\cdots\mu_n}$ is twice that of $\epsilon_{\mu_1\cdots\mu_{n-1}}$. The final degrees of freedom $N_0(d,n)$ of this field in $d$ dimensions is then
\begin{align}
    N_0(d,n) &= {d+n-1\choose n}-{d+n-5\choose n-4}-2\Bigg({d+n-2\choose n-1}-{d+n-4\choose n-3}\Bigg) \nonumber\\
    &= \frac{2n+d-4}{d-4} {d+n-5\choose n}
\end{align}
which reduces to the $2$ helicities of a massless spin $n$ particle when $d=4$. 

\sm

The Bianchi identity (\ref{eq:mless2ndBianchi}) implies that some components of (\ref{eq:mlesseom}) offer constraints on $\phi_{\mu_1\cdots\mu_n}$. Indeed, the following components of the gauge invariant field strength
\begin{align}
    \mathcal{F}_{0j_1\cdots j_{n-1}},\quad \mathcal{F}^0_{\;\;0 j_1\cdots j_{n-2}} - \mathcal{F}^i_{\;\;i j_1\cdots j_{n-2}},
\end{align}
where the $i,j$ indices run over spatial components, are first order in time. These constraints do not affect the counting of $N_0(d,n)$, because in the gauge (\ref{eq:indexdeDonder}), these components turn into dynamical equations. In other gauges the counting of these constraints may become relevant.

\sm

In the aim of constructing a gauge invariant, quadratic in derivatives action for $\phi_{\mu_1\cdots\mu_n}$, one must introduce an object $\mathcal{G}_{\mu_1\cdots\mu_n}$ that is linear in $\mathcal{F}_{\mu_1\cdots\mu_n}$ whose divergence is a pure trace. The appropriate object is $\mathcal{G}_{\mu_1\cdots\mu_n} = \mathcal{F}_{\mu_1\cdots\mu_n}-\frac{1}{4}n(n-1)\eta_{(\mu_1\mu_2}\mathcal{F}^{\lambda}_{\;\;\lambda\mu_3\cdots\mu_n)} $. This object may be thought of as the spin $n$ generalization of the linearized Einstein tensor $G_{\mu\nu}$. The divergence of $\mathcal{G}_{\mu_1\cdots\mu_n}$ is
\begin{align}
\label{eq:bianchi}    \partial^{\lambda}\mathcal{G}_{\lambda\mu_2\cdots\mu_n}
    =-\frac{1}{4}n(n-1)(n-2)\eta_{(\mu_2\mu_3}\partial^{\lambda}\mathcal{F}^{\omega}_{\;\;\omega\lambda\mu_4\cdots\mu_n)}
\end{align}
This is a direct consequence of (\ref{eq:mless2ndBianchi}). Using this, one can construct a gauge invariant action that generates the equations of motion (\ref{eq:mlesseom}), which we will write in terms of $\mathcal{F}_{\mu_1\cdots\mu_n}$
\begin{align}
\label{eq:mlessindexaction}
    S_n = \frac{1}{2}\int d^dX\phi^{\mu_1\cdots\mu_n}\Big(\mathcal{F}_{\mu_1\cdots\mu_n}-\frac{1}{4}n(n-1)\eta_{(\mu_1\mu_2}\mathcal{F}^{\lambda}_{\;\;\lambda\mu_3\cdots\mu_n)}\Big)
\end{align}
And hence the equations of motion derived from this is the spin $n$ generalization of the linearized Einstein field equations
\begin{align}
    \mathcal{G}_{\mu_1\cdots\mu_n} = \mathcal{F}_{\mu_1\cdots\mu_n}-\frac{1}{4}n(n-1)\eta_{(\mu_1\mu_2}\mathcal{F}^{\lambda}_{\;\;\lambda\mu_3\cdots\mu_n)} = 0
\end{align}
Which is equivalent to (\ref{eq:mlesseom}). From (\ref{eq:mlessindexaction}) it is in principle possible to calculate the propagator $\langle \phi_{\mu_1\cdots\mu_n}(X)\phi_{\nu_1\cdots\nu_n}(Y)\rangle$. However, the many indices and symmetrizations involved makes such a computation prohibitive for general $n$.

\subsection{Transition to hyperfields}
The transition to hyperfields is straightforward, and we simply list some of the various formulas above in terms of $\Phi_n(X,s)$.
\begin{align} &\phi^{\lambda\;\;\omega}_{\;\;\lambda\;\;\omega\mu_5\cdots\mu_n} =0 \qquad && \longrightarrow \qquad && (\partial_s^2)^2\Phi_n(X,s) = 0
\\
\,\nonumber
\\
&\delta\phi_{\mu_1\cdots\mu_n} = n\partial_{(\mu_1}\epsilon_{\mu_2\cdots\mu_n)} \qquad && \longrightarrow \qquad && \delta\Phi_n(X,s) = i^{-1/2} s\cdot\partial_X\epsilon_{n-1}(X,s)
\\
\,\nonumber
\\
&\epsilon^{\lambda}_{\;\;\lambda\mu_3\cdots\mu_{n-1}} = 0 \qquad  &&\longrightarrow \qquad && \partial_s^2\epsilon_{n-1}(X,s) = 0
\end{align}
\begin{align}
    \mathcal{F}_{\mu_1\cdots\mu_n} = 
    \partial^2\phi_{\mu_1\cdots\mu_n}-n\partial_{(\mu_1}\partial^{\lambda}&\phi_{\lambda\mu_2\cdots\mu_n)}+\frac{1}{2}n(n-1)\partial_{(\mu_1}\partial_{\mu_2}\phi^{\lambda}_{\;\;\lambda\mu_3\cdots\mu_n)} = 0 \nonumber
    \\
    &\downarrow
    \\
    \mathcal{F}_n(X,s) = \partial_X^2\Phi_n(X,s) - s\cdot\partial_X&\partial_s\cdot\partial_X\Phi_n(X,s)+\frac{1}{2}(s\cdot\partial_X)^2\partial_s^2\Phi_n(X,s) = 0 \nonumber
\end{align}
The action for $\Phi_n(X,s)$ then takes the form
\begin{align}
\label{eq:mlesshyperaction}
    S_n &= \frac{1}{2}n!\int d^dX\frac{d^dsd^ds'}{(2\pi)^d}e^{is\cdot s'} \Phi_n(X,s)\big(1-\frac{1}{4}s'^2\partial_{s'}^2\big)\mathcal{F}_n(X,s')
\end{align}
The messy algebraic combinations of $\phi_{\mu_1\cdots\mu_n}$ needed to describe massless particles are now recast as differential operators in the $2d$ dimensional space $(X^{\mu},s^{\mu})$.

\sm

Gauge invariance of (\ref{eq:mlesshyperaction}) is guaranteed by the identity (\ref{eq:mless2ndBianchi})
\begin{align}
\label{eq:hyperbianchi}
    \partial_s\cdot\partial_X\mathcal{F}_n-\frac{1}{2}s\cdot\partial_X\partial_s^2\mathcal{F}_n=0
\end{align}

\subsection{All propagators}
\label{sec:mlessprop}
We now compute the propagator $\langle\Phi_n(X,s)\Phi_n(Y,t)\rangle$ from (\ref{eq:mlesshyperaction}), which will give the expressions for all massless integer spin propagators $\langle\phi_{\mu_1\cdots\mu_n}(X)\phi_{\nu_1\cdots\nu_n}(Y)\rangle$.

\sm

This action is gauge invariant under $\delta\Phi_n=i^{-1/2}s\cdot\partial_X\epsilon_{n-1}$, and so there will be an ambiguity in how we define the propagator. To see this, consider adding an external hyperfield source $\mathcal{J}_n(X,s)$ to the equations of motion
\begin{align}
\label{eq:mlesssource}
    \big(1-\frac{1}{4}s^2\partial_s^2\big)\mathcal{F}_n = -\mathcal{J}_n
\end{align}
The left hand side is double traceless, and satisfies (\ref{eq:bianchi}). Consistent external sources $\mathcal{J}_n$ must be such that these constraints are maintained. This imposes constraints on $\mathcal{J}_n$ 
\begin{align}
\label{eq:sourceconstr}
    (\partial_s^2)^2\mathcal{J}_n=0 && \partial_s\cdot\partial_X\mathcal{J}_n = -i^{1/2} s^2\mathcal{K}_{n-3}
\end{align}
for some traceless hyperfield $\mathcal{K}_{n-3}$. Solving (\ref{eq:mlesssource}) for $\Phi_n(X,s)$, we expect a formula of the sort
\begin{align}
\label{eq:mlesssoln}
    \Phi_n(X,s) = i \int d^dY\frac{d^dtd^dt'}{(2\pi)^d}e^{it\cdot t'}\mathcal{G}(X-Y,s,t)\mathcal{J}_n(Y,t')
\end{align}
The transformation $\delta\mathcal{G}(X-Y,s,t) = i^{-1/2}s\cdot\partial_X\Pi(X-Y,s,t)$, where $\Pi(X-Y,s,t)$ is a rank $n-1$ traceless hyperfield in $s$ and a rank $n$ double traceless hyperfield in $t$, amounts to a gauge transformation on $\Phi_n(X,s)$, so $\mathcal{G}(X-Y,s,t)$ itself has a gauge redundancy. For the same reason, $\mathcal{G}(X-Y,s,t)$ cannot be interpreted as a Green's function for (\ref{eq:mlesssource}). Because of the underlying gauge invariance, the kinetic operator has a zero mode, so it is not possible to solve the equation
\begin{align}
\label{eq:mlessgreen}
    \big(1-\frac{1}{4}s^2\partial_s^2\big)\big(\partial_X^2-s\cdot\partial_X\partial_s\cdot\partial_X+\frac{1}{2}(s\cdot\partial_X)^2\partial_s^2\big)\mathcal{G}(X-Y,s,t) = \frac{1}{n!}i\delta^d(X-Y)P_{n,\eta^2}^d(s,t)
\end{align}
where $P_{n,\eta^2}^d(s,t)$ is the projection operator onto rank $n$ double traceless hyperfields, which is
\begin{align}
    P_{n,\eta^2}^d(s,t) = \frac{1}{(\frac{d-2}{2})_n}\Big(-\frac{i}{2}\sqrt{s^2t^2}\Big)^n \Bigg(C_n^{\frac{d-2}{2}}\Big(\frac{s\cdot t}{\sqrt{s^2t^2}}\Big)+\Big(\frac{d-2}{2}+n-2\Big)C_{n-2}^{\frac{d-2}{2}}\Big(\frac{s\cdot t}{\sqrt{s^2t^2}}\Big)\Bigg)
\end{align}
Instead, it is possible to find a $\mathcal{G}(X-Y,s,t)$ that satisfies
\begin{align}
\label{eq:mlesssourceamb}
    \big(1-\frac{1}{4}s^2\partial_s^2\big)\big(\partial_X^2-s\cdot\partial_X\partial_s&\cdot\partial_X+\frac{1}{2}(s\cdot\partial_X)^2\partial_s^2\big)\mathcal{G}(X-Y,s,t) \nonumber\\
    &= \frac{1}{n!}i\delta^d(X-Y)P_{n,\eta^2}^d(s,t) + i^{-1/2}t\cdot\partial_Y\Omega(X-Y,s,t)
\end{align}
for some function $\Omega(X-Y,s,t)$ which is a rank $n$ double traceless hyperfield in $s$ and a rank $n-1$ traceless hyperfield in~$t$. The $\mathcal{G}(X-Y,s,t)$ which solves (\ref{eq:mlesssourceamb}) provides a consistent solution to (\ref{eq:mlesssource}) of the form (\ref{eq:mlesssoln}) because 
\begin{align}
\label{eq:mlessGreenamb}
    \int d^dY\frac{d^dtd^dt'}{(2\pi)^d}e^{it\cdot t'}i^{-1/2}t\cdot\partial_Y\Omega(X-Y,s,t)\mathcal{J}_n(Y,t') = 0
\end{align}
for a source $\mathcal{J}_n$ which satisfies (\ref{eq:sourceconstr}). In total then, $\mathcal{G}(X-Y,s,t)$ is ambiguous up to transformations
\begin{align}
\label{eq:mlesspropamb}
    \delta\mathcal{G}(X-Y,s,t) = i^{-1/2}s\cdot\partial_X\Pi(X-Y,s,t) + i^{-1/2}t\cdot\partial_Y\Omega(X-Y,s,t)
\end{align}

The discussion above holds equally true for the propagator $\langle\Phi_n(X,s)\Phi_n(Y,t)\rangle$, i.e. it must satisfy the equation (\ref{eq:mlesssourceamb}) and is ambiguous up to transformations (\ref{eq:mlesspropamb}). The particular form of $\Omega(X-Y,s,t)$ in (\ref{eq:mlesssourceamb}) amounts to a partial gauge choice.

\sm

In practice, we can gauge fix $\Phi_n(X,s)$ so that the propagator is unambiguous. For integer spins there is a Lorentz covariant gauge choice which makes the calculation of the propagator simple. The gauge choice we make is (\ref{eq:indexdeDonder})
\begin{align}
\label{eq:deDonder}
    \big(\partial_s\cdot\partial_X-\frac{1}{2}(s\cdot\partial_X)\partial_s^2\big)\Phi_n(X,s) = 0
\end{align}
In this gauge, we have $\mathcal{F}_n(X,s) = \partial_X^2\Phi_n(X,s)$, and the action becomes 
\begin{align}
    S_n = \frac{1}{2}n!\int d^dX\frac{d^dsd^ds'}{(2\pi)^d}e^{is\cdot s'} \Phi_n(X,s)\big(1-\frac{1}{4}s'^2\partial_{s'}^2\big)\partial_X^2\Phi_n(X,s')
\end{align}
To find the propagator, we must find the inverse of $(1-\frac{1}{4}s^2\partial_{s}^2)\partial_X^2$ in the subspace of double traceless hyperfields $(\partial_s^2)^2\Phi_n = 0$. To make this simpler, we decompose $\Phi_n$ into its traceless components $\Phi_n(X,s) = A_n(X,s) + s^2 B_{n-2}(X,s)$, with $\partial_s^2A_n = \partial_s^2 B_{n-2} = 0$. In this gauge, $A_n$ and $B_{n-2}$ decouple
\begin{align}
    S_n = \frac{1}{2}n!\int d^dX\frac{d^dsd^ds'}{(2\pi)^d}e^{is\cdot s'}&\Big( A_n(X,s)\partial_X^2A_n(X,s') \nonumber \\
    &+ B_{n-2}(X,s)(d + 2(n-2))(d+2(n-3))\partial_X^2B_{n-2}(X,s')
    \Big)
\end{align}
The propagators for $A_n$ and $B_{n-2}$ satisfy the equations
\begin{align}
    \partial_X^2\langle A_n(X,s)A_n(Y,t)\rangle &= \frac{1}{n!}i\delta^d(X-Y) P_{n,\eta}^d(s,t) \\
    (d + 2(n-2))(d+2(n-3))\partial_X^2\langle B_{n-2}(X,s)B_{n-2}(Y,t)\rangle &= \frac{1}{n!}i\delta^d(X-Y) P_{n,\eta}^d(s,t)
\end{align}
where the insertion of $P_{_n\eta}^d$ from (\ref{eq:eproj}) on the right hand sides are necessary to maintain tracelessness. These two equations can be solved, and when combined back into $\Phi_n$ gives 
\begin{align}
\label{eq:mlesspropagator}
    \langle\Phi_n(X,s)\Phi_n(Y,t)\rangle &= G_0(X-Y)\frac{1}{n!} P_{n,\eta}^{d-2}(s,t) \\
    &=G_0(X-Y)\frac{1}{n!}\frac{1}{(\frac{d-4}{2})_n}\Big(-\frac{i}{2}\sqrt{s^2t^2}\Big)^n C_n^{\frac{d-4}{2}}\Big(\frac{s\cdot t}{\sqrt{s^2t^2}}\Big)
\end{align}
where $G_0(X-Y)$ is the standard massless spin $0$ propagator. This formula follows directly from the recursive relation $(n+\alpha)C^\alpha_n = \alpha(C^{\alpha+1}_n-C^{\alpha+1}_{n-2})$. This propagator agrees with previous work \cite{Francia:2007qt,Ponomarev:2016jqk}.

\sm

For $d=4$, the Gegenbauer polynomials showing up in the propagator reduce to Chebyshev polynomials of the first kind $T_n(x)$
\begin{align}
    \lim_{d\to 4}\frac{1}{(\frac{d-4}{2})_n}C_n^{\frac{d-4}{2}}(x) = \begin{cases}
        1 &  \quad n = 0 \\
        \frac{2}{n!}T_n(x) & \quad n > 0 
    \end{cases}
\end{align}

\sm 

The massless spin $n$ propagators $\langle\phi_{\mu_1\cdots\mu_n}(X)\phi_{\nu_1\cdots\nu_n}(Y)\rangle$ from (\ref{eq:mlessindexaction}) can be obtained from $\langle\Phi_n(X,s)\Phi_n(Y,t)\rangle$ by applying derivatives
\begin{align}
    \langle\phi_{\mu_1\cdots\mu_n}(X)\phi_{\nu_1\cdots\nu_n}(Y)\rangle = i^n\frac{\partial^n}{\partial_s^{\mu_1}\cdots\partial_s^{\mu_n}}\frac{\partial^n}{\partial_t^{\nu_1}\cdots\partial_t^{\nu_n}}\langle\Phi_n(X,s)\Phi_n(Y,t)\rangle
\end{align}
Since the $(s,t)$ dependence of the propagator is nicely factored out of the spacetime dependence, we list the index dependence of the first few propagators
\begin{align}
    &n=0: && 1\\
    &n=1: && \eta_{\mu_1\nu_1}\\
    &n=2: && \frac{1}{2}\big(\eta_{\mu_1\nu_1}\eta_{\nu_2\mu_2}+\eta_{\mu_1\nu_2}\eta_{\nu_1\mu_2}\big)-\frac{1}{d-2}\eta_{\mu_1\mu_2}\eta_{\nu_1\nu_2}\\
    &n=3: && \frac{1}{6}\big(\eta_{\mu_1\nu_1}\eta_{\nu_2\mu_2}\eta_{\mu_3\nu_3}+\eta_{\mu_1\nu_1}\eta_{\nu_2\mu_3}\eta_{\mu_2\nu_3}+\eta_{\mu_2\nu_1}\eta_{\nu_2\mu_3}\eta_{\mu_1\nu_3} \nonumber \\
    & && \phantom{+}+ \eta_{\mu_1\nu_2}\eta_{\nu_1\mu_2}\eta_{\mu_3\nu_3}+\eta_{\mu_1\nu_2}\eta_{\nu_1\mu_3}\eta_{\mu_2\nu_3}+\eta_{\mu_2\nu_2}\eta_{\nu_1\mu_3}\eta_{\mu_1\nu_3}\big) \nonumber \\
    & &&-\frac{1}{3d}\big(\eta_{\mu_1\mu_2}\eta_{\mu_3\nu_1}\eta_{\nu_2\nu_3}+\eta_{\mu_1\mu_2}\eta_{\mu_3\nu_2}\eta_{\nu_3\nu_1}+\eta_{\mu_1\mu_2}\eta_{\mu_3\nu_3}\eta_{\nu_1\nu_2} \nonumber\\    & &&\hspace{1 cm}+\eta_{\mu_2\mu_3}\eta_{\mu_1\nu_1}\eta_{\nu_2\nu_3}+\eta_{\mu_2\mu_3}\eta_{\mu_1\nu_2}\eta_{\nu_3\nu_1}+\eta_{\mu_2\mu_3}\eta_{\mu_1\nu_3}\eta_{\nu_1\nu_2} \nonumber\\    & &&\hspace{1 cm}+\eta_{\mu_3\mu_1}\eta_{\mu_2\nu_1}\eta_{\nu_2\nu_3}+\eta_{\mu_3\mu_1}\eta_{\mu_2\nu_2}\eta_{\nu_3\nu_1}+\eta_{\mu_3\mu_1}\eta_{\mu_2\nu_3}\eta_{\nu_1\nu_2}\big)
\end{align} 
The number of terms in the propagator with exposed indices grows factorially with $n$, making it impractical to do calculations with them. Instead, calculations can be done strictly in terms of the propagator (\ref{eq:mlesspropagator}), using the operations (\ref{eq:trace}), (\ref{eq:diverge}), (\ref{eq:symmderiv}), and (\ref{eq:contract}).

\sm

Possibly the most one gains from the expression (\ref{eq:mlesspropagator}) is the ability to have full control of the large spin limit.
Classical Gegenbauer polynomials have well studied large $n$ asymptotic limits \cite{nla.cat-vn2358422}, allowing $\langle\Phi_n(X,s)\Phi_n(Y,t)\rangle$ to be written in terms of simple trigonometric functions at large $n$
\begin{align}
    \langle\Phi_n(X,s)\Phi_n(Y,t)\rangle \sim G_0(X-Y)\frac{1}{2^n(n!)^2}(-i\sqrt{s^2t^2})^n\frac{2\cos\big((n+\frac{d-4}{2})\theta-\frac{\pi}{2}\frac{d-4}{2}\big)}{(2\sin\theta)^{\frac{d-4}{2}}}
\end{align}
as $n\to\infty$, where $\cos\theta = s\cdot t/\sqrt{s^2t^2}$, for $0 < \theta < \pi$.
\subsection{Other gauges}
Often it is useful to compute observables using propagators in different gauges as a consistency check. Furthermore, there may be gauges other than (\ref{eq:mlesspropagator}) which make computations simpler. For instance, using the gauge transformation (\ref{eq:mlesspropamb}) one may always in momentum space add a term to the propagator like
\begin{align}
  \xi(p^2) \frac{-i}{p^2} \frac{s\cdot p\, t\cdot p}{p^2}\frac{1}{(\frac{d-\alpha-2}{2})_{n-1}}\frac{1}{n!}\Big(\frac{1}{2}\sqrt{s\cdot A\cdot s\,t\cdot A\cdot t}\Big)^{n-1}C_{n-1}^{\frac{d-\alpha-2}{2}}\Big(\frac{s\cdot A\cdot t}{\sqrt{s\cdot A\cdot s\,t\cdot A\cdot t}}\Big)
\end{align}
where $\xi(p^2)$ is an arbitrary function, and $A_{\mu\nu}(p)$ is a projection matrix of rank $d-\alpha$. Some covariant choices of $A_{\mu\nu}(p)$ include e.g. $\eta_{\mu\nu}$, or $\eta_{\mu\nu}-p_{\mu}p_{\nu}/p^2$. One special modification of the propagator is
\begin{align}
    &\langle\Phi_n(p,s)\Phi_n(-p,t)\rangle =\nonumber\\
    &\frac{-i}{p^2}\frac{1}{n!}\Bigg(\frac{1}{(\frac{d-4}{2})_n}\Big(\frac{1}{2}\sqrt{s^2t^2}\Big)^n C_n^{\frac{d-4}{2}}\Big(\frac{s\cdot t}{\sqrt{s^2t^2}}\Big)-\frac{s\cdot p\, t\cdot p}{p^2}\frac{1}{(\frac{d-2}{2})_{n-1}}\Big(\frac{1}{2}\sqrt{s^2t^2}\Big)^{n-1} C_{n-1}^{\frac{d-2}{2}}\Big(\frac{s\cdot t}{\sqrt{s^2t^2}}\Big)\hspace{-.1cm}\Bigg)
\end{align}
Indeed, this choice may be thought of as the spin $n$ generalization of the spin 1 Landau gauge, because
\begin{align}
    (p\cdot\partial_s-\frac{1}{2}s\cdot p\,\partial_s^2)\langle\Phi_n(p,s)\Phi_n(-p,t)\rangle = 0
\end{align}
Though because this propagator satisfies the Bianchi identity (\ref{eq:deDonder}), we refer to this as the Bianchi gauge.

\section{Massless half integer spins}
\label{sec4}
In this section, we will use the hyperfield formalism of \autoref{sec2} to review massless half integer spin fields. In order to display uniform results for any dimension, the discussion will be restricted to Dirac spinors, similar results can be found for other spinor representations. We will again find compact, closed form expressions for their covariant actions and propagators. First, we review the standard description of massless half integer spin fields $\psi_{\mu_1\cdots\mu_n}(X)$.

\sm

The covariant formulation of massless half integer spin fields $\psi_{\mu_1\cdots\mu_n}(X)$ was first worked out in \cite{Fang:1978wz}. In this description, Fang and Fronsdal found that it was possible to write down a covariant action for $\psi_{\mu_1\cdots\mu_n}$ which is linear in derivatives, provided that $\psi_{\mu_1\cdots\mu_n}$ is a symmetric tensor that satisfies the unusual \textit{triple $\gamma$ traceless} condition $\gamma^{\mu_1}\gamma^{\mu_2}\gamma^{\mu_3}\psi_{\mu_1\cdots\mu_n}=\gamma^{\omega}\psi^{\lambda}_{\;\;\lambda\omega\mu_4\cdots\mu_n}=0$. The equations of motion are
\begin{align}
\label{eq:mlesshalfeom}
    \mathcal{S}_{\mu_1\cdots\mu_n} = \fsl{\partial}\psi_{\mu_1\cdots\mu_n}-n\partial_{(\mu_1}\gamma^{\lambda}\psi_{\lambda\mu_2\cdots\mu_n)}=0
\end{align}
where we will call $\mathcal{S}_{\mu_1\cdots\mu_n}$ the Fang field strength. $\mathcal{S}_{\mu_1\cdots\mu_n}$ satisfies a fermionic analog of the Bianchi identity (\ref{eq:mless2ndBianchi})
\begin{align}
\label{eq:mlesshalf2ndBianchi}
\gamma^{\lambda}\fsl{\partial}\mathcal{S}_{\lambda\mu_1\cdots\mu_{n-1}}-(n-1)\partial_{(\mu_1}\mathcal{S}^{\lambda}_{\;\;\lambda\mu_2\cdots\mu_{n-1})}=0
\end{align}

\sm

As with Rarita-Schwinger theory, these equations of motion (\ref{eq:mlesshalfeom}) have a gauge redundancy. Under the gauge transformation
\begin{align}
    \psi_{\mu_1\cdots\mu_n}(X)' = \psi_{\mu_1\cdots\mu_n}(X) + 
    n\partial_{(\mu_1}\epsilon_{\mu_2\cdots\mu_n)}(X)
\end{align}
the Fang field strength is invariant provided that $\epsilon_{\mu_1\cdots\mu_{n-1}}$ is a symmetric and $\gamma$ traceless, but otherwise arbitrary function, i.e.
\begin{align}
    \delta\mathcal{S}_{\mu_1\cdots\mu_n}=-n(n-1)\partial_{(\mu_1}\partial_{\mu_2}\gamma^{\lambda}\epsilon_{\lambda\mu_3\cdots\mu_n)}=0
\end{align}

\sm

This gauge redundancy allows us to impose a gauge fixing condition. The natural choice is the fermionic analog of (\ref{eq:indexdeDonder})
\begin{align}
\label{eq:indexdeDirac}
\gamma^{\lambda}\fsl{\partial}\psi_{\lambda\mu_1\cdots\mu_{n-1}}-(n-1)\partial_{(\mu_1}\psi^{\lambda}_{\;\;\lambda\mu_2\cdots\mu_{n-1})}=0
\end{align}
In this gauge, the $\gamma$ trace of the equations of motion (\ref{eq:mlesshalfeom}) reduces to
\begin{align}
    \fsl{\partial}(\gamma^{\lambda}\psi_{\lambda\mu_1\cdots\mu_{n-1}})=0
\end{align}
Applying $\fsl{\partial}$ on to $\mathcal{S}_{\mu_1\cdots\mu_n}$ in this gauge then implies
\begin{align}
    \fsl{\partial}\mathcal{S}_{\mu_1\cdots\mu_n} = \partial^2\psi_{\mu_1\cdots\mu_n} = 0
\end{align}
and so the particles described by $\psi_{\mu_1\cdots\mu_n}$ are indeed massless. Just as in the bosonic case, in this gauge there remains a residual gauge symmetry under a gauge parameter $\epsilon_{\mu_1\cdots\mu_{n-1}}$ satisfying $\partial^2\epsilon_{\mu_1\cdots\mu_{n-1}}=0$. Finally, one should consider the constraints on $\psi_{\mu_1\cdots\mu_n}$ which are a result of the Bianchi identity (\ref{eq:mlesshalf2ndBianchi}). The following components of the gauge invariant field strength
\begin{align}
    \gamma^0\mathcal{S}_{0j_1\cdots j_{n-1}}-\gamma^{i}\mathcal{S}_{i j_1\cdots j_{n-1}}
\end{align}
contain only spatial derivatives. In contrast to the bosonic case, these remain as constraints in the gauge (\ref{eq:indexdeDirac}) we picked. In total then, the number of degrees of freedom removed from $\psi_{\mu_1\cdots\mu_n}$ is thrice that of $\epsilon_{\mu_1\cdots\mu_{n-1}}$. The final degrees of freedom $N_0(d,n+1/2)$ of this field in $d$ dimensions is then
\sm
\begin{align}
    N_0(d,n+1/2) =& \,2^{\floor{d/2}}{d+n-1\choose n}-2^{\floor{d/2}}{d+n-4\choose n-3}\nonumber\\
    &-3\Bigg(2^{\floor{d/2}}{d+n-2\choose n-1}-2^{\floor{d/2}}{d+n-3\choose n-2}\Bigg) \nonumber\\
    =& \,2^{\floor{d/2}} {d+n-4\choose n}
\end{align}
which reduces to 4, corresponding to the 2 helicities of a massless spin $n+1/2$ particle and anti-particle when $d=4$.

\sm

Again, in order to construct a gauge invariant, linear in derivatives action for $\psi_{\mu_1\cdots\mu_n}$, one needs a object $\mathcal{T}_{\mu_1\cdots\mu_n}$ that is linear in $\mathcal{S}_{\mu_1\cdots\mu_n}$ whose divergence is a pure $\gamma$ trace. The appropriate object is $\mathcal{T}_{\mu_1\cdots\mu_n} = \mathcal{S}_{\mu_1\cdots\mu_n}-\frac{1}{2}n\gamma_{(\mu_1}\gamma^{\lambda}\mathcal{S}_{\lambda\mu_2\cdots\mu_n)}-\frac{1}{4}n(n-1)\eta_{(\mu_1\mu_2}\mathcal{S}^{\lambda}_{\;\;\lambda\mu_3\cdots\mu_n)}$, and its divergence is
\begin{align}
\label{eq:halfbianchi}
\partial^{\lambda}\mathcal{T}_{\lambda\mu_2\cdots\mu_n} = -\frac{1}{2}n(n-1)\gamma_{(\mu_2}\gamma^{\lambda}\partial^{\omega}\mathcal{S}_{\omega\lambda\mu_3\cdots\mu_n)}-\frac{1}{4}n(n-1)(n-2)\eta_{(\mu_2\mu_3}\partial^{\lambda}\mathcal{S}^{\omega}_{\;\;\omega\lambda\mu_4\cdots\mu_n)}
\end{align}
This is a direct consequence of (\ref{eq:mlesshalf2ndBianchi}). Using this, one can construct a gauge invariant action that generates the equations of motion (\ref{eq:mlesshalfeom}), which we will write in terms of $\mathcal{S}_{\mu_1\cdots\mu_n}$
\begin{align}
\label{eq:mlesshalfindexaction}
    S_{n+1/2} = -\int d^dX\overline{\psi}^{\mu_1\cdots\mu_n}\Big(\mathcal{S}_{\mu_1\cdots\mu_n}-\frac{1}{2}n\gamma_{(\mu_1}\gamma^{\lambda}\mathcal{S}_{\lambda\mu_2\cdots\mu_n)}-\frac{1}{4}n(n-1)\eta_{(\mu_1\mu_2}\mathcal{S}^{\lambda}_{\;\;\lambda\mu_3\cdots\mu_n)}\Big)
\end{align}
where $\overline{\psi}^{\mu_1\cdots\mu_n} = (\psi^{\mu_1\cdots\mu_n})^\dagger i \gamma^0$. The equations of motion derived from this action are therefore
\begin{align}
    \mathcal{T}_{\mu_1\cdots\mu_n} =  \mathcal{S}_{\mu_1\cdots\mu_n}-\frac{1}{2}n\gamma_{(\mu_1}\gamma^{\lambda}\mathcal{S}_{\lambda\mu_2\cdots\mu_n)}-\frac{1}{4}n(n-1)\eta_{(\mu_1\mu_2}\mathcal{S}^{\lambda}_{\;\;\lambda\mu_3\cdots\mu_n)} = 0
\end{align}
which is equivalent to (\ref{eq:mlesshalfeom}).

\subsection{Transition to hyperfields}
We now transition to hyperfields, listing some of the various formulas above in terms of $\Psi_n(X,s)$.
\begin{align} &\gamma^{\omega}\psi^{\lambda}_{\;\;\lambda\omega\mu_4\cdots\mu_n} =0 \qquad && \longrightarrow \qquad && (\fsl{\partial}_s)^3\Psi_n(X,s) = 0
\\
\,\nonumber
\\
&\delta\psi_{\mu_1\cdots\mu_n} = n\partial_{(\mu_1}\epsilon_{\mu_2\cdots\mu_n)} \qquad && \longrightarrow \qquad && \delta\Psi_n(X,s) = i^{-1/2} s\cdot\partial_X\epsilon_{n-1}(X,s)
\\
\,\nonumber
\\
&\gamma^{\lambda}\epsilon_{\lambda\mu_2\cdots\mu_{n-1}} = 0 \qquad  &&\longrightarrow \qquad && \fsl{\partial}_s\epsilon_{n-1}(X,s) = 0
\end{align}
\begin{align}
    \mathcal{S}_{\mu_1\cdots\mu_n} = \fsl{\partial}\psi_{\mu_1\cdots\mu_n}-& n\partial_{(\mu_1}\gamma^{\lambda}\psi_{\lambda\mu_2\cdots\mu_n)}=0 \nonumber
    \\
    &\downarrow
    \\
    \mathcal{S}_n(X,s) = \fsl{\partial}_X\Psi_n(X,s) &-  s\cdot\partial_X\fsl{\partial}_s\Psi_n(X,s) = 0 \nonumber
\end{align}
The action for $\Psi_n(X,s)$ then takes the form
\begin{align}
\label{eq:mlesshalfaction}
    S_{n+1/2} &= -n!\int d^dX\frac{d^dsd^ds'}{(2\pi)^d}e^{is\cdot s'}\overline{\Psi}_n(X,s)\big(1-\frac{1}{2}\fsl{s'}\fsl{\partial}_{s'}-\frac{1}{4}s'^2\partial^2_{s'}\big)\mathcal{S}_n(X,s') 
\end{align}
where $\overline{\Psi}_n(X,s) = \frac{1}{n!}i^{-n/2}\overline{\psi}_{\mu_1\cdots\mu_n}s^{\mu_1}\cdots s^{\mu_n}$. Gauge invariance of (\ref{eq:mlesshalfaction}) is guaranteed by the identity
\begin{align}
\label{eq:hyperhalfbianchi}
    (\fsl{\partial}_s\fsl{\partial}_X-s\cdot\partial_X\partial_s^2)\mathcal{S}_n(X,s) = 0
\end{align}
\subsection{All propagators}
We now compute the propagator $\langle\Psi_n(X,s)\overline{\Psi}_n(Y,t)\rangle$ from (\ref{eq:mlesshalfaction}). The discussion in \autoref{sec:mlessprop} on the gauge ambiguity of the propagator equally applies in this case. The propagator $\langle\Psi_n(X,s)\overline{\Psi}_n(Y,t)\rangle$ will solve the equation
\begin{align}
\label{eq:mlesshalfpropeom}
    \big(1-\frac{1}{2}\fsl{s}\fsl{\partial}_s-\frac{1}{4}s^2\partial_s^2\big)&\big(\fsl{\partial}_X-s\cdot\partial_X\fsl{\partial}_s\big)\langle\Psi_n(X,s)\overline{\Psi}_n(Y,t)\rangle \nonumber\\
    &= -\frac{1}{n!}i\delta^d(X-Y)P_{n,\gamma^3}^d(s,t)+i^{-1/2}t\cdot\partial_Y\Omega(X-Y,s,t)
\end{align}
where $P_{n,\gamma^3}^d(s,t)$ is the projection operator onto rank $n$ triple $\gamma$ traceless hyperfields, which is
\begin{align}
    P_{n,\gamma^3}^d(s,t) = \frac{1}{(\frac{d}{2})_n}\Big(-\frac{i}{2}\sqrt{s^2t^2}\Big)^n \Bigg(C_n^{\frac{d}{2}}\Big(\frac{s\cdot t}{\sqrt{s^2t^2}}\Big)+\Big(\frac{d}{2}+n-2\Big)C_{n-2}^{\frac{d}{2}}\Big(\frac{s\cdot t}{\sqrt{s^2t^2}}\Big)\Bigg) \nonumber\\
    +\; i\frac{\fsl{s}\fsl{t}}{2}\frac{1}{(\frac{d}{2})_{n-1}}\Big(-\frac{i}{2}\sqrt{s^2t^2}\Big)^{n-1} C_{n-3}^{\frac{d}{2}}\Big(\frac{s\cdot t}{\sqrt{s^2t^2}}\Big)
\end{align}
and $\Omega(X-Y,s,t)$ is some function which is a rank $n$ triple $\gamma$ traceless hyperfield in $s$ and a rank $n-1$ $\gamma$ traceless hyperfield in $t$ $(\fsl{\partial}_s)^3\Omega ,\, \Omega \overleftarrow{\fsl{\partial}}_t=0$. Because of this, the propagator is ambiguous up to transformations
\begin{align}
\label{eq:mlesshalfpropamb}
    \delta\langle\Psi_n(X,s)\overline{\Psi}_n(Y,t)\rangle =  i^{-1/2}s\cdot\partial_X\Pi(X-Y,s,t) + i^{-1/2}t\cdot\partial_Y\Omega(X-Y,s,t)
\end{align}

\sm

To find the propagator, one might proceed as in the integer spin case by gauge fixing $\Psi_n(X,s)$. A natural gauge to pick is (\ref{eq:indexdeDirac})
\begin{align}
\label{eq:deDirac}
    (\fsl{\partial}_s\fsl{\partial}_X-s\cdot\partial_X\partial_s^2)\Psi_n(X,s) = 0
\end{align}
In the case of half integer spins however, this choice does not simplify the analysis, and there does not appear to be a gauge choice which simplifies the analysis in an analogous manner to the gauge choice for integer spins (\ref{eq:deDonder}), so we instead proceed directly by finding the simplest solution to (\ref{eq:mlesshalfpropeom}). Working in momentum space, one can classify all possible structures which have a simple pole at $p^2=0$, have a numerator linear in momentum, and are not of the form (\ref{eq:mlesshalfpropamb}), which are triple $\gamma$ traceless in both $s$ and $t$, and make an ansatz that $\langle\Psi_n(X,s)\overline{\Psi}_n(Y,t)\rangle$ is a linear combination of these in momentum space. These structures can be obtained from the three basic structures
\begin{align}
    \frac{\fsl{p}}{p^2}(s\cdot t)^n,\quad
    \frac{\fsl{s}\,\fsl{p}\,\fsl{t}}{p^2}(s\cdot t)^n,\quad
    s^2t^2\frac{\fsl{p}}{p^2}(s\cdot t)^n
\end{align}
and applying various combinations of $P_{n,\gamma}^d$, $P_{n,\eta}^d$, and $P_{n,\gamma^3}^d$ on either side. A tedious but straighforward calculation reveals that this ansatz is sufficient and the propagator can be taken to be
\begin{align}
\label{eq:mlesshalfprop}
    \langle\Psi_n(X,s)\overline{\Psi}_n(Y,t)\rangle &=  \Delta_0(X-Y)\frac{1}{n!}\frac{1}{\big(\frac{d-2}{2}\big)_n}\Big(-\frac{i}{2}\sqrt{s^2t^2}\Big)^nC_{n}^{\frac{d-2}{2}}\Big(\frac{s\cdot t}{\sqrt{s^2t^2}}\Big) \nonumber \\
    &-\frac{i}{2}\fsl{s}\,\Delta_0(X-Y)\,\fsl{t}\frac{1}{n!}\frac{1}{\big(\frac{d-2}{2}\big)_{n}}\Big(-\frac{i}{2}\sqrt{s^2t^2}\Big)^{n-1}C_{n-1}^{\frac{d-2}{2}}\Big(\frac{s\cdot t}{\sqrt{s^2t^2}}\Big)
\end{align}
where $\Delta_0(X-Y)$ is the standard massless spin $1/2$ propagator. This propagator agrees with previous work \cite{Francia:2007qt}. As in the integer spin case, the massless spin $n+1/2$ propagators $\langle\psi_{\mu_1\cdots\mu_n}(X)\overline{\psi}_{\nu_1\cdots\nu_n}(Y)\rangle$ from (\ref{eq:mlesshalfindexaction}) can be obtained from $\langle\Psi_n(X,s)\overline{\Psi}_n(Y,t)\rangle$ straightforwardly by applying derivatives
\begin{align}
    \langle\psi_{\mu_1\cdots\mu_n}(X)\overline{\psi}_{\nu_1\cdots\nu_n}(Y)\rangle = i^n\frac{\partial^n}{\partial_s^{\mu_1}\cdots\partial_s^{\mu_n}}\frac{\partial^n}{\partial_t^{\nu_1}\cdots\partial_t^{\nu_n}}\langle\Psi_n(X,s)\overline{\Psi}_n(Y,t)\rangle
\end{align}
We list here the numerator of the first few propagators in momentum space
\begin{align}
    & n + 1/2 = 1/2: && \fsl{p} \\
    & n + 1/2 = 3/2: && \fsl{p}\,\eta_{\mu_1\nu_1}+\frac{\gamma_{\mu_1}\,\fsl{p}\,\gamma_{\nu_1}}{d-2} \\ 
    & n + 1/2 = 5/2: && \fsl{p}\,\Big(\frac{1}{2}\big(\eta_{\mu_1\nu_1}\eta_{\nu_2\mu_2}+\eta_{\mu_1\nu_2}\eta_{\nu_1\mu_2}\big)-\frac{1}{d}\eta_{\mu_1\mu_2}\eta_{\nu_1\nu_2}\Big) \nonumber \\
    & && +\frac{1}{2d}\big(\eta_{\mu_1\nu_2}\gamma_{\mu_2}\,\fsl{p}\,\gamma_{\nu_1}+\eta_{\mu_2\nu_1}\gamma_{\mu_1}\,\fsl{p}\,\gamma_{\nu_2}+\eta_{\mu_1\nu_1}\gamma_{\mu_2}\,\fsl{p}\,\gamma_{\nu_2}+\eta_{\mu_2\nu_2}\gamma_{\mu_1}\,\fsl{p}\,\gamma_{\nu_1}\big)
\end{align}
\begin{align}
 & n + 1/2 = 7/2: && \fsl{p}\Big(\frac{1}{6}\big(\eta_{\mu_1\nu_1}\eta_{\nu_2\mu_2}\eta_{\mu_3\nu_3}+\eta_{\mu_1\nu_1}\eta_{\nu_2\mu_3}\eta_{\mu_2\nu_3}+\eta_{\mu_2\nu_1}\eta_{\nu_2\mu_3}\eta_{\mu_1\nu_3} \nonumber \\
    & && \phantom{+}+ \eta_{\mu_1\nu_2}\eta_{\nu_1\mu_2}\eta_{\mu_3\nu_3}+\eta_{\mu_1\nu_2}\eta_{\nu_1\mu_3}\eta_{\mu_2\nu_3}+\eta_{\mu_2\nu_2}\eta_{\nu_1\mu_3}\eta_{\mu_1\nu_3}\big) \nonumber \\
    & &&-\frac{1}{3(d+2)}\big(\eta_{\mu_1\mu_2}\eta_{\mu_3\nu_1}\eta_{\nu_2\nu_3}+\eta_{\mu_1\mu_2}\eta_{\mu_3\nu_2}\eta_{\nu_3\nu_1}+\eta_{\mu_1\mu_2}\eta_{\mu_3\nu_3}\eta_{\nu_1\nu_2} \nonumber\\    & &&\hspace{2 cm}+\eta_{\mu_2\mu_3}\eta_{\mu_1\nu_1}\eta_{\nu_2\nu_3}+\eta_{\mu_2\mu_3}\eta_{\mu_1\nu_2}\eta_{\nu_3\nu_1}+\eta_{\mu_2\mu_3}\eta_{\mu_1\nu_3}\eta_{\nu_1\nu_2} \nonumber\\    & &&\hspace{2 cm}+\eta_{\mu_3\mu_1}\eta_{\mu_2\nu_1}\eta_{\nu_2\nu_3}+\eta_{\mu_3\mu_1}\eta_{\mu_2\nu_2}\eta_{\nu_3\nu_1}+\eta_{\mu_3\mu_1}\eta_{\mu_2\nu_3}\eta_{\nu_1\nu_2}\big)\Big) \nonumber \\
    & && +\frac{1}{3(d+2)}\Big(\big(\frac{1}{2}(\eta_{\mu_2\nu_2}\eta_{\mu_3\nu_3}+\eta_{\mu_2\nu_3}\eta_{\mu_3\nu_2}) -\frac{1}{d}\eta_{\mu_2\mu_3}\eta_{\nu_2\nu_3}\big)\gamma_{\mu_1}\,\fsl{p}\,\gamma_{\nu_1} \nonumber\\
    & && \hspace{2.25 cm}+ \big(\frac{1}{2}(\eta_{\mu_2\nu_1}\eta_{\mu_3\nu_3}+\eta_{\mu_3\nu_1}\eta_{\mu_2\nu_3})-\frac{1}{d}\eta_{\mu_2\mu_3}\eta_{\nu_1\nu_3}\big)\gamma_{\mu_1}\,\fsl{p}\,\gamma_{\nu_2} \nonumber\\
    & && \hspace{2.25 cm}+ \big(\frac{1}{2}(\eta_{\mu_1\nu_2}\eta_{\mu_3\nu_3}+\eta_{\mu_3\nu_2}\eta_{\mu_1\nu_3})-\frac{1}{d}\eta_{\mu_1\mu_3}\eta_{\nu_2\nu_3}\big)\gamma_{\mu_2}\,\fsl{p}\,\gamma_{\nu_1}\nonumber\\
    & && \hspace{2.25 cm}+ \big(\frac{1}{2}(\eta_{\mu_2\nu_1}\eta_{\mu_3\nu_2}+\eta_{\mu_3\nu_1}\eta_{\mu_2\nu_2})-\frac{1}{d}\eta_{\mu_2\mu_3}\eta_{\nu_1\nu_2}\big)\gamma_{\mu_1}\,\fsl{p}\,\gamma_{\nu_3}\nonumber\\
    & && \hspace{2.25 cm}+ \big(\frac{1}{2}(\eta_{\mu_1\nu_2}\eta_{\mu_2\nu_3}+\eta_{\mu_2\nu_2}\eta_{\mu_1\nu_3})-\frac{1}{d}\eta_{\mu_1\mu_2}\eta_{\nu_2\nu_3}\big)\gamma_{\mu_3}\,\fsl{p}\,\gamma_{\nu_1}\nonumber\\
    & && \hspace{2.25 cm}+ \big(\frac{1}{2}(\eta_{\mu_1\nu_1}\eta_{\mu_3\nu_3}+\eta_{\mu_3\nu_1}\eta_{\mu_1\nu_3})-\frac{1}{d}\eta_{\mu_1\mu_3}\eta_{\nu_1\nu_3}\big)\gamma_{\mu_2}\,\fsl{p}\,\gamma_{\nu_2}\nonumber\\
    & && \hspace{2.25 cm}+ \big(\frac{1}{2}(\eta_{\mu_1\nu_1}\eta_{\mu_3\nu_2}+\eta_{\mu_3\nu_1}\eta_{\mu_1\nu_2})-\frac{1}{d}\eta_{\mu_1\mu_3}\eta_{\nu_1\nu_2}\big)\gamma_{\mu_2}\,\fsl{p}\,\gamma_{\nu_3}\nonumber\\
    & && \hspace{2.25 cm}+ \big(\frac{1}{2}(\eta_{\mu_1\nu_1}\eta_{\mu_2\nu_3}+\eta_{\mu_2\nu_1}\eta_{\mu_1\nu_3})-\frac{1}{d}\eta_{\mu_1\mu_2}\eta_{\nu_1\nu_3}\big)\gamma_{\mu_3}\,\fsl{p}\,\gamma_{\nu_2}\nonumber\\
    & && \hspace{2.25 cm}+ \big(\frac{1}{2}(\eta_{\mu_1\nu_1}\eta_{\mu_2\nu_2}+\eta_{\mu_2\nu_1}\eta_{\mu_1\nu_2})-\frac{1}{d}\eta_{\mu_1\mu_2}\eta_{\nu_1\nu_2}\big)\gamma_{\mu_3}\,\fsl{p}\,\gamma_{\nu_3}\Big)
\end{align}

As before, this propagator has a well defined large $n$ asymptotic limit
\begin{align}
    &\langle \Psi_n(X,s)\overline{\Psi}_n(Y,t) \rangle \sim \nonumber \\
    &\Big(\Delta_0(X-Y) + \frac{s\cdot t}{s^2t^2}\,\fsl{s}\,\Delta_0(X-Y)\,\fsl{t}\Big)\frac{1}{2^n(n!)^2}(-i\sqrt{s^2t^2})^n\frac{2\cos\big((n+\frac{d-2}{2})\theta-\frac{\pi}{2}\frac{d-2}{2}\big)}{(2\sin\theta)^{\frac{d-2}{2}}} \nonumber \\
    &+\frac{\sqrt{s^2t^2-(s\cdot t)^2}}{s^2t^2}\,\fsl{s}\,\Delta_0(X-Y)\,\fsl{t}\frac{1}{2^n(n!)^2}(-i\sqrt{s^2t^2})^n\frac{2\sin\big((n+\frac{d-2}{2})\theta-\frac{\pi}{2}\frac{d-2}{2}\big)}{(2\sin\theta)^{\frac{d-2}{2}}}
\end{align}
as $n\to\infty$, where $\cos\theta = s\cdot t/\sqrt{s^2t^2}$, for $0 < \theta < \pi$.
\subsection{Other gauges}
Just as for the integer spin propagator, it may be useful to perform computations in different gauges as a consistency check. Using the gauge transformation (\ref{eq:mlesshalfpropamb}), possible terms one may add to the propagator (\ref{eq:mlesshalfprop}) in momentum space include
\begin{align}
   \frac{\xi(p^2)}{p^2}\frac{1}{(\frac{d-\alpha}{2})_{n-1}}\frac{1}{n!} &\Bigg((s\cdot p\,\tilde{\fsl{t}} + t\cdot p \,\tilde{\fsl{s}})\Big(\frac{1}{2}\sqrt{s\cdot A\cdot s\, t\cdot A\cdot t}\Big)^{n-1}C_{n-1}^{\frac{d-\alpha}{2}}\Big(\frac{s\cdot A\cdot t}{\sqrt{s\cdot A\cdot s\, t\cdot A\cdot t}}\Big)  \nonumber \\
   &-\frac{1}{2}\tilde{\fsl{s}}(s\cdot p\,\tilde{\fsl{t}} + t\cdot p \,\tilde{\fsl{s}})\tilde{t}\Big(\frac{1}{2}\sqrt{s\cdot A\cdot s\, t\cdot A\cdot t}\Big)^{n-2}C_{n-2}^{\frac{d-\alpha}{2}}\Big(\frac{s\cdot A\cdot t}{\sqrt{s\cdot A\cdot s\, t\cdot A\cdot t}}\Big) \Bigg) 
\end{align}
where $\xi(p^2)$ is an arbitrary function, $A_{\mu\nu}(p)$ is a projection matrix of rank $d-\alpha$, and $\tilde{\fsl{s}}$ for instance stands for contracting $s^{\mu}$ with $A_{\mu\nu}\gamma^{\nu}$. 

\sm

The propagator (\ref{eq:mlesshalfprop}) is already in what may be thought of as the fermionic Bianchi gauge, in the sense that in momentum space
\begin{align}
    (\fsl{\partial}_s\fsl{p}-s\cdot p\partial_s^2)\langle \Psi_n(p,s)\overline{\Psi}_n(-p,t) \rangle =0
\end{align}

\section{Massive integer spins}
\label{sec5}
In this section, we will use the hyperfield formalism of \autoref{sec2} to study massive integer spin fields. We will find compact, closed form expressions for their covariant actions and propagators.

\sm

 Historically it was realized that the following system of equations is sufficient for describing a freely propagating spin $n$ massive particle with positive definite energy \cite{Fierz:1939zz,Fierz:1939ix}
\begin{align}
\label{eq:Klein}
    (\partial^2-m^2)\phi_{\mu_1\cdots\mu_n} = 0 \\
\label{eq:transverse}
    \partial^{\lambda}\phi_{\lambda\mu_2\cdots\mu_{n}} = 0 \\
\label{eq:traceless}
    \phi^\lambda_{\;\;\lambda\mu_3\cdots\mu_n} = 0
\end{align}
for a symmetric rank $n$ tensor. We will refer to the system of equations (\ref{eq:Klein})--(\ref{eq:traceless}) as a Fierz-Pauli system. The transverse and traceless conditions ensure that the correct number $N_m(d,n)$ of degrees of freedom for a spin $n$ particle in $d$ spacetime dimensions propagate with mass $m$ 
\begin{align}
\label{eq:mN}
    N_m(d,n) &= {d + n - 1\choose n} - {d + n - 3\choose n - 2} - \Bigg({d + n - 2\choose n - 1} - {d + n - 4\choose n - 3}\Bigg) \nonumber \\
    &= \frac{2n+d-3}{d-3} {d+n-4\choose n}
\end{align}
which reduces to $2n+1$ when $d=4$. Attempts to construct a covariant action whose equations of motion imposes (\ref{eq:Klein})--(\ref{eq:traceless}) was met with considerable difficulty for over three decades. It was noted in \cite{Fierz:1939ix} that it is impossible to construct an action principle which produces a Fierz-Pauli system out of just a symmetric and traceless field $\phi_{\mu_1\cdots\mu_n}$. Additional auxiliary fields are in general necessary to impose the transverse constraint (\ref{eq:transverse}). How these auxiliary fields are introduced is not unique, but there is a ``minimal" set of auxiliary fields necessary in this construction. 

\sm

A minimal covariant Lagrangian formulation of massive integer spin fields $\phi_{\mu_1\cdots\mu_n}$ in four spacetime dimensions was first worked out in \cite{Singh:1974qz}. In it, Singh and Hagen needed to introduce, apart from the symmetric and traceless rank $n$ tensor, symmetric and traceless tensors of rank $0,1,2,\dots,n-2$ which intricately couple so that all fields of rank lower than $n$ vanish on shell, and the rank $n$ tensor made up a Fierz-Pauli system. Note that this field content is equivalent to using symmetric rank $n$ and $n-3$ tensors, with no tracelessness conditions. 

\sm

Because this formulation only works in four spacetime dimensions, we will not present the action, and will opt in the following subsection instead to construct a formulation which works in any dimension. What will continue to be same is the minimal number of auxiliary fields needed in the action. Interestingly, the massless limit of \cite{Singh:1974qz} decouples all fields of rank $0,\dots,n-3$, enabling one to consider only the rank $n$ and $n-2$ fields. These two fields can be combined into a single, double traceless rank $n$ field $\phi_{\mu_1\cdots\mu_n}$, resulting in precisely Fronsdal's formulation \cite{Fronsdal:1978rb}, which \textit{is} valid for any dimension.

\subsection{Massive particles from dimensional reduction}
\label{sec:dimredint}

Instead of constructing a covariant action for a massive spin $n$ from the ground up, we note as in \cite{Aragone:1987dtt} that one reliable way of getting a massive particle of mass $m$ in $d$ dimensions is by starting with a massless particle in $d+1$ dimensions, and compactifying one of the spatial directions to be a circle of radius $2\pi/m$. Compactifying in this way will generate an infinite tower of massive particles, corresponding to different windings around the circle. If the $d+1$ dimensional theory is free as in our case, the particles with varying masses do not interact, and we may freely throw away all particles except for the one with mass $m$ without any problems. This procedure is different from compactification, and so instead we refer to it as ``dimensional reduction". 

\sm

As a simple check that dimensional reduction of a massless spin $n$ particle in $d+1$ dimensions yields a massive spin $n$ particle in $d$ dimensions, we note that this procedure is equivalent to fixing the magnitude of one component of the massless particle's momentum, and that $N_0(d+1,n)=N_m(d,n)$. The degrees of freedom of a dimensionally reduced $d+1$ dimensional massless spin $n$ particle is therefore the same as that of two $d$ dimensional massive spin $n$ particles, one for each sign of the fixed momentum component.

\sm

The benefit of this starting point is that it is valid for any spacetime dimension $d$, making it possible to easily write down results as a function of $d$. Furthermore, the apparent non-uniqueness of the massive description is manifestly understood to arise from the $d+1$ dimensional massless gauge symmetry that is inherited. At the level of the $d$ dimensional massive theory, the gauge symmetry presents itself as a St{\"u}ckelberg gauge symmetry, and different gauges amount to a different choice of auxiliary fields.

\sm

We therefore start with the massless hyperfield action (\ref{eq:mlesshyperaction}) in $d+1$ dimensions, with coordinates $(X^{\mu},X_d)$ and an auxiliary $d+1$ dimensional vector $(s^{\mu},s_d)$ for $\mu = 0,\dots,d-1$, and enforce that the real hyperfield $\Phi_n(X,X_d,s,s_d)$ has the following dependence on the $X_d$ coordinate
\begin{align}
    \Phi_n(X,X_d,s,s_d) = e^{imX_d}\Phi_m(X,s,s_d) + e^{-imX_d}\tilde{\Phi}_m(X,s,s_d)
\end{align}
where $\Phi_m(X,s,s_d)$ will generate the fields necessary to covariantly describe a massive spin $n$ particle in $d$ dimensions. The additional auxiliary component $s_d$ has the effect of generating $n+1$ $d$ dimensional hyperfields
\begin{align}
\label{eq:mdindependent}
    \Phi_m(X,s,s_d) = \sum_{k=0}^{n}\frac{1}{k!}i^{-k/2}(s_d)^k\Phi_{n-k}(X,s)
\end{align}
At the level of an individual $d+1$ dimensional massless rank $n$ field $\phi_{\mu_1\cdots\mu_n}$, this is the same as decomposing $\phi_{\mu_1\cdots\mu_n}$ into its $d$ dimensional tensors $\phi_{\mu_1\cdots\mu_n}$, $\phi_{d\mu_1\cdots\mu_{n-1}}$, $\dots,\,\phi_{dd\cdots d}$. The $d$ dimensional hyperfields $\Phi_{n-k}(X,s)$ are not all independent of each other, because they descend from a $d+1$ dimensional massless hyperfield which satisfies the $d+1$ dimensional double traceless condition $(\partial_s^2+\partial_d^2)^2\Phi_n=0$, where $\partial_d$ is a derivative with respect to $s_d$. All $\Phi_{n-k}$ with $k>3$ can be written in terms of $\Phi_{n}$, $\Phi_{n-1}$, $\Phi_{n-2}$, and $\Phi_{n-3}$
\begin{align}
    \Phi_m(X,s,s_d) &= \sum_{k=0}^{\floor{n/2}}\frac{(-1)^k}{(2k)!}(s_d)^{2k}\Big((1-k)\partial_s^{2k}\Phi_{n} + ik \partial_s^{2(k-1)}\Phi_{n-2}\Big) \nonumber\\
    &+ i^{-1/2}\sum_{k=0}^{\floor{(n-1)/2}}\frac{(-1)^k}{(2k+1)!}(s_d)^{2k+1}\Big((1-k)\partial_s^{2k}\Phi_{n-1} + ik \partial_s^{2(k-1)}\Phi_{n-3}\Big)
\end{align}
where $\Phi_n,\dots,\Phi_{n-3}$ are unconstrained $d$ dimensional hyperfields. The gauge symmetry of the massless theory acts on $\Phi_m(X,s,s_d)$ via 
\begin{align}
    \delta\Phi_m = i^{-1/2}s\cdot\partial_X\epsilon_m(X,s,s_d) + i^{1/2}s_d m\epsilon_m(X,s,s_d)
\end{align}
for some $d+1$ dimensional traceless hyperfield gauge parameter $\epsilon_m(X,s,s_d)$, which may be written in terms of its independent, unconstrained, $d$ dimensional hyperfield components $\epsilon_{n-1}(X,s)$ and $\epsilon_{n-2}(X,s)$
\begin{align}
    \epsilon_m(X,s,s_d) = \sum_{k=0}^{\floor{(n-1)/2}}\frac{(-1)^k}{(2k)!}(s_d)^{2k}\partial_s^{2k}\epsilon_{n-1} + i^{-1/2}\sum_{k=0}^{\floor{(n-2)/2}}\frac{(-1)^k}{(2k+1)!}(s_d)^{2k+1}\partial_s^{2k}\epsilon_{n-2}
\end{align}
The St{\"u}ckelberg gauge symmetry may then be expressed in terms of the gauge parameters $\epsilon_{n-1}$ and $\epsilon_{n-2}$, acting on $\Phi_n,\dots,\Phi_{n-3}$
\begin{align}
\label{eq:Stueck}
    &\delta\Phi_n = i^{-1/2} s\cdot\partial_X \epsilon_{n-1} && \delta\Phi_{n-1} = i^{-1/2} s\cdot \partial_X \epsilon_{n-2} + im\,\epsilon_{n-1} \nonumber\\
    &\delta\Phi_{n-2} = -i^{1/2} s\cdot\partial_X\partial_s^2\epsilon_{n-1} +2im\,\epsilon_{n-2} && \delta\Phi_{n-3} = -i^{1/2} s\cdot\partial_X\partial_s^2\epsilon_{n-2} +3m\partial_s^2\epsilon_{n-1}
\end{align}

The $d$ dimensional massive action is written in terms of $\Phi_n,\dots,\Phi_{n-3}$, and the independent components $\mathcal{F}_n,\dots,\mathcal{F}_{n-3}$ of the $d+1$ dimensional double traceless Fronsdal field strength $\mathcal{F}_n(X,X_d,s,s_d)$ 
\begin{align}
         S_n = \frac{1}{2}n!\int d^{d+1}X\frac{d^{d+1}sd^{d+1}s'}{(2\pi)^{d+1}}e^{is\cdot s'} \Phi_n(X,s)\big(1-\frac{1}{4}s'^2\partial_{s'}^2\big)\mathcal{F}_n(X,s') \nonumber
\end{align}
\begin{align}
\label{eq:mhyperaction}
     S_n = n!\int& d^dX\frac{d^dsd^ds'}{(2\pi)^d}e^{is\cdot s'}\times \nonumber\\
     \Bigg\{ &\sum_{k = 0}^{\floor{n/2}} \frac{(-1)^k}{(2k)!}\Bigg(&&\hspace{-1.25cm}\Big(1-\frac{3k}{2}\Big)\partial_{s}^{2k}\tilde{\Phi}_n\partial_{s'}^{2k}\mathcal{F}_n+i\frac{k}{2}\partial_{s}^{2k}\tilde{\Phi}_n\partial_{s'}^{2(k-1)}\mathcal{F}_{n-2} \nonumber\\
    & && \hspace{-.75cm}+i\frac{k}{2}\partial_{s}^{2(k-1)}\tilde{\Phi}_{n-2}\partial_{s'}^{2k}\mathcal{F}_n -\frac{k}{2}\partial_{s}^{2(k-1)}\tilde{\Phi}_{n-2}\partial_{s'}^{2(k-1)}\mathcal{F}_{n-2}\Bigg) \nonumber\\
    & \hspace{-.5 cm}+ \sum_{k = 0}^{\floor{(n-1)/2}} \frac{(-1)^k}{(2k+1)!}\Bigg(&&\hspace{-.25cm}\Big(1-\frac{5k}{2}\Big)\partial_{s}^{2k}\tilde{\Phi}_{n-1}\partial_{s'}^{2k}\mathcal{F}_{n-1}+i\frac{3k}{2}\partial_{s}^{2k}\tilde{\Phi}_{n-1}\partial_{s'}^{2(k-1)}\mathcal{F}_{n-3} \nonumber \\
    & &&\hspace{-.4cm}+i\frac{3k}{2}\partial_{s}^{2(k-1)}\tilde{\Phi}_{n-3}\partial_{s'}^{2k}\mathcal{F}_{n-1} +\frac{k}{2}\partial_{s}^{2(k-1)}\tilde{\Phi}_{n-3}\partial_{s'}^{2(k-1)}\mathcal{F}_{n-3}\Bigg)\Bigg\}
\end{align}
where
\begin{align}
    \mathcal{F}_n =& \big(\partial_X^2-m^2-s\cdot\partial_X\partial_s\cdot\partial_X+\frac{1}{2}(s\cdot\partial_X)^2\partial_s^2\big)\Phi_n-\frac{i}{2}(s\cdot\partial_X)^2\Phi_{n-2} - i^{1/2}m \,s\cdot\partial_X\Phi_{n-1} \\
    \mathcal{F}_{n-1} =& \big(\partial_X^2-s\cdot\partial_X\partial_s\cdot\partial_X+\frac{1}{2}(s\cdot\partial_X)^2\partial_s^2\big)\Phi_{n-1} - \frac{i}{2}(s\cdot\partial_X)^2\Phi_{n-3}\nonumber \\
    & + i^{-1/2}m\partial_s\cdot\partial_X\Phi_n - i^{-1/2} m\, s\cdot\partial_X\partial_s^2\Phi_n \\
    \mathcal{F}_{n-2} =& \big(\partial_X^2-s\cdot\partial_X\partial_s\cdot\partial_X-\frac{1}{2}(s\cdot\partial_X)^2\partial_s^2\big)\Phi_{n-2}- \frac{i}{2} (s\cdot\partial_X)^2\partial_s^4\Phi_n - im^2\partial_s^2\Phi_n \nonumber\\
    & + 2i^{-1/2}m\partial_s\cdot\partial_X\Phi_{n-1} - 2i^{-1/2} m\, s\cdot\partial_X\partial_s^2\Phi_{n-1} + i^{1/2}m\, s\cdot\partial_X\Phi_{n-3} \\
    \mathcal{F}_{n-3}=&\big(\partial_X^2-m^2-s\cdot\partial_X\partial_s\cdot\partial_X-\frac{1}{2}(s\cdot\partial_X)^2\partial_s^2\big)\Phi_{n-3} -\frac{i}{2}(s\cdot\partial_X)^2\partial_s^4\Phi_{n-1} - 3im^2\partial_s^2\Phi_{n-1} \nonumber \\
    & +2i^{1/2}m\, s\cdot\partial_X\partial_s^4\Phi_n +3i^{-1/2}m\partial_s\cdot\partial_X\Phi_{n-2} + i^{-1/2} m\, s\cdot\partial_X\partial_s^2\Phi_{n-2}
\end{align}
That (\ref{eq:mhyperaction}) is invariant under the gauge transformation (\ref{eq:Stueck}) is guaranteed by the fact that $\mathcal{F}_n,\dots,\mathcal{F}_{n-3}$ are themselves gauge invariant, and that they satisfy the identities
\begin{align}
    &\partial_s\cdot\partial_X\mathcal{F}_n-\frac{1}{2}s\cdot\partial_X\partial_s^2\mathcal{F}_n + \frac{i}{2}s\cdot\partial_X\mathcal{F}_{n-2} + i^{1/2}m\mathcal{F}_{n-1} = 0   \\
    &\partial_s\cdot\partial_X\mathcal{F}_{n-1} - \frac{1}{2}s\cdot\partial_X\partial_s^2\mathcal{F}_{n-1} + \frac{i}{2}s\cdot\partial_X\mathcal{F}_{n-3} +\frac{1}{2}i^{-1/2}m\partial_s^2\mathcal{F}_n +\frac{1}{2}i^{1/2}m\mathcal{F}_{n-2} = 0 
\end{align}
which is a direct consequence of (\ref{eq:hyperbianchi}).

\sm

The St{\"u}ckelberg gauge symmetry (\ref{eq:Stueck}) can be used to set for instance $\Phi_{n-1}$ and $\Phi_{n-2}$ to zero, leaving $\Phi_n$ and $\Phi_{n-3}$ unconstrained. This is precisely the minimal field content of \cite{Singh:1974qz}. Massive spin 0, 1, and 2 particles then do not require additional fields to describe them off-shell. It is only for spins higher than 2 that additional fields are necessary.

\sm

The action (\ref{eq:mhyperaction}) is for complex massive integer spin particles. For real massive integer spin particles, it is sufficient to impose the reality conditions $\tilde{\Phi}_n=\Phi_n$, $\tilde{\Phi}_{n-1}=-\Phi_{n-1}$, $\tilde{\Phi}_{n-2}=\Phi_{n-2}$, and $\tilde{\Phi}_{n-3}=-\Phi_{n-3}$.

\subsection{Equations of motion}
\label{sec:eqnint}
By virtue of the fact that the $d+1$ dimensional field strength $\mathcal{F}_n(X,X_d,s,s_d)$ is set to zero by the equations of motion, the $d$ dimensional massive equations of motion are
\begin{align}
\label{eq:meom}
    \mathcal{F}_n = 0,\quad \mathcal{F}_{n-1} = 0, \quad \mathcal{F}_{n-2} = 0,\quad \mathcal{F}_{n-3} = 0 
\end{align}
That these equations describe massive integer spin particles amounts to showing that they imply that $\Phi_{n-1}$, $\Phi_{n-2}$, and $\Phi_{n-3}$ can always be set to zero, and that $\Phi_n$ is a Fierz-Pauli system in hyperspace
\begin{align}
\label{eq:hyperFP}
    (\partial_X^2-m^2)\Phi_n &= 0 \nonumber\\
    \partial_s\cdot\partial_X\Phi_n &=0 \nonumber\\
    \partial_s^2\Phi_n &=0 
\end{align}

The gauge symmetry (\ref{eq:Stueck}) may be used to set $\Phi_{n-1}=0$ and $\Phi_{n-2}=0$ before imposing the equations of motion, so (\ref{eq:meom}) becomes
\begin{align}
\label{eq:F0}
    \mathcal{F}_n =& \big(\partial_X^2-m^2-s\cdot\partial_X\partial_s\cdot\partial_X+\frac{1}{2}(s\cdot\partial_X)^2\partial_s^2\big)\Phi_n = 0\\
\label{eq:F1}
    \mathcal{F}_{n-1} =& - \frac{i}{2}(s\cdot\partial_X)^2\Phi_{n-3} + i^{-1/2}m\partial_s\cdot\partial_X\Phi_n - i^{-1/2} m\, s\cdot\partial_X\partial_s^2\Phi_n = 0 \\
\label{eq:F2}
    \mathcal{F}_{n-2} =& - \frac{i}{2} (s\cdot\partial_X)^2\partial_s^4\Phi_n - im^2\partial_s^2\Phi_n + i^{1/2}m\, s\cdot\partial_X\Phi_{n-3} = 0 \\
\label{eq:F3}
    \mathcal{F}_{n-3}=&\big(\partial_X^2-m^2-s\cdot\partial_X\partial_s\cdot\partial_X-\frac{1}{2}(s\cdot\partial_X)^2\partial_s^2\big)\Phi_{n-3} +2i^{1/2}m\, s\cdot\partial_X\partial_s^4\Phi_n = 0
\end{align}
A quick inspection of these gauge fixed equations of motion shows that if $\Phi_{n-3}=0$, then $\Phi_n$ satisfies (\ref{eq:hyperFP}). What remains is to show that $\Phi_{n-3}$ may always be set to zero via the residual gauge symmetry of this system
\begin{align}
    \label{eq:mintresid}
    \delta\Phi_n = \frac{1}{m}(s\cdot\partial_X)^2\epsilon_{n-2},\quad \delta\Phi_{n-3} = 2i^{1/2}\partial_s\cdot\partial_X\epsilon_{n-2} + 2i^{1/2}\partial_s^2(s\cdot\partial_X\epsilon_{n-2})    
\end{align}
These transformations keep $\Phi_{n-1}=\Phi_{n-2}=0$ provided that $\epsilon_{n-2}$ satisfies\footnote{Note that this residual gauge symmetry does not exist when $n\leq 2$.}
\begin{align}
\label{eq:mintresideq}
s\cdot\partial_X\partial_s^2(s\cdot\partial_X\epsilon_{n-2})-2m^2\epsilon_{n-2}=0
\end{align}

This is indeed the case, as guaranteed by dimensional reduction. To demonstrate this, we decompose $\Phi_n$ into its transverse and traceless part $\Phi_n'$, and a part which is not transverse or not traceless $\Delta\Phi_n$
\begin{align}
    \Phi_n = \Phi_n' + \Delta\Phi_n
\end{align}
and note that in order for (\ref{eq:F1}) to be consistent, $\Delta\Phi_n$ cannot have a term like $A_n$ or $s\cdot\partial_XB_{n-1}$ for $A_n$ and $B_{n-1}$ transverse $\partial_s\cdot\partial_X A_n = \partial_s\cdot\partial_X B_{n-1} = 0$. Hence we may write $\Delta\Phi_n$ suggestively as
\begin{align}
    \Delta\Phi_n = \frac{1}{m}(s\cdot\partial_X)^2\epsilon_{n-2}
\end{align}
for some hyperfield $\epsilon_{n-2}$. Solving for $\Phi_{n-3}$ in terms of $\epsilon_{n-2}$ using (\ref{eq:F1}) yields
\begin{align}
    \Phi_{n-3} = 2i^{1/2}\partial_s\cdot\partial_X\epsilon_{n-2} + 2i^{1/2}\partial_s^2(s\cdot\partial_X\epsilon_{n-2})
\end{align}
Plugging these forms into (\ref{eq:F0}), we find $(\partial_X^2-m^2)\Phi_n'=0$ and an $\epsilon_{n-2}$ satisfying (\ref{eq:mintresideq}). $\Delta\Phi_n$ and $\Phi_{n-3}$ can therefore be gauged away, leaving a massive integer spin field $\Phi_n'$ satisfying (\ref{eq:hyperFP}).

\subsection{All propagators}
\label{sec:mhyperprop}
We now compute the correlation functions $\langle\Phi_{n-i}(X,s)\tilde{\Phi}_{n-j}(Y,t)\rangle$, which will give expressions for all correlation functions $\langle\phi_{\mu_1\cdots\mu_{n-i}}(X)\phi^*_{\nu_1\cdots\nu_{n-j}}(Y)\rangle$ with $i,j = 0,\dots,3$. These 16 correlation functions are needed to fully specify the physical integer spin $n$ propagator.

\sm

First, we detail the ambiguity in the correlation functions arising from the St{\"u}ckelberg gauge symmetry (\ref{eq:Stueck}). Coupling $\Phi_n$, $\Phi_{n-1}$, $\Phi_{n-2}$, and $\Phi_{n-3}$ to external sources $\mathcal{J}_n$, $\mathcal{J}_{n-1}$, $\mathcal{J}_{n-2}$, and $\mathcal{J}_{n-3}$, respectively, will lead to inconsistencies unless it is done in a gauge invariant way. The gauge invariant coupling of $\Phi_{n-i}$ to $\mathcal{J}_{n-i}$ is
\begin{align}
    \Delta S = n!\int d^dX\frac{d^dsd^ds'}{(2\pi)^d}e^{is\cdot s'}\Big(\tilde{\Phi}_n\mathcal{J}_n + \tilde{\Phi}_{n-1}\mathcal{J}_{n-1} + \tilde{\Phi}_{n-2}\mathcal{J}_{n-2} + \tilde{\Phi}_{n-3}\mathcal{J}_{n-3} + \text{c.c.} \Big)
\end{align}
provided that the sources satisfy
\begin{align}
    &\partial_s\cdot\partial_X\mathcal{J}_n + is^2\partial_s\cdot\partial_X\mathcal{J}_{n-2} + i^{1/2}m\mathcal{J}_{n-1} -3i^{-1/2}ms^2\mathcal{J}_{n-3}=0\\
    &\partial_s\cdot\partial_X\mathcal{J}_{n-1} + is^2\partial_s\cdot\partial_X\mathcal{J}_{n-3} + 2i^{1/2}m\mathcal{J}_{n-2} =0
\end{align}
We expect the solution to the equations of motion for $\Phi_{n-i}$ when coupled to this source to be of the form
\begin{align}
    \Phi_{n-i}(X,s) = i\int d^dY\frac{d^dtd^dt'}{(2\pi)^d}e^{it\cdot t'}\mathcal{G}_{ij}(X-Y,s,t)\mathcal{J}_{n-j}(Y,t')
\end{align}
where the sum over $j$ is implied. The transformation $\delta\mathcal{G}_{ij}(X-Y,s,t) = \mathcal{N}_{ij}(X-Y,s,t)$, where
\begin{align}
    &\mathcal{N}_{0j} = i^{-1/2} s\cdot\partial_X \Pi_{1j} && \mathcal{N}_{1j} = i^{-1/2} s\cdot \partial_X \Pi_{2j} + im\,\Pi_{1j} \nonumber\\
    &\mathcal{N}_{2j} = -i^{1/2} s\cdot\partial_X\partial_s^2\Pi_{1j} + 2im\,\Pi_{2j} && \mathcal{N}_{3j} = -i^{1/2} s\cdot\partial_X\partial_s^2\Pi_{2j} + 3m\partial_s^2\Pi_{1j}
\end{align}
and $\Pi_{1j}(X-Y,s,t)$, $\Pi_{2j}(X-Y,s,t)$ are arbitrary rank $n-1$ and $n-2$ hyperfields in $s$, and rank $n-j$ hyperfields in $t$, respectively, amounts to a St{\"u}ckelberg gauge transformation on $\Phi_{n-i}(X,s)$, and so $\mathcal{G}_{ij}(X-Y,s,t)$ itself has a gauge redundancy. $\mathcal{G}_{ij}(X-Y,s,t)$ is further ambiguous up to transformations $\delta\mathcal{G}_{ij}(X-Y,s,t) = \mathcal{M}_{ij}$, where
\begin{align}
    &\mathcal{M}_{i0} = i^{-1/2} t\cdot\partial_Y \Omega_{i1} && \mathcal{M}_{i1} = i^{-1/2} t\cdot \partial_Y \Omega_{i2} - im\,\Omega_{i1} \nonumber\\
    &\mathcal{M}_{i2} = -i^{1/2} t\cdot\partial_Y\partial_t^2\Omega_{i1} -2im\,\Omega_{i2} && \mathcal{M}_{i3} = -i^{1/2} t\cdot\partial_Y\partial_t^2\Omega_{i2} -3m\partial_t^2\Omega_{i1}
\end{align}
for similarly arbitrary hyperfields $\Omega_{i1}(X-Y,s,t)$ and $\Omega_{i2}(X-Y,s,t)$, precisely because
\begin{align}
    i\int d^dY\frac{d^dtd^dt'}{(2\pi)^d}e^{it\cdot t'}\mathcal{M}_{ij}(X-Y,s,t)\mathcal{J}_{n-j}(Y,t')=0
\end{align}
In total then, $\mathcal{G}_{ij}(X-Y,s,t)$ is ambiguous up to transformations
\begin{align}
\label{eq:mintpropamb}
    \delta\mathcal{G}_{ij}(X-Y,s,t) = \mathcal{N}_{ij}(X-Y,s,t) + \mathcal{M}_{ij}(X-Y,s,t)
\end{align}
This ambiguity equally applies to the correlation functions $\langle\Phi_{n-i}(X,s)\tilde{\Phi}_{n-j}(Y,t)\rangle$, i.e. if we change them by $\delta\langle\Phi_{n-i}(X,s)\tilde{\Phi}_{n-j}(Y,t)\rangle = \mathcal{N}_{ij}+\mathcal{M}_{ij}$, they represent the same massive particle. 

\sm

Instead of pursuing the exact form of the massive propagators directly from (\ref{eq:mhyperaction}), we recognize that the massive propagator can be straightforwardly obtained from the massless propagator (\ref{eq:mlesspropagator}) in $d+1$ dimensions. In the gauge (\ref{eq:deDonder}), the dimensionally reduced action may be written as 
\begin{align}
    &S_n =\nonumber\\
    &n!\int d^dX\frac{d^{d+1}sd^{d+1}s'}{(2\pi)^{d+1}}e^{is\cdot s'}\tilde{\Phi}_m(X,s,s_d)\big(1-\frac{1}{4}(s'^2 + s_d'^2)(\partial_{s'}^2+\partial_{d'}^2)\big)\big(\partial_X^2-m^2\big)\Phi_m(X,s',s_d')
\end{align}
The propagator for $\Phi_m(X,s,s_d)$ in this gauge can be obtained in the same way it was in \autoref{sec:mlessprop}, it is
\begin{align}
    \langle\Phi_m(X,s,s_d)\tilde{\Phi}_m(Y,t,t_d)\rangle = G_m(X-Y)\frac{1}{(\frac{d-3}{2})_n}\Big(-\frac{i}{2}\sqrt{\overline{s}^2\overline{t}^2}\Big)^nC_n^{\frac{d-3}{2}}\Big(\frac{\overline{s}\cdot\overline{t}}{\sqrt{\overline{s}^2\overline{t}^2}}\Big)
\end{align}
where $G_m(X-Y)$ is the standard massive spin 0 propagator, and $\overline{s}=(s,s_d)$ and $\overline{t} = (t,t_d)$ are $d+1$ dimensional auxiliary vectors. All that remains is to decompose this propagator into its independent $d$ dimensional components $\langle\Phi_{n-i}(X,s)\tilde{\Phi}_{n-j}(Y,t)\rangle$ using (\ref{eq:mdindependent})
\begin{align}
\label{eq:mprop1}
    &\langle\Phi_{n}(X,s)\tilde{\Phi}_{n}(Y,t)\rangle = G_m\frac{1}{(\frac{d-3}{2})_n}\frac{1}{n!}\Big(-\frac{i}{2}\sqrt{s^2t^2}\Big)^n C_n^{\frac{d-3}{2}}\Big(\frac{s\cdot t}{\sqrt{s^2t^2}}\Big) \\
     &\langle\Phi_{n}(X,s)\tilde{\Phi}_{n-2}(Y,t)\rangle = iG_m\frac{1}{2}s^2\frac{1}{(\frac{d-1}{2})_{n-1}}\frac{1}{n!}\Big(-\frac{i}{2}\sqrt{s^2t^2}\Big)^{n-2} C_{n-2}^{\frac{d-1}{2}}\Big(\frac{s\cdot t}{\sqrt{s^2t^2}}\Big)\\
     &\langle\Phi_{n-2}(X,s)\tilde{\Phi}_{n-2}(Y,t)\rangle = \nonumber\\
     &G_m \frac{2}{n!}\Big(-\frac{i}{2}\sqrt{s^2t^2}\Big)^{n-2}\Bigg(\frac{1}{(\frac{d-1}{2})_{n-1}}\Big(\frac{d-6}{2}+n\Big)C_{n-2}^{\frac{d-1}{2}}\Big(\frac{s\cdot t}{\sqrt{s^2t^2}}\Big) + \frac{3}{2}\frac{1}{(\frac{d+1}{2})_{n-2}}C_{n-4}^{\frac{d+1}{2}}\Big(\frac{s\cdot t}{\sqrt{s^2t^2}}\Big)\hspace{-0.1cm}\Bigg)
     \end{align}
\begin{align}
     &\langle\Phi_{n-1}(X,s)\tilde{\Phi}_{n-1}(Y,t)\rangle = G_m\frac{1}{(\frac{d-1}{2})_{n-1}}\frac{1}{n!}\Big(-\frac{i}{2}\sqrt{s^2t^2}\Big)^{n-1} C_{n-1}^{\frac{d-1}{2}}\Big(\frac{s\cdot t}{\sqrt{s^2t^2}}\Big) \\
     &\langle\Phi_{n-1}(X,s)\tilde{\Phi}_{n-3}(Y,t)\rangle = iG_m\frac{3}{2}s^2\frac{1}{(\frac{d+1}{2})_{n-2}}\frac{1}{n!}\Big(-\frac{i}{2}\sqrt{s^2t^2}\Big)^{n-3} C_{n-3}^{\frac{d+1}{2}}\Big(\frac{s\cdot t}{\sqrt{s^2t^2}}\Big) \\
     &\langle\Phi_{n-3}(X,s)\tilde{\Phi}_{n-3}(Y,t)\rangle =  \nonumber\\
\label{eq:mprop6}    
     &G_m \frac{6}{n!}\Big(-\frac{i}{2}\sqrt{s^2t^2}\Big)^{n-3}\Bigg(\frac{1}{(\frac{d+1}{2})_{n-2}}\Big(\frac{d-8}{2}+n\Big)C_{n-3}^{\frac{d+1}{2}}\Big(\frac{s\cdot t}{\sqrt{s^2t^2}}\Big) + \frac{5}{2}\frac{1}{(\frac{d+3}{2})_{n-3}}C_{n-5}^{\frac{d+3}{2}}\Big(\frac{s\cdot t}{\sqrt{s^2t^2}}\Big)\hspace{-0.1cm}\Bigg)
\end{align}
and all other correlation functions are either zero or related via Hermitian conjugation. One can straightforwardly consider the $n\to\infty$ asymptotic limit of these expressions, as we did for the massless propagators. We omit the explicit asymptotic limits for the sake of brevity.

\sm

Recall that the hyperfields $\Phi_{n-i}(X,s)$ correspond to fields $\phi_{\mu_1\cdots\mu_{n-i}}$ for $i = 0,\dots,3$. In a simplified notation, the massive spin $n$ action in terms of these fields is
\begin{align}
\label{eq:mindexaction}
     S_n = \int d^dX\Bigg\{ &\sum_{k = 0}^{\floor{n/2}} {n\choose 2k}\Bigg(&&\hspace{-1.625cm}\Big(1-\frac{3k}{2}\Big)\phi_0^{*(k)}\cdot\mathcal{F}_0^{(k)}-\frac{k}{2}\phi_0^{*(k)}\cdot\mathcal{F}_2^{(k-1)} \nonumber\\
    & && \hspace{-0cm}-\frac{k}{2}\phi_2^{*(k-1)}\cdot\mathcal{F}_0^{(k)}+\frac{k}{2}\phi_2^{*(k-1)}\cdot\mathcal{F}_2^{(k-1)}\Bigg) \nonumber\\
    & \hspace{-.5 cm}+ \sum_{k = 0}^{\floor{(n-1)/2}} {n\choose 2k+1}\Bigg(&&\hspace{-.5cm}\Big(1-\frac{5k}{2}\Big)\phi_1^{*(k)}\cdot\mathcal{F}_1^{(k)}-\frac{3k}{2}\phi_1^{*(k)}\cdot\mathcal{F}_3^{(k-1)}\nonumber \\
    & &&\hspace{-0cm}-\frac{3k}{2}\phi_3^{*(k-1)}\cdot\mathcal{F}_1^{(k)}-\frac{k}{2}\phi_3^{*(k-1)}\cdot\mathcal{F}_3^{(k-1)}\Bigg)\Bigg\}
\end{align}
where $\mathcal{O}_i^{(k)}$ denotes the $k$-fold trace of $\mathcal{O}_{\mu_1\cdots\mu_{n-i}}$, and the dot product $\cdot$ contracts the remaining indices. The massive spin $n$ correlation functions $\langle\phi_{\mu_1\cdots\mu_{n-i}}(X)\phi^*_{\nu_1\cdots\nu_{n-j}}(Y)\rangle$ from (\ref{eq:mindexaction}) can be obtained from $\langle\Phi_{n-i}(X,s)\tilde{\Phi}_{n-j}(Y,t)\rangle$ by applying derivatives
\begin{align}
    \langle\phi_{\mu_1\cdots\mu_{n-i}}(X)\phi^*_{\nu_1\cdots\nu_{n-j}}(Y)\rangle = i^{n-(i+j)/2}\frac{\partial^{n-i}}{\partial_s^{\mu_1}\cdots\partial_s^{\mu_{n-i}}}\frac{\partial^{n-j}}{\partial_t^{\nu_1}\cdots\partial_t^{\nu_{n-j}}}\langle\Phi_{n-i}(X,s)\tilde{\Phi}_{n-j}(Y,t)\rangle
\end{align}  

\section{Massive half integer spins}
\label{sec6}
In this section, we will use the hyperfield formalism of \autoref{sec2} to study massive half integer spin fields. We will find compact, closed form expressions for their covariant actions and propagators.

\sm

The half integer spin case was also considered in \cite{Fierz:1939zz,Fierz:1939ix}, but an alternative formulation was given in \cite{PhysRev.60.61}. In it, Rarita and Schwinger noted that the following system of equations is sufficient for describing a freely propagating spin $n+1/2$ massive particle
\begin{align}
    \label{eq:Dirac}
    (\fsl{\partial}+m)\psi_{\mu_1\cdots\mu_n}=0 \\
    \label{eq:gtraceless}
    \gamma^{\lambda}\psi_{\lambda\mu_2\cdots\mu_n}=0
\end{align}
for a symmetric rank $n$ tensor $\psi_{\mu_1\cdots\mu_n}$, with an additional implicit Dirac index. We will refer to the system of equations (\ref{eq:Dirac}) and (\ref{eq:gtraceless}) as a Rarita-Schwinger system. Any Rarita-Schwinger system is also a Fierz-Pauli system. The transverse and $\gamma$ traceless conditions ensure that the correct number $N_m(d,n+1/2)$ of degrees of freedom for a spin $n+1/2$ particle in $d$ spacetime dimensions propagate with mass $m$
\begin{align}
\label{eq:mhalfN}
    N_m(d,n+1/2) = &2^{\floor{d/2}}{d+n-1\choose n}-2^{\floor{d/2}}{d+n-2\choose n-1}\nonumber\\
    &-\Bigg(2^{\floor{d/2}}{d+n-2\choose n-1}-2^{\floor{d/2}}{d+n-3\choose n-2}\Bigg) \nonumber \\
    =& 2^{\floor{d/2}}{d+n-3\choose n}
\end{align}
which reduces to $2(2n+2)$ when $d=4$.

\sm

A minimal covariant Lagrangian formulation of massive half integer spin fields $\psi_{\mu_1\cdots\mu_n}$ in four spacetime dimensions that imposes (\ref{eq:Dirac})-(\ref{eq:gtraceless}) was first worked out in \cite{Singh:1974rc}. In it, Singh and Hagen needed to introduce, apart from the symmetric and $\gamma$ traceless rank $n$ Dirac tensor, a symmetric and $\gamma$ traceless rank $n-1$ Dirac tensor, and two of each symmetric and $\gamma$ traceless Dirac tensors of rank $0,1,2,\dots,n-2$ which couple so that all fields of lower rank than $n$ vanish on shell, and the rank $n$ field made up a Rarita-Schwinger system. Note that this field content is equivalent to using symmetric rank $n$ and $n-2$ Dirac tensors, with no tracelessness conditions.
\subsection{Massive particles from dimensional reduction}
In this section, we will construct an action for massive half integer spins which work in any spacetime dimension $d$ using dimensional reduction. The dimensional reduction procedure will be similar to the integer spin case, but will differ in some basic ways. 

\sm

An immediate difference is that $N_0(d+1,n+1/2)=N_m(d,n+1/2)$ only when $d$ is even, and so dimensional reduction appears not to work for odd $d$. The origin of this mismatch is that the spinor representation being used in the $d$ dimensional theory will always have a $\gamma_d$ with $\gamma_d^2=1$ and $\{\gamma_d,\gamma_{\mu}\}=0$ for $\mu = 0,\dots,d-1$. For the Dirac representation, this is only possible when $d$ is even. When $d$ is odd then, the $d$ dimensional fields are not in the Dirac representation, but in some larger one. The mismatch is therefore a difference in choice of spinor representation, which does not affect the spin of the particle. Shortly, we will write down a suitable field redefinition which removes all instances of $\gamma_d$ in the formulation, so that even for odd $d$, we may choose fields in the Dirac representation. As we will see, the result of performing this field redefinition is a formulation of massive half integer spin fields whose equations of motion are independent of $d$, and is conveniently well defined for the Dirac representation in $d$ spacetime dimensions, whether even or odd. For odd $d$ then, we may replace the larger representation obtained from dimensional reduction with the irreducible Dirac representation. A straightforward analysis of the equations of motion in \autoref{sec:eqnhalf} will reveal that this formulation has the correct degrees of freedom for a massive spin $n+1/2$ Dirac particle in any dimension $d$.

\sm

We therefore start with the massless hyperfield action (\ref{eq:mlesshalfaction}) in $d+1$ dimensions, and enforce that the hyperfield $\Psi_n(X,X_d,s,s_d)$ has the following dependence on the $X_d$ coordinate
\begin{align}
    \Psi_n(X,X_d,s,s_d) = e^{imX_d}\Psi_m(X,s,s_d)
\end{align}
The additional auxiliary component $s_d$ generates $n+1$ $d$ dimensional hyperfields 
\begin{align}
    \Psi_m(X,s,s_d) = \sum_{k=0}^{n}\frac{1}{k!}i^{-k/2}(s_d)^k\Psi_{n-k}(X,s)
\end{align}
The hyperfields $\Psi_{n-k}(X,s)$ are not all independent of each other, because they descend from a $d+1$ dimensional massless hyperfield satisfying the $d+1$ dimensional triple $\gamma$ traceless condition $(\fsl{\partial}_s+\gamma_d\partial_d)^3\Psi_n=0$. All $\Psi_{n-k}$ with $k>2$ can be written in terms of $\Psi_n$, $\Psi_{n-1}$, and $\Psi_{n-2}$
\begin{align}
    \Psi_m(X,s,s_d) &= \sum_{k=0}^{\floor{n/2}}\frac{(-1)^k}{(2k)!}(s_d)^{2k}\Big((1-k)\fsl{\partial}_s^{2k}\Psi_n + ik\fsl{\partial}_s^{2(k-1)}\Psi_{n-2}\Big) \nonumber \\
    +&\sum_{k=0}^{\floor{(n-1)/2}}\frac{(-1)^k}{(2k+1)!}(s_d)^{2k+1}\Big(k\gamma_d\,\fsl{\partial}_s^{2k+1}\Psi_n+i^{-1/2}\fsl{\partial}_s^{2k}\Psi_{n-1}-ik\gamma_d\,\fsl{\partial}_s^{2k-1}\Psi_{n-2}\Big)
\end{align}
where $\Psi_n,\Psi_{n-1},\Psi_{n-2}$ are unconstrained $d$ dimensional hyperfields. We would like to get rid of the $\gamma_d$'s in this expression. This is achieved by making the field redefinitions $\Psi_{n-i}\to e^{-i\pi\gamma_d/4}\Psi_{n-i}$.
\begin{align}
    \Psi_m(X,s,s_d) &= e^{-i\pi\gamma_d/4}\sum_{k=0}^{\floor{n/2}}\frac{(-1)^k}{(2k)!}(s_d)^{2k}\Big((1-k)\fsl{\partial}_s^{2k}\Psi_n + ik\fsl{\partial}_s^{2(k-1)}\Psi_{n-2}\Big) \nonumber \\
    +e^{-i\pi\gamma_d/4}&\sum_{k=0}^{\floor{(n-1)/2}}\frac{(-1)^k}{(2k+1)!}(s_d)^{2k+1}\Big(ik\fsl{\partial}_s^{2k+1}\Psi_n+i^{-1/2}\fsl{\partial}_s^{2k}\Psi_{n-1}+k\fsl{\partial}_s^{2k-1}\Psi_{n-2}\Big)
\end{align}

The $d+1$ dimensional $\gamma$ traceless hyperfield gauge parameter $\epsilon_m(X,s,s_d)$ may also be written in terms of an unconstrained, $d$ dimensional hyperfield component $\epsilon_{n-1}(X,s)$
\begin{align}
    &\epsilon_m(X,s,s_d) = \nonumber \\
&e^{-i\pi\gamma_d/4}\Big(\sum_{k=0}^{\floor{(n-1)/2}}\frac{(-1)^k}{(2k)!}(s_d)^{2k}\fsl{\partial}_s^{2k}\epsilon_{n-1}-i\sum_{k=0}^{\floor{(n-1)/2}}\frac{(-1)^k}{(2k+1)!}(s_d)^{2k+1}\fsl{\partial}_s^{2k+1}\epsilon_{n-1}\Big)
\end{align}
The St{\"u}ckelberg gauge symmetry may then be expressed in terms of the gauge parameter $\epsilon_{n-1}$ acting on $\Psi_n,\,\Psi_{n-1},\,\Psi_{n-2}$
\begin{align}
\label{eq:halfStueck}
&\delta\Psi_n=i^{-1/2}s\cdot\partial_X\epsilon_{n-1} \nonumber\\
&\delta\Psi_{n-1} = -i s\cdot\partial_X\fsl{\partial}_s\epsilon_{n-1}+im\epsilon_{n-1} \nonumber\\
&\delta\Psi_{n-2} = -i^{1/2}s\cdot\partial_X\partial_s^2\epsilon_{n-1}+2i^{1/2}m\fsl{\partial}_s\epsilon_{n-1}
\end{align}

The $d$ dimensional massive action is written in terms of $\Psi_n,\Psi_{n-1},\Psi_{n-2}$, and the independent components $\mathcal{S}_n,\,\mathcal{S}_{n-1},\,\mathcal{S}_{n-2}$ of the $d+1$ dimensional Fang field strength $\mathcal{S}_n(X,X_d,s,s_d)$
\begin{align}
    &S_{n+1/2} = -n!\int d^{d+1}X\frac{d^{d+1}sd^{d+1}s'}{(2\pi)^{d+1}}e^{is\cdot s'}\overline{\Psi}_n(X,s)\big(1-\frac{1}{2}\fsl{s'}\fsl{\partial}_{s'}-\frac{1}{4}s'^2\partial^2_{s'}\big)\mathcal{S}_n(X,s')
\end{align}
\begin{align}
\label{eq:mhalfhyperaction}
S_{n+1/2} &=- n!\int d^dX\frac{d^dsd^ds'}{(2\pi)^d}e^{is\cdot s'}\times\nonumber \\
\Bigg\{ &\sum_{k = 0}^{\floor{n/2}} \frac{(-1)^k}{(2k)!}\Bigg(&&\hspace{-2.75cm}\Big(1-\frac{3k}{2}\Big)\partial_{s}^{2k}\overline{\Psi}_n\partial_{s'}^{2k}\mathcal{S}_n+i\frac{k}{2}\partial_{s}^{2k}\overline{\Psi}_n\partial_{s'}^{2(k-1)}\mathcal{S}_{n-2}\nonumber\\
    & && \hspace{-3cm}+i\frac{k}{2}\partial_{s}^{2(k-1)}\overline{\Psi}_{n-2}\partial_{s'}^{2k}\mathcal{S}_n +\frac{k}{2}\partial_{s}^{2(k-1)}\overline{\Psi}_{n-2}\partial_{s'}^{2(k-1)}\mathcal{S}_{n-2}\nonumber\\
    & && \hspace{-3.5cm} -i^{-1/2}k\partial_s^{2(k-1)}\overline{\Psi}_{n-1}\overleftarrow{\fsl{\partial}_s}\partial_{s'}^{2(k-1)}\mathcal{S}_{n-2} + i^{-1/2}k\partial_s^{2(k-1)}\overline{\Psi}_{n-2}\fsl{\partial}_{s'}\partial_{s'}^{2(k-1)}\mathcal{S}_{n-1} \nonumber \\
    & && \hspace{2.75cm}+ ik\partial_{s}^{2(k-1)}\overline{\Psi}_{n-1}\overleftarrow{\fsl{\partial}_s}\fsl{\partial}_{s'}\partial_{s'}^{2(k-1)}\mathcal{S}_{n-1}\Bigg)
    \nonumber\\
    & \hspace{-1cm}+ \sum_{k = 0}^{\floor{(n-1)/2}} \frac{(-1)^k}{(2k+1)!}\Bigg(&&\hspace{-2.5cm}-i\Big(\frac{1}{2}+\frac{3k}{2}\Big)\partial_{s}^{2k}\overline{\Psi}_n\overleftarrow{\fsl{\partial}_s}\fsl{\partial}_{s'}\partial_{s'}^{2k}\mathcal{S}_n-\frac{k}{2}\partial_{s}^{2k}\overline{\Psi}_n\overleftarrow{\fsl{\partial}_s}\fsl{\partial}_{s'}\partial_{s'}^{2(k-1)}\mathcal{S}_{n-2} \nonumber \\
    & &&\hspace{-2.5cm}-\frac{k}{2}\partial_{s}^{2(k-1)}\overline{\Psi}_{n-2}\overleftarrow{\fsl{\partial}_s}\fsl{\partial}_{s'}\partial_{s'}^{2k}\mathcal{S}_n + i\frac{k}{2}\partial_{s}^{2(k-1)}\overline{\Psi}_{n-2}\overleftarrow{\fsl{\partial}_s}\fsl{\partial}_{s'}\partial_{s'}^{2(k-1)}\mathcal{S}_{n-2} \nonumber\\
    & &&\hspace{-1.25cm}+\frac{1}{2}i^{-1/2}\partial_s^{2k}\overline{\Psi}_n\overleftarrow{\fsl{\partial}_s}\partial_{s'}^{2k}\mathcal{S}_{n-1} - \frac{1}{2}i^{-1/2}\partial_s^{2k}\overline{\Psi}_{n-1}\fsl{\partial}_{s'}\partial_{s'}^{2k}\mathcal{S}_n \nonumber\\
    & && \hspace{-2.25cm}+i^{1/2}k\partial_s^{2(k-1)}\overline{\Psi}_{n-2}\overleftarrow{\fsl{\partial}_s}\partial_{s'}^{2k}\mathcal{S}_{n-1}-i^{1/2}k\partial_s^{2k}\overline{\Psi}_{n-1}\fsl{\partial}_{s'}\partial_{s'}^{2(k-1)}\mathcal{S}_{n-2} \nonumber \\
     & && \hspace{3.375cm}+ \Big(\frac{1}{2}-k\Big)\partial_s^{2k}\overline{\Psi}_{n-1}\partial_{s'}^{2k}\mathcal{S}_{n-1} \Bigg)\Bigg\}
\end{align}
where
\begin{align}
    \mathcal{S}_n &= (\fsl{\partial}_X+m-s\cdot\partial_X\fsl{\partial}_s)\Psi_n+i^{1/2}s\cdot\partial_X\Psi_{n-1} \\
    \mathcal{S}_{n-1} &= (\fsl{\partial}_X-s\cdot\partial_X\fsl{\partial}_s)\Psi_{n-1} + i^{1/2}s\cdot\partial_X\Psi_{n-2} + i^{-1/2}m\fsl{\partial}_s\Psi_n \\
    \mathcal{S}_{n-2} &= (\fsl{\partial}_X-m)\Psi_{n-2}+is\cdot\partial_X\fsl{\partial}_s^3\Psi_n + i^{-1/2}s\cdot\partial_X\partial_s^2\Psi_{n-1} + 2i^{-1/2}m\fsl{\partial}_s\Psi_{n-1}
\end{align}
That (\ref{eq:mhalfhyperaction}) is invariant under the gauge transformation (\ref{eq:halfStueck}) is guaranteed by the fact that $\mathcal{S}_n,\,\mathcal{S}_{n-1},\,\mathcal{S}_{n-2}$ are themselves gauge invariant, and satisfy the identity
\begin{align}
\fsl{\partial}_s(\fsl{\partial}_X-m)\mathcal{S}_n-s\cdot\partial_X\partial_s^2\mathcal{S}_n-i^{1/2}(\fsl{\partial}_X-m)\mathcal{S}_{n-1} + i\,s\cdot\partial_X\mathcal{S}_{n-2} = 0
\end{align}
which is a direct consequence of (\ref{eq:hyperhalfbianchi}). As promised (\ref{eq:mhalfhyperaction}) has no instance of $\gamma_d$, making this action valid for the Dirac representation in any spacetime dimension. 

\sm

The St{\"u}ckelberg gauge symmetry (\ref{eq:halfStueck}) can be used to set $\Psi_{n-1}$ to zero, leaving the fields $\Psi_{n}$ and $\Psi_{n-2}$, consistent with the minimal field content of \cite{Singh:1974rc}.
\subsection{Equations of motion}
\label{sec:eqnhalf}
The massive equations of motion are equivalent to setting the $d+1$ dimensional field strength $\mathcal{S}_n(X,X_d,s,s_d)$ to zero
\begin{align}
\label{eq:mhalfeom}
    \mathcal{S}_n = 0,\quad \mathcal{S}_{n-1} = 0,\quad \mathcal{S}_{n-2} = 0
\end{align}
That these equations describe massive half integer spin particles amounts to showing that they imply that $\Psi_{n-1}$ and $\Psi_{n-2}$ can always be set to zero, and that $\Psi_n$ is a Rarita-Schwinger system in hyperspace
\begin{align}
\label{eq:hyperRaritaSchwinger}
    (\fsl{\partial}_X + m)\Psi_n&=0\nonumber\\
    \fsl{\partial}_s\Psi_n&=0
\end{align}

\sm

The gauge symmetry (\ref{eq:halfStueck}) may be used to set $\Psi_{n-1}=0$ before imposing the equations of motion, so (\ref{eq:mhalfeom}) becomes
\begin{align}
\label{eq:S0}
     &\mathcal{S}_n = (\fsl{\partial}_X+m-s\cdot\partial_X\fsl{\partial}_s)\Psi_n \\
\label{eq:S1}
    &\mathcal{S}_{n-1} = i^{1/2}s\cdot\partial_X\Psi_{n-2} + i^{-1/2}m\fsl{\partial}_s\Psi_n \\
\label{eq:S2}
    &\mathcal{S}_{n-2} = (\fsl{\partial}_X-m)\Psi_{n-2}+is\cdot\partial_X\fsl{\partial}_s^3\Psi_n
\end{align}
In this gauge, if $\Psi_{n-2}=0$, then $\Psi_n$ satisfies (\ref{eq:hyperRaritaSchwinger}). What remains is to show that $\Psi_{n-2}$ may always be set to zero via the residual gauge symmetry of this system
\begin{align}
    \delta\Psi_n = i^{-1/2}s\cdot\partial_X\epsilon_{n-1},\quad \delta\Psi_{n-2}=-i^{1/2}s\cdot\partial_X\partial_s^2\epsilon_{n-1}+2i^{1/2}m\fsl{\partial}_s\epsilon_{n-1}
\end{align}
These transformations keep $\Psi_{n-1}=0$ provided that $\epsilon_{n-1}$ satisfies\footnote{Note that this residual gauge symmetry does not exist when $n\leq 1$.}
\begin{align}  
\label{eq:mhalfresid}
s\cdot\partial_X\fsl{\partial}_s\epsilon_{n-1}-m\epsilon_{n-1}=0 
\end{align}

\sm

To show that this is the case, we decompose $\Psi_n$ into a transverse and $\gamma$ traceless part $\Psi_n'$ and a part $\Delta\Psi_n$ which is not transverse or not $\gamma$ traceless 
\begin{align}
    \Psi_n = \Psi_n'+\Delta\Psi_n
\end{align}
We may always write $\Delta\Psi_n$ suggestively as $\Delta\Psi_n= i^{-1/2}s\cdot\partial_X\epsilon_{n-1}$, for some hyperfield $\epsilon_{n-1}$. 
Plugging this into (\ref{eq:S0}), we find $(\fsl{\partial}_X+m)\Psi_n'=0$, and an $\epsilon_{n-1}$ satisfying (\ref{eq:mhalfresid}). Finally, we solve for $\Psi_{n-2}$ using (\ref{eq:S1}) and (\ref{eq:mhalfresid})
\begin{align}
    \Psi_{n-2} = -i^{1/2}s\cdot\partial_X\partial_s^2\epsilon_{n-1}+2i^{1/2}m\fsl{\partial}_s\epsilon_{n-1}
\end{align}
Hence, $\Delta\Psi_n$ and $\Psi_{n-2}$ can be gauged away, leaving a massive half integer spin field $\Psi_n'$ satisfying (\ref{eq:hyperRaritaSchwinger}).
\subsection{All propagators}
\label{sec:mhalfprop}
We now compute the correlation functions $\langle\Psi_{n-i}(X,s)\overline{\Psi}_{n-j}(Y,t)\rangle$, which will give expressions for all correlation functions $\langle\psi_{\mu_1\cdots\mu_{n-i}}(X)\overline{\psi}_{\nu_1\cdots\nu_{n-j}}(Y)\rangle$ with $i,j=0,1,2$. These 9 correlation functions are needed to fully specify the physical half integer spin $n+1/2$ propagator.

\sm

The discussion in \autoref{sec:mhyperprop} on the gauge ambiguity of the correlation functions equally applies here. Coupling $\overline{\Psi}_n$, $\overline{\Psi}_{n-1}$, and $\overline{\Psi}_{n-2}$ to external sources $\mathcal{Q}_n$, $\mathcal{Q}_{n-1}$, and $\mathcal{Q}_{n-2}$ via
\begin{align}
    \Delta S = n!\int d^dX\frac{d^dsd^ds'}{(2\pi)^d}e^{is\cdot s'}\Big(\overline{\Psi}_n\mathcal{Q}_n+\overline{\Psi}_{n-1}\mathcal{Q}_{n-1}+\overline{\Psi}_{n-2}\mathcal{Q}_{n-2}\Big)
\end{align}
is gauge invariant provided that the sources satisfy
\begin{align}
    \partial_s\cdot\partial_X\mathcal{Q}_n-i^{1/2}\fsl{s}\partial_s\cdot\partial_X\mathcal{Q}_{n-1}+i^{1/2}m\mathcal{Q}_{n-1}+is^2\partial_s\cdot\partial_X\mathcal{Q}_{n-2}-2im\fsl{s}\mathcal{Q}_{n-2}=0
\end{align}
Because of this, the correlation functions $\langle\Psi_{n-i}(X,s)\overline{\Psi}_{n-j}(Y,t)\rangle$ are ambiguous up to the transformations 
\begin{align}
\label{eq:mhalfpropamb}
    \delta\langle\Psi_{n-i}(X,s)\overline{\Psi}_{n-j}(Y,t)\rangle = \mathcal{N}_{ij}(X-Y,s,t) + \mathcal{M}_{ij}(X-Y,s,t)
\end{align}
where
\begin{align}
    &\mathcal{N}_{0j}=i^{-1/2}s\cdot\partial_X\Pi_{1j} &&\mathcal{M}_{i0} = i^{-1/2}t\cdot\partial_Y\overline{\Omega}_{i1} \nonumber \\
    &\mathcal{N}_{1j} = -i\, s\cdot\partial_X\fsl{\partial}_s\Pi_{1j} + im \Pi_{1j} && \mathcal{M}_{i1} = -i\,t\cdot\partial_Y\overline{\Omega}_{i1}\overleftarrow{\fsl{\partial}_t} - im\overline{\Omega}_{i1} \nonumber \\
    &\mathcal{N}_{2j} = -i^{1/2}s\cdot\partial_X\partial_s^2\Pi_{1j}+2i^{1/2}m\fsl{\partial}_s\Pi_{1j} && \mathcal{M}_{i2} = -i^{1/2}t\cdot\partial_Y\partial_t^2\overline{\Omega}_{i1}-2i^{1/2}m\overline{\Omega}_{i1}\overleftarrow{\fsl{\partial}_t}
\end{align}
for arbitrary hyperfields $\Pi_{1j}(X-Y,s,t)$ and $\overline{\Omega}_{i1}(X-Y,s,t)$. 

\sm

The massive correlation functions in a specific gauge are obtained straightforwardly from the $d+1$ dimensional massless propagator (\ref{eq:mlesshalfprop}). After dimensionally reducing, the $\Psi_m(X,s,s_d)$ propagator is
\begin{align}
    \langle\Psi_m(X,s,s_d)\overline{\Psi}_m(Y,t,t_d)\rangle &=  \tilde{\Delta}_m(X-Y)\frac{1}{\big(\frac{d-1}{2}\big)_n}\Big(-\frac{i}{2}\sqrt{\overline{s}^2\overline{t}^2}\Big)^nC_n^{\frac{d-1}{2}}\Big(\frac{\overline{s}\cdot\overline{t}}{\sqrt{\overline{s}^2\overline{t}^2}}\Big) \nonumber \\
    &-\frac{i}{2}\overline{\fsl{s}}\,\tilde{\Delta}_m(X-Y)\,\overline{\fsl{t}}\frac{1}{\big(\frac{d-1}{2}\big)_{n}}\Big(-\frac{i}{2}\sqrt{\overline{s}^2\overline{t}^2}\Big)^{n-1}C_{n-1}^{\frac{d-1}{2}}\Big(\frac{\overline{s}\cdot\overline{t}}{\sqrt{\overline{s}^2\overline{t}^2}}\Big)
\end{align}
where $\tilde{\Delta}_m(X-Y)$ is the massive spin 1/2 propagator, whose mass is proportional to $\gamma_d$
\begin{align}
\label{eq:chiralprop}
    \tilde{\Delta}_m(X-Y) = -\int\frac{d^dp}{(2\pi)^d}\frac{\fsl{p}+m\gamma_d}{p^2+m^2-i\epsilon}e^{ip\cdot(X-Y)}
\end{align}
and $\overline{s}=(s,s_d)$ and $\overline{t}=(t,t_d)$ are $d+1$ dimensional auxiliary vectors. We now decompose this propagator into its independent $d$ dimensional components $\langle\Psi_{n-i}(X,s)\overline{\Psi}_{n-j}(Y,t)\rangle$, remembering also to perform the field redefinition $\Psi_{n-i}\to e^{-i\pi\gamma_d/4}\Psi_{n-i}$
\begin{align}
\label{eq:mhalfprop1}
    \langle\Psi_n(X,s)\overline{\Psi}_n(Y,t)\rangle &=  \Delta_m\frac{1}{\big(\frac{d-1}{2}\big)_n}\frac{1}{n!}\Big(-\frac{i}{2}\sqrt{s^2t^2}\Big)^nC_{n}^{\frac{d-1}{2}}\Big(\frac{s\cdot t}{\sqrt{s^2t^2}}\Big) \nonumber \\
    &-\frac{i}{2}\fsl{s}\,\Delta_{-m}\,\fsl{t}\frac{1}{\big(\frac{d-1}{2}\big)_{n}}\frac{1}{n!}\Big(-\frac{i}{2}\sqrt{s^2t^2}\Big)^{n-1}C_{n-1}^{\frac{d-1}{2}}\Big(\frac{s\cdot t}{\sqrt{s^2t^2}}\Big)\\
    \langle\Psi_n(X,s)\overline{\Psi}_{n-1}(Y,t)\rangle &=\frac{1}{2}i^{1/2}\fsl{s}\Delta_{-m}\frac{1}{\big(\frac{d-1}{2}\big)_{n}}\frac{1}{n!}\Big(-\frac{i}{2}\sqrt{s^2t^2}\Big)^{n-1}C_{n-1}^{\frac{d-1}{2}}\Big(\frac{s\cdot t}{\sqrt{s^2t^2}}\Big) \\
    \langle\Psi_n(X,s)\overline{\Psi}_{n-2}(Y,t)\rangle &=\frac{1}{2}is^2\Delta_m\frac{1}{\big(\frac{d+1}{2}\big)_{n-1}}\frac{1}{n!}\Big(-\frac{i}{2}\sqrt{s^2t^2}\Big)^{n-2}C_{n-2}^{\frac{d+1}{2}}\Big(\frac{s\cdot t}{\sqrt{s^2t^2}}\Big) \nonumber \\
    &+\frac{1}{4}s^2\fsl{s}\Delta_{-m}\fsl{t}\frac{1}{\big(\frac{d+1}{2}\big)_{n-1}}\frac{1}{n!}\Big(-\frac{i}{2}\sqrt{s^2t^2}\Big)^{n-3}C_{n-3}^{\frac{d+1}{2}}\Big(\frac{s\cdot t}{\sqrt{s^2t^2}}\Big) \\
     \langle\Psi_{n-1}(X,s)\overline{\Psi}_{n-1}(Y,t)\rangle &= \Delta_m\frac{1}{\big(\frac{d+1}{2}\big)_{n-1}}\frac{1}{n!}\Big(-\frac{i}{2}\sqrt{s^2t^2}\Big)^{n-1}C_{n-1}^{\frac{d+1}{2}}\Big(\frac{s\cdot t}{\sqrt{s^2t^2}}\Big) \nonumber \\
     &-\frac{1}{2}\Delta_{-m}\frac{1}{\big(\frac{d-1}{2}\big)_{n}}\frac{1}{n!}\Big(-\frac{i}{2}\sqrt{s^2t^2}\Big)^{n-1}C_{n-1}^{\frac{d-1}{2}}\Big(\frac{s\cdot t}{\sqrt{s^2t^2}}\Big) \nonumber \\
     & -\frac{i}{2}\fsl{s}\Delta_{-m}\fsl{t}\frac{1}{\big(\frac{d+1}{2}\big)_{n-1}}\frac{1}{n!}\Big(-\frac{i}{2}\sqrt{s^2t^2}\Big)^{n-2}C_{n-2}^{\frac{d+1}{2}}\Big(\frac{s\cdot t}{\sqrt{s^2t^2}}\Big) \\
     \langle\Psi_{n-1}(X,s)\overline{\Psi}_{n-2}(Y,t)\rangle &= i^{1/2}\fsl{s}\Delta_{-m}\frac{1}{\big(\frac{d+1}{2}\big)_{n-1}}\frac{1}{n!}\Big(-\frac{i}{2}\sqrt{s^2t^2}\Big)^{n-2}C_{n-2}^{\frac{d+1}{2}}\Big(\frac{s\cdot t}{\sqrt{s^2t^2}}\Big) \nonumber \\
     & -\frac{1}{4}i^{-1/2}s^2\Delta_{-m}\fsl{t}\frac{1}{\big(\frac{d+1}{2}\big)_{n-1}}\frac{1}{n!}\Big(-\frac{i}{2}\sqrt{s^2t^2}\Big)^{n-3}C_{n-3}^{\frac{d+1}{2}}\Big(\frac{s\cdot t}{\sqrt{s^2t^2}}\Big)
     \end{align}
     \begin{align}
\label{eq:mhalfprop6}
&\langle\Psi_{n-2}(X,s)\overline{\Psi}_{n-2}(Y,t)\rangle =-2\Delta_{-m}\frac{1}{\big(\frac{d+1}{2}\big)_{n-1}}\frac{1}{n!}\Big(-\frac{i}{2}\sqrt{s^2t^2}\Big)^{n-2}C_{n-2}^{\frac{d+1}{2}}\Big(\frac{s\cdot t}{\sqrt{s^2t^2}}\Big) \nonumber \\
      &+\frac{2\Delta_m}{n!}\Big(-\frac{i}{2}\sqrt{s^2t^2}\Big)^{n-2}\nonumber \\
& \hspace{3.cm}\times\Bigg(\frac{1}{\big(\frac{d+1}{2}\big)_{n-1}}\Big(\frac{d-4}{2}+n\Big)C_{n-2}^{\frac{d+1}{2}}\Big(\frac{s\cdot t}{\sqrt{s^2t^2}}\Big)+\frac{3}{2}\frac{1}{\big(\frac{d+3}{2}\big)_{n-2}}C_{n-4}^{\frac{d+3}{2}}\Big(\frac{s\cdot t}{\sqrt{s^2t^2}}\Big)\Bigg) \nonumber\\
      &-i\frac{\fsl{s}\Delta_{-m}\fsl{t}}{n!}\Big(-\frac{i}{2}\sqrt{s^2t^2}\Big)^{n-3}\nonumber \\
      & \hspace{3.cm}\times\Bigg(\frac{1}{\big(\frac{d+1}{2}\big)_{n-1}}\Big(\frac{d-6}{2}+n\Big)C_{n-3}^{\frac{d+1}{2}}\Big(\frac{s\cdot t}{\sqrt{s^2t^2}}\Big)+\frac{3}{2}\frac{1}{\big(\frac{d+3}{2}\big)_{n-2}}C_{n-5}^{\frac{d+3}{2}}\Big(\frac{s\cdot t}{\sqrt{s^2t^2}}\Big)\Bigg)
\end{align}
where now $\Delta_m(X-Y)$ is the standard massive spin 1/2 propagator, obtained by replacing $m\gamma_d$ with $im$ in (\ref{eq:chiralprop}). All other correlation functions are related via Hermitian conjugation. Again, we omit their asymptotic $n\to \infty$ limits for the sake of brevity.

\sm

Recall that the hyperfields $\Psi_{n-i}(X,s)$ correspond to fields $\psi_{\mu_1\cdots\mu_{n-i}}$ for $i=0,1,2$. In a simplified notation, the massive spin $n+1/2$ action in terms of these fields is
\begin{align}
\label{eq:mhalfindexaction}
S_{n+1/2} =- \int d^dX\Bigg\{ &\sum_{k = 0}^{\floor{n/2}} {n\choose 2k}\Bigg(&&\hspace{-.75cm}\Big(1-\frac{3k}{2}\Big)\overline{\psi}_0^{(2k)}\cdot\mathcal{S}_0^{(2k)}-\frac{k}{2}\overline{\psi}_0^{(2k)}\cdot\mathcal{S}_2^{(2k-2)} \nonumber\\
    & && \hspace{-.5cm}-\frac{k}{2}\overline{\psi}_2^{(2k-2)}\cdot\mathcal{S}_0^{(2k)} -\frac{k}{2}\overline{\psi}_2^{(2k-2)}\cdot\mathcal{S}_2^{(2k-2)}\nonumber\\
    & && \hspace{-.5cm} -ik\overline{\psi}_1^{(2k-1)}\cdot\mathcal{S}_2^{(2k-2)} + ik\overline{\psi}_2^{(2k-2)}\cdot\mathcal{S}_1^{(2k-1)} \nonumber \\
    & && \hspace{3cm}-k\overline{\psi}_1^{(2k-1)}\cdot\mathcal{S}_1^{(2k-1)}\Bigg)
    \nonumber\\
    & \hspace{-2.5 cm}+ \sum_{k = 0}^{\floor{(n-1)/2}} {n\choose 2k+1}\Bigg(&&\hspace{-1.5cm}-\Big(\frac{1}{2}+\frac{3k}{2}\Big)\overline{\psi}_0^{(2k+1)}\cdot\mathcal{S}_0^{(2k+1)}-\frac{k}{2}\overline{\psi}_0^{(2k+1)}\cdot\mathcal{S}_2^{(2k-1)} \nonumber \\
    & &&\hspace{-1cm}-\frac{k}{2}\overline{\psi}_2^{(2k-1)}\cdot\mathcal{S}_0^{(2k+1)} -\frac{k}{2}\overline{\psi}_2^{(2k-1)}\cdot\mathcal{S}_2^{(2k-1)} \nonumber\\
    & &&\hspace{-.5cm}-\frac{i}{2}\overline{\psi}_0^{(2k+1)}\cdot\mathcal{S}_1^{(2k)} + \frac{i}{2}\overline{\psi}_1^{(2k)}\cdot\mathcal{S}_0^{(2k+1)} \nonumber\\
    & && \hspace{-.5cm}+ik\overline{\psi}_2^{(2k-1)}\cdot\mathcal{S}_1^{(2k)}-ik\overline{\psi}_1^{(2k)}\cdot\mathcal{S}_2^{(2k-1)} \nonumber \\
     & && \hspace{2cm}+ \Big(\frac{1}{2}-k\Big)\overline{\psi}_1^{(2k)}\cdot\mathcal{S}_1^{(2k)} \Bigg)\Bigg\}
\end{align}
where now $\mathcal{O}^{(k)}_i$ denotes the $k$-fold $\gamma$ trace of $\mathcal{O}_{\mu_1\cdots\mu_{n-i}}$ applied from the left, and $\overline{\mathcal{O}}^{(k)}_i$ denotes the $k$-fold $\gamma$ trace of $\overline{\mathcal{O}}_{\mu_1\cdots\mu_{n-i}}$ applied from the right. The massive spin $n+1/2$ correlation functions $\langle\psi_{\mu_1\cdots\mu_{n-i}}(X)\overline{\psi}_{\nu_1\cdots\nu_{n-j}}(Y)\rangle$ from (\ref{eq:mhalfindexaction}) can be obtained from $\langle\Psi_{n-i}(X,s)\overline{\Psi}_{n-j}(Y,t)\rangle$ by applying derivatives
\begin{align}
    \langle\psi_{\mu_1\cdots\mu_{n-i}}(X)\overline{\psi}_{\nu_1\cdots\nu_{n-j}}(Y)\rangle = i^{n-(i+j)/2}\frac{\partial^{n-i}}{\partial_s^{\mu_1}\cdots\partial_s^{\mu_{n-i}}}\frac{\partial^{n-j}}{\partial_t^{\nu_1}\cdots\partial_t^{\nu_{n-j}}}\langle\Psi_{n-i}(X,s)\overline{\Psi}_{n-j}(Y,t)\rangle
\end{align}

\section{Discussion}
\label{sec:disc}
In this paper, we have outlined the covariant formulation of free particles with any mass $m$ and spin $s$, in any spacetime dimension $d$. For massless particles with integer spins, their covariant action is (\ref{eq:mlessindexaction}), first written down in \cite{Fronsdal:1978rb}, and their propagator is (\ref{eq:mlesspropagator}), in agreement with \cite{Ponomarev:2016jqk}. For massless particles with half integer spins, their covariant action is (\ref{eq:mlesshalfindexaction}), first written down in \cite{Fang:1978wz}, and their propagator is (\ref{eq:mlesshalfprop}). For massive particles we arrived at their covariant actions (\ref{eq:mindexaction}), (\ref{eq:mhalfindexaction}) and their corresponding propagators (\ref{eq:mprop1}) -- (\ref{eq:mprop6}), (\ref{eq:mhalfprop1}) -- (\ref{eq:mhalfprop6}) by following the prescription of dimensional reduction of massless particles in $d+1$ dimensions first described in \cite{Aragone:1987dtt}. 

\sm

The formulation of massive particles exhibits a gauge symmetry (\ref{eq:Stueck}), (\ref{eq:halfStueck}). Because of this, we were able to arrive at propagators which grow like $\sim p^{-2}$, and $\sim p^{-1}$, respectively, at the cost of including unphysical polarizations. Because of the gauge invariance, there must exist a gauge which only includes physical polarizations. The physical gauges for integer and half integer spins are obtained by completely exhausting the gauge freedom. An example of a physical gauge is respectively
\begin{align}
    &\text{Integer spin:} &&P^d_{\eta}\Big((1-k)\partial_s^{2k}\Phi_n + ik\partial_s^{2(k-1)}\Phi_{n-2}\Big) = 0 &&& k \geq 1 \nonumber \\
    & && P^d_{\eta}\Big((1-k)\partial_s^{2k}\Phi_{n-1} + ik\partial_s^{2(k-1)}\Phi_{n-3}\Big) = 0 &&& k \geq 0 \\
    &\text{Half integer spin:} && P^d_{\gamma}\Big((1-k)\fsl{\partial}_s^{2k}\Psi_n + ik\fsl{\partial}_s^{2(k-1)}\Psi_{n-2}\Big) = 0 &&& k\geq 1 \nonumber \\
    & && P^d_{\gamma}\Big(ik\fsl{\partial}_s^{2k+1}\Psi_n+i^{-1/2}\fsl{\partial}_s^{2k}\Psi_{n-1}+k\fsl{\partial}_s^{2k-1}\Psi_{n-2}\Big) = 0 &&& k\geq 0
\end{align}
where $P^d_{\eta}$ and $P^d_{\gamma}$ are the projection operators onto traceless and $\gamma$ traceless hyperfields, respectively, defined in \autoref{subsec:Hyperbasics}. In these physical gauges, we are left in the action with traceless rank $n$, $n-2$, $n-3$, $\dots$, 0 hyperfields for integer spins, and $\gamma$ traceless rank $n$, $n-1$ hyperfields, and two series of $\gamma$ traceless rank $n-2$, $n-3$, $\dots$, 0 hyperfields for half integer spins. These are precisely the minimal auxiliary field contents used in the Singh-Hagen actions \cite{Singh:1974qz,Singh:1974rc}. The equations of motion in this gauge imply that $\Phi_n$ and $\Psi_n$ Fierz-Pauli and Rarita-Schwinger systems (\ref{eq:hyperFP}), (\ref{eq:hyperRaritaSchwinger}), respectively, while the remaining hyperfields vanish. Operators which vanish on-shell make correlation functions vanish after insertions except at coincident points. The physical gauges are then such that correlations functions with $\Phi_{n-1}$, $\Phi_{n-2}$, $\Phi_{n-3}$, $(\partial_X^2-m^2)\Phi_n$ , $\partial_s\cdot\partial_X\Phi_n$, $\partial_s^2\Phi_n$, $\Psi_{n-1}$, $\Psi_{n-2}$,  $(\fsl{\partial}_X+m)\Psi_n$, or $\fsl{\partial}_s\Psi_n$ inserted vanish except at coincident points. In momentum space, this forces the correlation functions to be of the form
\begin{align}
\label{eq:unitaryint}
    &\langle\Phi_n(p,s)\tilde{\Phi}_n(-p,t)\rangle =\nonumber\\
    &G_m(p)\frac{1}{\big(\frac{d-3}{2}\big)_n}\frac{1}{n!}\Big(-\frac{i}{2}\sqrt{s\cdot P\cdot s\,t\cdot P\cdot t}\Big)^n C_n^{\frac{d-3}{2}}\Big(\frac{s\cdot P\cdot t}{\sqrt{s\cdot P\cdot s\,t\cdot P\cdot t}}\Big) + \text{polynomial} \\
    &\langle\Phi_{n-i}(p,s)\tilde{\Phi}_{n-j}(-p,t)\rangle = \text{polynomial} \quad\quad j=1,2,3 \\
\label{eq:unitaryhalf}
    &\langle\Psi_n(p,s)\overline{\Psi}_n(-p,t)\rangle =\frac{1}{\big(\frac{d-1}{2}\big)_n}\frac{1}{n!}\times\nonumber\\
    &\Bigg(\Delta_m(p)\Big(-\frac{i}{2}\sqrt{s\cdot P\cdot s\,t\cdot P\cdot t}\Big)^n C_n^{\frac{d-1}{2}}\Big(\frac{s\cdot P\cdot t}{\sqrt{s\cdot P\cdot s\,t\cdot P\cdot t}}\Big) \nonumber \\
    &-\frac{i}{2}\big(\fsl{s}-i\frac{s\cdot p}{m}\big)\Delta_{-m}(p)\big(\fsl{t}-i\frac{t\cdot p}{m}\big)\Big(-\frac{i}{2}\sqrt{s\cdot P\cdot s\,t\cdot P\cdot t}\Big)^{n-1} C_{n-1}^{\frac{d-1}{2}}\Big(\frac{s\cdot P\cdot t}{\sqrt{s\cdot P\cdot s\,t\cdot P\cdot t}}\Big)\Bigg) \nonumber \\
    &\hspace{12cm}+ \text{polynomial} \\
    \label{eq:unitaryprop6}
    &\langle\Psi_{n-i}(p,s)\overline{\Psi}_{n-j}(-p,t)\rangle = \text{polynomial} \quad\quad j=1,2
\end{align}
where $P_{\mu\nu} = \eta_{\mu\nu} + p_{\mu}p_{\nu}/m^2$. The physical poles in (\ref{eq:unitaryint}), (\ref{eq:unitaryhalf}) are uniquely determined by demanding that $\partial_s^2\langle\Phi_n(p,s)\tilde{\Phi}_n(-p,t)\rangle$, $p\cdot\partial_s\langle\Phi_n(p,s)\tilde{\Phi}_n(-p,t)\rangle$, and $\fsl{\partial}_s\langle\Psi_n(p,s)\overline{\Psi}_n(-p,t)\rangle$, as well as the equivalent expressions in $t$ are polynomials in $p$. The normalization is determined such that there exists gauge transformations (\ref{eq:mintpropamb}), (\ref{eq:mhalfpropamb}) between this gauge and the expressions (\ref{eq:mprop1}) and (\ref{eq:mhalfprop1}). It would be interesting to find the specific polynomials which appear in (\ref{eq:unitaryint}) -- (\ref{eq:unitaryprop6}).

\sm

The exact forms of the propagators in this physical gauge are however not necessary to have. Indeed, as long as interactions are introduced in a gauge invariant way, the propagators (\ref{eq:mprop1}) -- (\ref{eq:mprop6}), (\ref{eq:mhalfprop1}) -- (\ref{eq:mhalfprop6}) are sufficient. In the case of spin 1 particles, gauge invariance implies generalized Ward identities which guarantee that all Feynman diagrams with external legs with unphysical polarizations vanish \cite{tHooft:1971akt, tHooft:1971qjg}. Thus whenever perturbation theory is valid, the gauge invariant interacting theory can continue to describe a spin 1 particle at the level of the $S$-matrix. A natural expectation is that this general result will continue to hold for higher spins, although an independent proof of this is necessary. One complication of such a proof is that when interactions are introduced, one generally needs to correspondingly deform the gauge symmetry, as is what happens for instance in Yang-Mills theory and general relativity. The exact form of the generalized Ward identities in turn depend on the gauge symmetry. One should therefore first find a consistent set of interactions and corresponding deformations of the gauge symmetry in the way outlined for instance in \cite{Berends:1984rq}. Previous work in this direction includes \cite{Zinoviev:2006im, Zinoviev:2008ck, Cortese:2013lda, Cangemi:2022bew}. A particularly important application is finding consistent interactions of massive higher spins with electromagnetism and gravity to model black hole binary dynamics \cite{Guevara:2017csg,Chung:2018kqs, Guevara:2018wpp, Arkani-Hamed:2019ymq,Guevara:2019fsj,Bern:2020buy, Bern:2020uwk, Bern:2022kto, Chiodaroli:2021eug, FebresCordero:2022jts, Aoude:2022thd, Cangemi:2022bew}. An analysis of introducing interactions within this formulation is postponed for future work.

\sm

Finally, with the explicit expressions for all propagators, this work may be thought of as ``solving" every Poincar{\'e} and parity invariant free theory when, importantly, $d=4$. When $d > 5$, there are more irreducible representations of the little group which are not covered by totally symmetric representations, corresponding to Young tableaux with more than one row. In the future it would be useful to do the same for more general representations with mixed symmetry, along the lines of the formulation of massless particles in \cite{Labastida:1987kw}, to complete the program of solving every free theory. Having a complete understanding of such representations, much more control is to be gained when constructing more general theories with the same field content as string field theory by using a hyperfield $\Phi(X,\{s_i\})$ as described in \autoref{sec2}, and allow one to find classes of interactions which result in UV complete amplitudes, and avoids the causality violations inherent with interacting massive particles \cite{PhysRev.186.1337,Camanho:2014apa, Afkhami-Jeddi:2018apj}.

\subsection*{Acknowledgements}
The author would like to thank Lucile Cangemi, Henrik Johansson, Callum Jones, Ruslan Metsaev, Mojtaba Najafizadeh, Paolo Pichini, Massimo Porrati, and Radu Roiban for sharing useful references. The author would also like to thank Amey Gaikwad, Trevor Scheopner, and E.T. Tomboulis for helpful discussions during the preparation of this work. This research is supported in part by the Mani L. Bhaumik Institute for Theoretical Physics.

\bibliography{Allspins}

\begin{thebibliography}{67}%
\makeatletter
\providecommand \@ifxundefined [1]{%
 \@ifx{#1\undefined}
}%
\providecommand \@ifnum [1]{%
 \ifnum #1\expandafter \@firstoftwo
 \else \expandafter \@secondoftwo
 \fi
}%
\providecommand \@ifx [1]{%
 \ifx #1\expandafter \@firstoftwo
 \else \expandafter \@secondoftwo
 \fi
}%
\providecommand \natexlab [1]{#1}%
\providecommand \enquote  [1]{``#1''}%
\providecommand \bibnamefont  [1]{#1}%
\providecommand \bibfnamefont [1]{#1}%
\providecommand \citenamefont [1]{#1}%
\providecommand \href@noop [0]{\@secondoftwo}%
\providecommand \href [0]{\begingroup \@sanitize@url \@href}%
\providecommand \@href[1]{\@@startlink{#1}\@@href}%
\providecommand \@@href[1]{\endgroup#1\@@endlink}%
\providecommand \@sanitize@url [0]{\catcode `\\12\catcode `\$12\catcode `\&12\catcode `\#12\catcode `\^12\catcode `\_12\catcode `\%12\relax}%
\providecommand \@@startlink[1]{}%
\providecommand \@@endlink[0]{}%
\providecommand \url  [0]{\begingroup\@sanitize@url \@url }%
\providecommand \@url [1]{\endgroup\@href {#1}{\urlprefix }}%
\providecommand \urlprefix  [0]{URL }%
\providecommand \Eprint [0]{\href }%
\providecommand \doibase [0]{http://dx.doi.org/}%
\providecommand \selectlanguage [0]{\@gobble}%
\providecommand \bibinfo  [0]{\@secondoftwo}%
\providecommand \bibfield  [0]{\@secondoftwo}%
\providecommand \translation [1]{[#1]}%
\providecommand \BibitemOpen [0]{}%
\providecommand \bibitemStop [0]{}%
\providecommand \bibitemNoStop [0]{.\EOS\space}%
\providecommand \EOS [0]{\spacefactor3000\relax}%
\providecommand \BibitemShut  [1]{\csname bibitem#1\endcsname}%
\let\auto@bib@innerbib\@empty
\bibitem [{\citenamefont {Dirac}(1936)}]{Dirac:1936tg}%
  \BibitemOpen
  \bibfield  {author} {\bibinfo {author} {\bibfnamefont {P.~A.~M.}\ \bibnamefont {Dirac}},\ }\href {\doibase 10.1098/rspa.1936.0111} {\bibfield  {journal} {\bibinfo  {journal} {Proc. Roy. Soc. Lond. A}\ }\textbf {\bibinfo {volume} {155}},\ \bibinfo {pages} {447} (\bibinfo {year} {1936})}\BibitemShut {NoStop}%
\bibitem [{\citenamefont {Fierz}\ and\ \citenamefont {Pauli}(1939)}]{Fierz:1939ix}%
  \BibitemOpen
  \bibfield  {author} {\bibinfo {author} {\bibfnamefont {M.}~\bibnamefont {Fierz}}\ and\ \bibinfo {author} {\bibfnamefont {W.}~\bibnamefont {Pauli}},\ }\href {\doibase 10.1098/rspa.1939.0140} {\bibfield  {journal} {\bibinfo  {journal} {Proc. Roy. Soc. Lond. A}\ }\textbf {\bibinfo {volume} {173}},\ \bibinfo {pages} {211} (\bibinfo {year} {1939})}\BibitemShut {NoStop}%
\bibitem [{\citenamefont {Singh}\ and\ \citenamefont {Hagen}(1974{\natexlab{a}})}]{Singh:1974qz}%
  \BibitemOpen
  \bibfield  {author} {\bibinfo {author} {\bibfnamefont {L.~P.~S.}\ \bibnamefont {Singh}}\ and\ \bibinfo {author} {\bibfnamefont {C.~R.}\ \bibnamefont {Hagen}},\ }\href {\doibase 10.1103/PhysRevD.9.898} {\bibfield  {journal} {\bibinfo  {journal} {Phys. Rev. D}\ }\textbf {\bibinfo {volume} {9}},\ \bibinfo {pages} {898} (\bibinfo {year} {1974}{\natexlab{a}})}\BibitemShut {NoStop}%
\bibitem [{\citenamefont {Singh}\ and\ \citenamefont {Hagen}(1974{\natexlab{b}})}]{Singh:1974rc}%
  \BibitemOpen
  \bibfield  {author} {\bibinfo {author} {\bibfnamefont {L.~P.~S.}\ \bibnamefont {Singh}}\ and\ \bibinfo {author} {\bibfnamefont {C.~R.}\ \bibnamefont {Hagen}},\ }\href {\doibase 10.1103/PhysRevD.9.910} {\bibfield  {journal} {\bibinfo  {journal} {Phys. Rev. D}\ }\textbf {\bibinfo {volume} {9}},\ \bibinfo {pages} {910} (\bibinfo {year} {1974}{\natexlab{b}})}\BibitemShut {NoStop}%
\bibitem [{\citenamefont {Fronsdal}(1978)}]{Fronsdal:1978rb}%
  \BibitemOpen
  \bibfield  {author} {\bibinfo {author} {\bibfnamefont {C.}~\bibnamefont {Fronsdal}},\ }\href {\doibase 10.1103/PhysRevD.18.3624} {\bibfield  {journal} {\bibinfo  {journal} {Phys. Rev. D}\ }\textbf {\bibinfo {volume} {18}},\ \bibinfo {pages} {3624} (\bibinfo {year} {1978})}\BibitemShut {NoStop}%
\bibitem [{\citenamefont {Fang}\ and\ \citenamefont {Fronsdal}(1978)}]{Fang:1978wz}%
  \BibitemOpen
  \bibfield  {author} {\bibinfo {author} {\bibfnamefont {J.}~\bibnamefont {Fang}}\ and\ \bibinfo {author} {\bibfnamefont {C.}~\bibnamefont {Fronsdal}},\ }\href {\doibase 10.1103/PhysRevD.18.3630} {\bibfield  {journal} {\bibinfo  {journal} {Phys. Rev. D}\ }\textbf {\bibinfo {volume} {18}},\ \bibinfo {pages} {3630} (\bibinfo {year} {1978})}\BibitemShut {NoStop}%
\bibitem [{\citenamefont {Weinberg}(1964{\natexlab{a}})}]{Weinberg:1964ew}%
  \BibitemOpen
  \bibfield  {author} {\bibinfo {author} {\bibfnamefont {S.}~\bibnamefont {Weinberg}},\ }\href {\doibase 10.1103/PhysRev.135.B1049} {\bibfield  {journal} {\bibinfo  {journal} {Phys. Rev.}\ }\textbf {\bibinfo {volume} {135}},\ \bibinfo {pages} {B1049} (\bibinfo {year} {1964}{\natexlab{a}})}\BibitemShut {NoStop}%
\bibitem [{\citenamefont {Weinberg}\ and\ \citenamefont {Witten}(1980)}]{Weinberg:1980kq}%
  \BibitemOpen
  \bibfield  {author} {\bibinfo {author} {\bibfnamefont {S.}~\bibnamefont {Weinberg}}\ and\ \bibinfo {author} {\bibfnamefont {E.}~\bibnamefont {Witten}},\ }\href {\doibase 10.1016/0370-2693(80)90212-9} {\bibfield  {journal} {\bibinfo  {journal} {Phys. Lett. B}\ }\textbf {\bibinfo {volume} {96}},\ \bibinfo {pages} {59} (\bibinfo {year} {1980})}\BibitemShut {NoStop}%
\bibitem [{\citenamefont {Porrati}(2008)}]{Porrati:2008rm}%
  \BibitemOpen
  \bibfield  {author} {\bibinfo {author} {\bibfnamefont {M.}~\bibnamefont {Porrati}},\ }\href {\doibase 10.1103/PhysRevD.78.065016} {\bibfield  {journal} {\bibinfo  {journal} {Phys. Rev. D}\ }\textbf {\bibinfo {volume} {78}},\ \bibinfo {pages} {065016} (\bibinfo {year} {2008})},\ \Eprint {http://arxiv.org/abs/0804.4672}{arXiv:0804.4672 [hep-th]}\BibitemShut {NoStop}%
\bibitem [{\citenamefont {Vasiliev}(1990)}]{Vasiliev:1990en}%
  \BibitemOpen
  \bibfield  {author} {\bibinfo {author} {\bibfnamefont {M.~A.}\ \bibnamefont {Vasiliev}},\ }\href {\doibase 10.1016/0370-2693(90)91400-6} {\bibfield  {journal} {\bibinfo  {journal} {Phys. Lett. B}\ }\textbf {\bibinfo {volume} {243}},\ \bibinfo {pages} {378} (\bibinfo {year} {1990})}\BibitemShut {NoStop}%
\bibitem [{\citenamefont {Workman}\ \emph {et~al.}(2022)\citenamefont {Workman} \emph {et~al.}}]{ParticleDataGroup:2022pth}%
  \BibitemOpen
  \bibfield  {author} {\bibinfo {author} {\bibfnamefont {R.~L.}\ \bibnamefont {Workman}} \emph {et~al.} (\bibinfo {collaboration} {Particle Data Group}),\ }\href {\doibase 10.1093/ptep/ptac097} {\bibfield  {journal} {\bibinfo  {journal} {PTEP}\ }\textbf {\bibinfo {volume} {2022}},\ \bibinfo {pages} {083C01} (\bibinfo {year} {2022})}\BibitemShut {NoStop}%
\bibitem [{\citenamefont {Velo}\ and\ \citenamefont {Zwanziger}(1969)}]{PhysRev.186.1337}%
  \BibitemOpen
  \bibfield  {author} {\bibinfo {author} {\bibfnamefont {G.}~\bibnamefont {Velo}}\ and\ \bibinfo {author} {\bibfnamefont {D.}~\bibnamefont {Zwanziger}},\ }\href {\doibase 10.1103/PhysRev.186.1337} {\bibfield  {journal} {\bibinfo  {journal} {Phys. Rev.}\ }\textbf {\bibinfo {volume} {186}},\ \bibinfo {pages} {1337} (\bibinfo {year} {1969})}\BibitemShut {NoStop}%
\bibitem [{\citenamefont {Camanho}\ \emph {et~al.}(2016)\citenamefont {Camanho}, \citenamefont {Edelstein}, \citenamefont {Maldacena},\ and\ \citenamefont {Zhiboedov}}]{Camanho:2014apa}%
  \BibitemOpen
  \bibfield  {author} {\bibinfo {author} {\bibfnamefont {X.~O.}\ \bibnamefont {Camanho}}, \bibinfo {author} {\bibfnamefont {J.~D.}\ \bibnamefont {Edelstein}}, \bibinfo {author} {\bibfnamefont {J.}~\bibnamefont {Maldacena}}, \ and\ \bibinfo {author} {\bibfnamefont {A.}~\bibnamefont {Zhiboedov}},\ }\href {\doibase 10.1007/JHEP02(2016)020} {\bibfield  {journal} {\bibinfo  {journal} {JHEP}\ }\textbf {\bibinfo {volume} {02}},\ \bibinfo {pages} {020} (\bibinfo {year} {2016})},\ \Eprint {http://arxiv.org/abs/1407.5597}{arXiv:1407.5597 [hep-th]}\BibitemShut {NoStop}%
\bibitem [{\citenamefont {Afkhami-Jeddi}\ \emph {et~al.}(2019)\citenamefont {Afkhami-Jeddi}, \citenamefont {Kundu},\ and\ \citenamefont {Tajdini}}]{Afkhami-Jeddi:2018apj}%
  \BibitemOpen
  \bibfield  {author} {\bibinfo {author} {\bibfnamefont {N.}~\bibnamefont {Afkhami-Jeddi}}, \bibinfo {author} {\bibfnamefont {S.}~\bibnamefont {Kundu}}, \ and\ \bibinfo {author} {\bibfnamefont {A.}~\bibnamefont {Tajdini}},\ }\href {\doibase 10.1007/JHEP04(2019)056} {\bibfield  {journal} {\bibinfo  {journal} {JHEP}\ }\textbf {\bibinfo {volume} {04}},\ \bibinfo {pages} {056} (\bibinfo {year} {2019})},\ \Eprint {http://arxiv.org/abs/1811.01952}{arXiv:1811.01952 [hep-th]}\BibitemShut {NoStop}%
\bibitem [{\citenamefont {Deser}\ and\ \citenamefont {Zumino}(1977)}]{Deser:1977uq}%
  \BibitemOpen
  \bibfield  {author} {\bibinfo {author} {\bibfnamefont {S.}~\bibnamefont {Deser}}\ and\ \bibinfo {author} {\bibfnamefont {B.}~\bibnamefont {Zumino}},\ }\href {\doibase 10.1103/PhysRevLett.38.1433} {\bibfield  {journal} {\bibinfo  {journal} {Phys. Rev. Lett.}\ }\textbf {\bibinfo {volume} {38}},\ \bibinfo {pages} {1433} (\bibinfo {year} {1977})}\BibitemShut {NoStop}%
\bibitem [{\citenamefont {Porrati}\ \emph {et~al.}(2011)\citenamefont {Porrati}, \citenamefont {Rahman},\ and\ \citenamefont {Sagnotti}}]{Porrati:2010hm}%
  \BibitemOpen
  \bibfield  {author} {\bibinfo {author} {\bibfnamefont {M.}~\bibnamefont {Porrati}}, \bibinfo {author} {\bibfnamefont {R.}~\bibnamefont {Rahman}}, \ and\ \bibinfo {author} {\bibfnamefont {A.}~\bibnamefont {Sagnotti}},\ }\href {\doibase 10.1016/j.nuclphysb.2011.01.007} {\bibfield  {journal} {\bibinfo  {journal} {Nucl. Phys. B}\ }\textbf {\bibinfo {volume} {846}},\ \bibinfo {pages} {250} (\bibinfo {year} {2011})},\ \Eprint {http://arxiv.org/abs/1011.6411}{arXiv:1011.6411 [hep-th]}\BibitemShut {NoStop}%
\bibitem [{\citenamefont {Singh}(1981)}]{Singh:1981aw}%
  \BibitemOpen
  \bibfield  {author} {\bibinfo {author} {\bibfnamefont {L.~P.~S.}\ \bibnamefont {Singh}},\ }\href {\doibase 10.1103/PhysRevD.23.2236} {\bibfield  {journal} {\bibinfo  {journal} {Phys. Rev. D}\ }\textbf {\bibinfo {volume} {23}},\ \bibinfo {pages} {2236} (\bibinfo {year} {1981})}\BibitemShut {NoStop}%
\bibitem [{\citenamefont {Weinberg}(1964{\natexlab{b}})}]{Weinberg:1964cn}%
  \BibitemOpen
  \bibfield  {author} {\bibinfo {author} {\bibfnamefont {S.}~\bibnamefont {Weinberg}},\ }\href {\doibase 10.1103/PhysRev.133.B1318} {\bibfield  {journal} {\bibinfo  {journal} {Phys. Rev.}\ }\textbf {\bibinfo {volume} {133}},\ \bibinfo {pages} {B1318} (\bibinfo {year} {1964}{\natexlab{b}})}\BibitemShut {NoStop}%
\bibitem [{\citenamefont {Sagnotti}\ and\ \citenamefont {Taronna}(2011)}]{Sagnotti:2010at}%
  \BibitemOpen
  \bibfield  {author} {\bibinfo {author} {\bibfnamefont {A.}~\bibnamefont {Sagnotti}}\ and\ \bibinfo {author} {\bibfnamefont {M.}~\bibnamefont {Taronna}},\ }\href {\doibase 10.1016/j.nuclphysb.2010.08.019} {\bibfield  {journal} {\bibinfo  {journal} {Nucl. Phys. B}\ }\textbf {\bibinfo {volume} {842}},\ \bibinfo {pages} {299} (\bibinfo {year} {2011})},\ \Eprint {http://arxiv.org/abs/1006.5242}{arXiv:1006.5242 [hep-th]}\BibitemShut {NoStop}%
\bibitem [{\citenamefont {Szeg\"{o}}(1939)}]{nla.cat-vn2358422}%
  \BibitemOpen
  \bibfield  {author} {\bibinfo {author} {\bibfnamefont {G.}~\bibnamefont {Szeg\"{o}}},\ }\href@noop {} {\emph {\bibinfo {title} {Orthogonal polynomials}}},\ \bibinfo {edition} {4th}\ ed.\ (\bibinfo  {publisher} {American Mathematical Society Providence},\ \bibinfo {year} {1939})\ pp.\ \bibinfo {pages} {xiii, 432 p. ;}\BibitemShut {NoStop}%
\bibitem [{\citenamefont {Guevara}(2019)}]{Guevara:2017csg}%
  \BibitemOpen
  \bibfield  {author} {\bibinfo {author} {\bibfnamefont {A.}~\bibnamefont {Guevara}},\ }\href {\doibase 10.1007/JHEP04(2019)033} {\bibfield  {journal} {\bibinfo  {journal} {JHEP}\ }\textbf {\bibinfo {volume} {04}},\ \bibinfo {pages} {033} (\bibinfo {year} {2019})},\ \Eprint {http://arxiv.org/abs/1706.02314}{arXiv:1706.02314 [hep-th]}\BibitemShut {NoStop}%
\bibitem [{\citenamefont {Chung}\ \emph {et~al.}(2019)\citenamefont {Chung}, \citenamefont {Huang}, \citenamefont {Kim},\ and\ \citenamefont {Lee}}]{Chung:2018kqs}%
  \BibitemOpen
  \bibfield  {author} {\bibinfo {author} {\bibfnamefont {M.-Z.}\ \bibnamefont {Chung}}, \bibinfo {author} {\bibfnamefont {Y.-T.}\ \bibnamefont {Huang}}, \bibinfo {author} {\bibfnamefont {J.-W.}\ \bibnamefont {Kim}}, \ and\ \bibinfo {author} {\bibfnamefont {S.}~\bibnamefont {Lee}},\ }\href {\doibase 10.1007/JHEP04(2019)156} {\bibfield  {journal} {\bibinfo  {journal} {JHEP}\ }\textbf {\bibinfo {volume} {04}},\ \bibinfo {pages} {156} (\bibinfo {year} {2019})},\ \Eprint {http://arxiv.org/abs/1812.08752}{arXiv:1812.08752 [hep-th]}\BibitemShut {NoStop}%
\bibitem [{\citenamefont {Guevara}\ \emph {et~al.}(2019{\natexlab{a}})\citenamefont {Guevara}, \citenamefont {Ochirov},\ and\ \citenamefont {Vines}}]{Guevara:2018wpp}%
  \BibitemOpen
  \bibfield  {author} {\bibinfo {author} {\bibfnamefont {A.}~\bibnamefont {Guevara}}, \bibinfo {author} {\bibfnamefont {A.}~\bibnamefont {Ochirov}}, \ and\ \bibinfo {author} {\bibfnamefont {J.}~\bibnamefont {Vines}},\ }\href {\doibase 10.1007/JHEP09(2019)056} {\bibfield  {journal} {\bibinfo  {journal} {JHEP}\ }\textbf {\bibinfo {volume} {09}},\ \bibinfo {pages} {056} (\bibinfo {year} {2019}{\natexlab{a}})},\ \Eprint {http://arxiv.org/abs/1812.06895}{arXiv:1812.06895 [hep-th]}\BibitemShut {NoStop}%
\bibitem [{\citenamefont {Arkani-Hamed}\ \emph {et~al.}(2020)\citenamefont {Arkani-Hamed}, \citenamefont {Huang},\ and\ \citenamefont {O'Connell}}]{Arkani-Hamed:2019ymq}%
  \BibitemOpen
  \bibfield  {author} {\bibinfo {author} {\bibfnamefont {N.}~\bibnamefont {Arkani-Hamed}}, \bibinfo {author} {\bibfnamefont {Y.-t.}\ \bibnamefont {Huang}}, \ and\ \bibinfo {author} {\bibfnamefont {D.}~\bibnamefont {O'Connell}},\ }\href {\doibase 10.1007/JHEP01(2020)046} {\bibfield  {journal} {\bibinfo  {journal} {JHEP}\ }\textbf {\bibinfo {volume} {01}},\ \bibinfo {pages} {046} (\bibinfo {year} {2020})},\ \Eprint {http://arxiv.org/abs/1906.10100}{arXiv:1906.10100 [hep-th]}\BibitemShut {NoStop}%
\bibitem [{\citenamefont {Guevara}\ \emph {et~al.}(2019{\natexlab{b}})\citenamefont {Guevara}, \citenamefont {Ochirov},\ and\ \citenamefont {Vines}}]{Guevara:2019fsj}%
  \BibitemOpen
  \bibfield  {author} {\bibinfo {author} {\bibfnamefont {A.}~\bibnamefont {Guevara}}, \bibinfo {author} {\bibfnamefont {A.}~\bibnamefont {Ochirov}}, \ and\ \bibinfo {author} {\bibfnamefont {J.}~\bibnamefont {Vines}},\ }\href {\doibase 10.1103/PhysRevD.100.104024} {\bibfield  {journal} {\bibinfo  {journal} {Phys. Rev. D}\ }\textbf {\bibinfo {volume} {100}},\ \bibinfo {pages} {104024} (\bibinfo {year} {2019}{\natexlab{b}})},\ \Eprint {http://arxiv.org/abs/1906.10071}{arXiv:1906.10071 [hep-th]}\BibitemShut {NoStop}%
\bibitem [{\citenamefont {Bern}\ \emph {et~al.}(2021{\natexlab{a}})\citenamefont {Bern}, \citenamefont {Luna}, \citenamefont {Roiban}, \citenamefont {Shen},\ and\ \citenamefont {Zeng}}]{Bern:2020buy}%
  \BibitemOpen
  \bibfield  {author} {\bibinfo {author} {\bibfnamefont {Z.}~\bibnamefont {Bern}}, \bibinfo {author} {\bibfnamefont {A.}~\bibnamefont {Luna}}, \bibinfo {author} {\bibfnamefont {R.}~\bibnamefont {Roiban}}, \bibinfo {author} {\bibfnamefont {C.-H.}\ \bibnamefont {Shen}}, \ and\ \bibinfo {author} {\bibfnamefont {M.}~\bibnamefont {Zeng}},\ }\href {\doibase 10.1103/PhysRevD.104.065014} {\bibfield  {journal} {\bibinfo  {journal} {Phys. Rev. D}\ }\textbf {\bibinfo {volume} {104}},\ \bibinfo {pages} {065014} (\bibinfo {year} {2021}{\natexlab{a}})},\ \Eprint {http://arxiv.org/abs/2005.03071}{arXiv:2005.03071 [hep-th]}\BibitemShut {NoStop}%
\bibitem [{\citenamefont {Bern}\ \emph {et~al.}(2021{\natexlab{b}})\citenamefont {Bern}, \citenamefont {Parra-Martinez}, \citenamefont {Roiban}, \citenamefont {Sawyer},\ and\ \citenamefont {Shen}}]{Bern:2020uwk}%
  \BibitemOpen
  \bibfield  {author} {\bibinfo {author} {\bibfnamefont {Z.}~\bibnamefont {Bern}}, \bibinfo {author} {\bibfnamefont {J.}~\bibnamefont {Parra-Martinez}}, \bibinfo {author} {\bibfnamefont {R.}~\bibnamefont {Roiban}}, \bibinfo {author} {\bibfnamefont {E.}~\bibnamefont {Sawyer}}, \ and\ \bibinfo {author} {\bibfnamefont {C.-H.}\ \bibnamefont {Shen}},\ }\href {\doibase 10.1007/JHEP05(2021)188} {\bibfield  {journal} {\bibinfo  {journal} {JHEP}\ }\textbf {\bibinfo {volume} {05}},\ \bibinfo {pages} {188} (\bibinfo {year} {2021}{\natexlab{b}})},\ \Eprint {http://arxiv.org/abs/2010.08559}{arXiv:2010.08559 [hep-th]}\BibitemShut {NoStop}%
\bibitem [{\citenamefont {Bern}\ \emph {et~al.}(2023)\citenamefont {Bern}, \citenamefont {Kosmopoulos}, \citenamefont {Luna}, \citenamefont {Roiban},\ and\ \citenamefont {Teng}}]{Bern:2022kto}%
  \BibitemOpen
  \bibfield  {author} {\bibinfo {author} {\bibfnamefont {Z.}~\bibnamefont {Bern}}, \bibinfo {author} {\bibfnamefont {D.}~\bibnamefont {Kosmopoulos}}, \bibinfo {author} {\bibfnamefont {A.}~\bibnamefont {Luna}}, \bibinfo {author} {\bibfnamefont {R.}~\bibnamefont {Roiban}}, \ and\ \bibinfo {author} {\bibfnamefont {F.}~\bibnamefont {Teng}},\ }\href {\doibase 10.1103/PhysRevLett.130.201402} {\bibfield  {journal} {\bibinfo  {journal} {Phys. Rev. Lett.}\ }\textbf {\bibinfo {volume} {130}},\ \bibinfo {pages} {201402} (\bibinfo {year} {2023})},\ \Eprint {http://arxiv.org/abs/2203.06202}{arXiv:2203.06202 [hep-th]}\BibitemShut {NoStop}%
\bibitem [{\citenamefont {Chiodaroli}\ \emph {et~al.}(2022)\citenamefont {Chiodaroli}, \citenamefont {Johansson},\ and\ \citenamefont {Pichini}}]{Chiodaroli:2021eug}%
  \BibitemOpen
  \bibfield  {author} {\bibinfo {author} {\bibfnamefont {M.}~\bibnamefont {Chiodaroli}}, \bibinfo {author} {\bibfnamefont {H.}~\bibnamefont {Johansson}}, \ and\ \bibinfo {author} {\bibfnamefont {P.}~\bibnamefont {Pichini}},\ }\href {\doibase 10.1007/JHEP02(2022)156} {\bibfield  {journal} {\bibinfo  {journal} {JHEP}\ }\textbf {\bibinfo {volume} {02}},\ \bibinfo {pages} {156} (\bibinfo {year} {2022})},\ \Eprint {http://arxiv.org/abs/2107.14779}{arXiv:2107.14779 [hep-th]}\BibitemShut {NoStop}%
\bibitem [{\citenamefont {Cordero}\ \emph {et~al.}(2023)\citenamefont {Cordero}, \citenamefont {Kraus}, \citenamefont {Lin}, \citenamefont {Ruf},\ and\ \citenamefont {Zeng}}]{FebresCordero:2022jts}%
  \BibitemOpen
  \bibfield  {author} {\bibinfo {author} {\bibfnamefont {F.~F.}\ \bibnamefont {Cordero}}, \bibinfo {author} {\bibfnamefont {M.}~\bibnamefont {Kraus}}, \bibinfo {author} {\bibfnamefont {G.}~\bibnamefont {Lin}}, \bibinfo {author} {\bibfnamefont {M.~S.}\ \bibnamefont {Ruf}}, \ and\ \bibinfo {author} {\bibfnamefont {M.}~\bibnamefont {Zeng}},\ }\href {\doibase 10.1103/PhysRevLett.130.021601} {\bibfield  {journal} {\bibinfo  {journal} {Phys. Rev. Lett.}\ }\textbf {\bibinfo {volume} {130}},\ \bibinfo {pages} {021601} (\bibinfo {year} {2023})},\ \Eprint {http://arxiv.org/abs/2205.07357}{arXiv:2205.07357 [hep-th]}\BibitemShut {NoStop}%
\bibitem [{\citenamefont {Aoude}\ \emph {et~al.}(2022)\citenamefont {Aoude}, \citenamefont {Haddad},\ and\ \citenamefont {Helset}}]{Aoude:2022thd}%
  \BibitemOpen
  \bibfield  {author} {\bibinfo {author} {\bibfnamefont {R.}~\bibnamefont {Aoude}}, \bibinfo {author} {\bibfnamefont {K.}~\bibnamefont {Haddad}}, \ and\ \bibinfo {author} {\bibfnamefont {A.}~\bibnamefont {Helset}},\ }\href {\doibase 10.1103/PhysRevLett.129.141102} {\bibfield  {journal} {\bibinfo  {journal} {Phys. Rev. Lett.}\ }\textbf {\bibinfo {volume} {129}},\ \bibinfo {pages} {141102} (\bibinfo {year} {2022})},\ \Eprint {http://arxiv.org/abs/2205.02809}{arXiv:2205.02809 [hep-th]}\BibitemShut {NoStop}%
\bibitem [{\citenamefont {Cangemi}\ \emph {et~al.}(2022)\citenamefont {Cangemi}, \citenamefont {Chiodaroli}, \citenamefont {Johansson}, \citenamefont {Ochirov}, \citenamefont {Pichini},\ and\ \citenamefont {Skvortsov}}]{Cangemi:2022bew}%
  \BibitemOpen
  \bibfield  {author} {\bibinfo {author} {\bibfnamefont {L.}~\bibnamefont {Cangemi}}, \bibinfo {author} {\bibfnamefont {M.}~\bibnamefont {Chiodaroli}}, \bibinfo {author} {\bibfnamefont {H.}~\bibnamefont {Johansson}}, \bibinfo {author} {\bibfnamefont {A.}~\bibnamefont {Ochirov}}, \bibinfo {author} {\bibfnamefont {P.}~\bibnamefont {Pichini}}, \ and\ \bibinfo {author} {\bibfnamefont {E.}~\bibnamefont {Skvortsov}},\ }\href@noop {} {\  (\bibinfo {year} {2022})},\ \Eprint {http://arxiv.org/abs/2212.06120}{arXiv:2212.06120 [hep-th]}\BibitemShut {NoStop}%
\bibitem [{\citenamefont {Klishevich}(2000{\natexlab{a}})}]{Klishevich:1998ng}%
  \BibitemOpen
  \bibfield  {author} {\bibinfo {author} {\bibfnamefont {S.~M.}\ \bibnamefont {Klishevich}},\ }\href {\doibase 10.1142/S0217751X00000264} {\bibfield  {journal} {\bibinfo  {journal} {Int. J. Mod. Phys. A}\ }\textbf {\bibinfo {volume} {15}},\ \bibinfo {pages} {535} (\bibinfo {year} {2000}{\natexlab{a}})},\ \Eprint {http://arxiv.org/abs/hep-th/9810228}{arXiv:hep-th/9810228}\BibitemShut {NoStop}%
\bibitem [{\citenamefont {Klishevich}(2000{\natexlab{b}})}]{Klishevich:1998yt}%
  \BibitemOpen
  \bibfield  {author} {\bibinfo {author} {\bibfnamefont {S.~M.}\ \bibnamefont {Klishevich}},\ }\href {\doibase 10.1142/S0217751X00000306} {\bibfield  {journal} {\bibinfo  {journal} {Int. J. Mod. Phys. A}\ }\textbf {\bibinfo {volume} {15}},\ \bibinfo {pages} {609} (\bibinfo {year} {2000}{\natexlab{b}})},\ \Eprint {http://arxiv.org/abs/hep-th/9811030}{arXiv:hep-th/9811030}\BibitemShut {NoStop}%
\bibitem [{\citenamefont {Zinoviev}(2001)}]{Zinoviev:2001dt}%
  \BibitemOpen
  \bibfield  {author} {\bibinfo {author} {\bibfnamefont {Y.~M.}\ \bibnamefont {Zinoviev}},\ }\href@noop {} {\  (\bibinfo {year} {2001})},\ \Eprint {http://arxiv.org/abs/hep-th/0108192}{arXiv:hep-th/0108192}\BibitemShut {NoStop}%
\bibitem [{\citenamefont {Metsaev}(2006)}]{Metsaev:2006zy}%
  \BibitemOpen
  \bibfield  {author} {\bibinfo {author} {\bibfnamefont {R.~R.}\ \bibnamefont {Metsaev}},\ }\href {\doibase 10.1016/j.physletb.2006.11.002} {\bibfield  {journal} {\bibinfo  {journal} {Phys. Lett. B}\ }\textbf {\bibinfo {volume} {643}},\ \bibinfo {pages} {205} (\bibinfo {year} {2006})},\ \Eprint {http://arxiv.org/abs/hep-th/0609029}{arXiv:hep-th/0609029}\BibitemShut {NoStop}%
\bibitem [{\citenamefont {Asano}(2019)}]{Asano:2019smc}%
  \BibitemOpen
  \bibfield  {author} {\bibinfo {author} {\bibfnamefont {M.}~\bibnamefont {Asano}},\ }\href {\doibase 10.1007/JHEP04(2019)051} {\bibfield  {journal} {\bibinfo  {journal} {JHEP}\ }\textbf {\bibinfo {volume} {04}},\ \bibinfo {pages} {051} (\bibinfo {year} {2019})},\ \Eprint {http://arxiv.org/abs/1902.05685}{arXiv:1902.05685 [hep-th]}\BibitemShut {NoStop}%
\bibitem [{\citenamefont {Buchbinder}\ \emph {et~al.}(2000)\citenamefont {Buchbinder}, \citenamefont {Gitman},\ and\ \citenamefont {Pershin}}]{Buchbinder:2000fy}%
  \BibitemOpen
  \bibfield  {author} {\bibinfo {author} {\bibfnamefont {I.~L.}\ \bibnamefont {Buchbinder}}, \bibinfo {author} {\bibfnamefont {D.~M.}\ \bibnamefont {Gitman}}, \ and\ \bibinfo {author} {\bibfnamefont {V.~D.}\ \bibnamefont {Pershin}},\ }\href {\doibase 10.1016/S0370-2693(00)01082-0} {\bibfield  {journal} {\bibinfo  {journal} {Phys. Lett. B}\ }\textbf {\bibinfo {volume} {492}},\ \bibinfo {pages} {161} (\bibinfo {year} {2000})},\ \Eprint {http://arxiv.org/abs/hep-th/0006144}{arXiv:hep-th/0006144}\BibitemShut {NoStop}%
\bibitem [{\citenamefont {Zinoviev}(2007)}]{Zinoviev:2006im}%
  \BibitemOpen
  \bibfield  {author} {\bibinfo {author} {\bibfnamefont {Y.~M.}\ \bibnamefont {Zinoviev}},\ }\href {\doibase 10.1016/j.nuclphysb.2007.02.005} {\bibfield  {journal} {\bibinfo  {journal} {Nucl. Phys. B}\ }\textbf {\bibinfo {volume} {770}},\ \bibinfo {pages} {83} (\bibinfo {year} {2007})},\ \Eprint {http://arxiv.org/abs/hep-th/0609170}{arXiv:hep-th/0609170}\BibitemShut {NoStop}%
\bibitem [{\citenamefont {Zinoviev}(2009)}]{Zinoviev:2008ck}%
  \BibitemOpen
  \bibfield  {author} {\bibinfo {author} {\bibfnamefont {Y.~M.}\ \bibnamefont {Zinoviev}},\ }\href {\doibase 10.1088/0264-9381/26/3/035022} {\bibfield  {journal} {\bibinfo  {journal} {Class. Quant. Grav.}\ }\textbf {\bibinfo {volume} {26}},\ \bibinfo {pages} {035022} (\bibinfo {year} {2009})},\ \Eprint {http://arxiv.org/abs/0805.2226}{arXiv:0805.2226 [hep-th]}\BibitemShut {NoStop}%
\bibitem [{\citenamefont {Aragone}\ \emph {et~al.}(1987)\citenamefont {Aragone}, \citenamefont {Deser},\ and\ \citenamefont {Yang}}]{Aragone:1987dtt}%
  \BibitemOpen
  \bibfield  {author} {\bibinfo {author} {\bibfnamefont {C.}~\bibnamefont {Aragone}}, \bibinfo {author} {\bibfnamefont {S.}~\bibnamefont {Deser}}, \ and\ \bibinfo {author} {\bibfnamefont {Z.}~\bibnamefont {Yang}},\ }\href {\doibase 10.1016/S0003-4916(87)80005-2} {\bibfield  {journal} {\bibinfo  {journal} {Annals Phys.}\ }\textbf {\bibinfo {volume} {179}},\ \bibinfo {pages} {76} (\bibinfo {year} {1987})}\BibitemShut {NoStop}%
\bibitem [{\citenamefont {Rindani}\ \emph {et~al.}(1989{\natexlab{a}})\citenamefont {Rindani}, \citenamefont {Sahdev},\ and\ \citenamefont {Sivakumar}}]{Rindani:1988gb}%
  \BibitemOpen
  \bibfield  {author} {\bibinfo {author} {\bibfnamefont {S.~D.}\ \bibnamefont {Rindani}}, \bibinfo {author} {\bibfnamefont {D.}~\bibnamefont {Sahdev}}, \ and\ \bibinfo {author} {\bibfnamefont {M.}~\bibnamefont {Sivakumar}},\ }\href {\doibase 10.1142/S0217732389000332} {\bibfield  {journal} {\bibinfo  {journal} {Mod. Phys. Lett. A}\ }\textbf {\bibinfo {volume} {4}},\ \bibinfo {pages} {265} (\bibinfo {year} {1989}{\natexlab{a}})}\BibitemShut {NoStop}%
\bibitem [{\citenamefont {Rindani}\ \emph {et~al.}(1989{\natexlab{b}})\citenamefont {Rindani}, \citenamefont {Sivakumar},\ and\ \citenamefont {Sahdev}}]{Rindani:1989ym}%
  \BibitemOpen
  \bibfield  {author} {\bibinfo {author} {\bibfnamefont {S.~D.}\ \bibnamefont {Rindani}}, \bibinfo {author} {\bibfnamefont {M.}~\bibnamefont {Sivakumar}}, \ and\ \bibinfo {author} {\bibfnamefont {D.}~\bibnamefont {Sahdev}},\ }\href {\doibase 10.1142/S0217732389000344} {\bibfield  {journal} {\bibinfo  {journal} {Mod. Phys. Lett. A}\ }\textbf {\bibinfo {volume} {4}},\ \bibinfo {pages} {275} (\bibinfo {year} {1989}{\natexlab{b}})}\BibitemShut {NoStop}%
\bibitem [{\citenamefont {Bekaert}\ \emph {et~al.}(2004)\citenamefont {Bekaert}, \citenamefont {Buchbinder}, \citenamefont {Pashnev},\ and\ \citenamefont {Tsulaia}}]{Bekaert:2003uc}%
  \BibitemOpen
  \bibfield  {author} {\bibinfo {author} {\bibfnamefont {X.}~\bibnamefont {Bekaert}}, \bibinfo {author} {\bibfnamefont {I.~L.}\ \bibnamefont {Buchbinder}}, \bibinfo {author} {\bibfnamefont {A.}~\bibnamefont {Pashnev}}, \ and\ \bibinfo {author} {\bibfnamefont {M.}~\bibnamefont {Tsulaia}},\ }\href {\doibase 10.1088/0264-9381/21/10/018} {\bibfield  {journal} {\bibinfo  {journal} {Class. Quant. Grav.}\ }\textbf {\bibinfo {volume} {21}},\ \bibinfo {pages} {S1457} (\bibinfo {year} {2004})},\ \Eprint {http://arxiv.org/abs/hep-th/0312252}{arXiv:hep-th/0312252}\BibitemShut {NoStop}%
\bibitem [{\citenamefont {Buchbinder}\ and\ \citenamefont {Galajinsky}(2008)}]{Buchbinder:2008ss}%
  \BibitemOpen
  \bibfield  {author} {\bibinfo {author} {\bibfnamefont {I.~L.}\ \bibnamefont {Buchbinder}}\ and\ \bibinfo {author} {\bibfnamefont {A.~V.}\ \bibnamefont {Galajinsky}},\ }\href {\doibase 10.1088/1126-6708/2008/11/081} {\bibfield  {journal} {\bibinfo  {journal} {JHEP}\ }\textbf {\bibinfo {volume} {11}},\ \bibinfo {pages} {081} (\bibinfo {year} {2008})},\ \Eprint {http://arxiv.org/abs/0810.2852}{arXiv:0810.2852 [hep-th]}\BibitemShut {NoStop}%
\bibitem [{\citenamefont {Bekaert}\ and\ \citenamefont {Mourad}(2006)}]{Bekaert:2005in}%
  \BibitemOpen
  \bibfield  {author} {\bibinfo {author} {\bibfnamefont {X.}~\bibnamefont {Bekaert}}\ and\ \bibinfo {author} {\bibfnamefont {J.}~\bibnamefont {Mourad}},\ }\href {\doibase 10.1088/1126-6708/2006/01/115} {\bibfield  {journal} {\bibinfo  {journal} {JHEP}\ }\textbf {\bibinfo {volume} {01}},\ \bibinfo {pages} {115} (\bibinfo {year} {2006})},\ \Eprint {http://arxiv.org/abs/hep-th/0509092}{arXiv:hep-th/0509092}\BibitemShut {NoStop}%
\bibitem [{\citenamefont {Ponomarev}\ and\ \citenamefont {Tseytlin}(2016)}]{Ponomarev:2016jqk}%
  \BibitemOpen
  \bibfield  {author} {\bibinfo {author} {\bibfnamefont {D.}~\bibnamefont {Ponomarev}}\ and\ \bibinfo {author} {\bibfnamefont {A.~A.}\ \bibnamefont {Tseytlin}},\ }\href {\doibase 10.1007/JHEP05(2016)184} {\bibfield  {journal} {\bibinfo  {journal} {JHEP}\ }\textbf {\bibinfo {volume} {05}},\ \bibinfo {pages} {184} (\bibinfo {year} {2016})},\ \Eprint {http://arxiv.org/abs/1603.06273}{arXiv:1603.06273 [hep-th]}\BibitemShut {NoStop}%
\bibitem [{\citenamefont {Roiban}\ and\ \citenamefont {Tseytlin}(2017)}]{Roiban:2017iqg}%
  \BibitemOpen
  \bibfield  {author} {\bibinfo {author} {\bibfnamefont {R.}~\bibnamefont {Roiban}}\ and\ \bibinfo {author} {\bibfnamefont {A.~A.}\ \bibnamefont {Tseytlin}},\ }\href {\doibase 10.1007/JHEP04(2017)139} {\bibfield  {journal} {\bibinfo  {journal} {JHEP}\ }\textbf {\bibinfo {volume} {04}},\ \bibinfo {pages} {139} (\bibinfo {year} {2017})},\ \Eprint {http://arxiv.org/abs/1701.05773}{arXiv:1701.05773 [hep-th]}\BibitemShut {NoStop}%
\bibitem [{\citenamefont {Najafizadeh}(2018)}]{Najafizadeh:2018cpu}%
  \BibitemOpen
  \bibfield  {author} {\bibinfo {author} {\bibfnamefont {M.}~\bibnamefont {Najafizadeh}},\ }\href {\doibase 10.1103/PhysRevD.98.125012} {\bibfield  {journal} {\bibinfo  {journal} {Phys. Rev. D}\ }\textbf {\bibinfo {volume} {98}},\ \bibinfo {pages} {125012} (\bibinfo {year} {2018})},\ \Eprint {http://arxiv.org/abs/1807.01124}{arXiv:1807.01124 [hep-th]}\BibitemShut {NoStop}%
\bibitem [{\citenamefont {Veneziano}(1968)}]{Veneziano:1968yb}%
  \BibitemOpen
  \bibfield  {author} {\bibinfo {author} {\bibfnamefont {G.}~\bibnamefont {Veneziano}},\ }\href {\doibase 10.1007/BF02824451} {\bibfield  {journal} {\bibinfo  {journal} {Nuovo Cim. A}\ }\textbf {\bibinfo {volume} {57}},\ \bibinfo {pages} {190} (\bibinfo {year} {1968})}\BibitemShut {NoStop}%
\bibitem [{\citenamefont {Virasoro}(1969)}]{Virasoro:1969me}%
  \BibitemOpen
  \bibfield  {author} {\bibinfo {author} {\bibfnamefont {M.~A.}\ \bibnamefont {Virasoro}},\ }\href {\doibase 10.1103/PhysRev.177.2309} {\bibfield  {journal} {\bibinfo  {journal} {Phys. Rev.}\ }\textbf {\bibinfo {volume} {177}},\ \bibinfo {pages} {2309} (\bibinfo {year} {1969})}\BibitemShut {NoStop}%
\bibitem [{\citenamefont {Coon}(1969)}]{Coon:1969yw}%
  \BibitemOpen
  \bibfield  {author} {\bibinfo {author} {\bibfnamefont {D.~D.}\ \bibnamefont {Coon}},\ }\href {\doibase 10.1016/0370-2693(69)90106-3} {\bibfield  {journal} {\bibinfo  {journal} {Phys. Lett. B}\ }\textbf {\bibinfo {volume} {29}},\ \bibinfo {pages} {669} (\bibinfo {year} {1969})}\BibitemShut {NoStop}%
\bibitem [{\citenamefont {Arkani-Hamed}\ \emph {et~al.}(2021)\citenamefont {Arkani-Hamed}, \citenamefont {Huang},\ and\ \citenamefont {Huang}}]{Arkani-Hamed:2017jhn}%
  \BibitemOpen
  \bibfield  {author} {\bibinfo {author} {\bibfnamefont {N.}~\bibnamefont {Arkani-Hamed}}, \bibinfo {author} {\bibfnamefont {T.-C.}\ \bibnamefont {Huang}}, \ and\ \bibinfo {author} {\bibfnamefont {Y.-t.}\ \bibnamefont {Huang}},\ }\href {\doibase 10.1007/JHEP11(2021)070} {\bibfield  {journal} {\bibinfo  {journal} {JHEP}\ }\textbf {\bibinfo {volume} {11}},\ \bibinfo {pages} {070} (\bibinfo {year} {2021})},\ \Eprint {http://arxiv.org/abs/1709.04891}{arXiv:1709.04891 [hep-th]}\BibitemShut {NoStop}%
\bibitem [{\citenamefont {Caron-Huot}\ \emph {et~al.}(2017)\citenamefont {Caron-Huot}, \citenamefont {Komargodski}, \citenamefont {Sever},\ and\ \citenamefont {Zhiboedov}}]{Caron-Huot:2016icg}%
  \BibitemOpen
  \bibfield  {author} {\bibinfo {author} {\bibfnamefont {S.}~\bibnamefont {Caron-Huot}}, \bibinfo {author} {\bibfnamefont {Z.}~\bibnamefont {Komargodski}}, \bibinfo {author} {\bibfnamefont {A.}~\bibnamefont {Sever}}, \ and\ \bibinfo {author} {\bibfnamefont {A.}~\bibnamefont {Zhiboedov}},\ }\href {\doibase 10.1007/JHEP10(2017)026} {\bibfield  {journal} {\bibinfo  {journal} {JHEP}\ }\textbf {\bibinfo {volume} {10}},\ \bibinfo {pages} {026} (\bibinfo {year} {2017})},\ \Eprint {http://arxiv.org/abs/1607.04253}{arXiv:1607.04253 [hep-th]}\BibitemShut {NoStop}%
\bibitem [{\citenamefont {Witten}(1986)}]{Witten:1985cc}%
  \BibitemOpen
  \bibfield  {author} {\bibinfo {author} {\bibfnamefont {E.}~\bibnamefont {Witten}},\ }\href {\doibase 10.1016/0550-3213(86)90155-0} {\bibfield  {journal} {\bibinfo  {journal} {Nucl. Phys. B}\ }\textbf {\bibinfo {volume} {268}},\ \bibinfo {pages} {253} (\bibinfo {year} {1986})}\BibitemShut {NoStop}%
\bibitem [{\citenamefont {Cheung}\ and\ \citenamefont {Remmen}(2023{\natexlab{a}})}]{Cheung:2022mkw}%
  \BibitemOpen
  \bibfield  {author} {\bibinfo {author} {\bibfnamefont {C.}~\bibnamefont {Cheung}}\ and\ \bibinfo {author} {\bibfnamefont {G.~N.}\ \bibnamefont {Remmen}},\ }\href {\doibase 10.1007/JHEP01(2023)122} {\bibfield  {journal} {\bibinfo  {journal} {JHEP}\ }\textbf {\bibinfo {volume} {01}},\ \bibinfo {pages} {122} (\bibinfo {year} {2023}{\natexlab{a}})},\ \Eprint {http://arxiv.org/abs/2210.12163}{arXiv:2210.12163 [hep-th]}\BibitemShut {NoStop}%
\bibitem [{\citenamefont {Cheung}\ and\ \citenamefont {Remmen}(2023{\natexlab{b}})}]{Cheung:2023adk}%
  \BibitemOpen
  \bibfield  {author} {\bibinfo {author} {\bibfnamefont {C.}~\bibnamefont {Cheung}}\ and\ \bibinfo {author} {\bibfnamefont {G.~N.}\ \bibnamefont {Remmen}},\ }\href@noop {} {\  (\bibinfo {year} {2023}{\natexlab{b}})},\ \Eprint {http://arxiv.org/abs/2302.12263}{arXiv:2302.12263 [hep-th]}\BibitemShut {NoStop}%
\bibitem [{\citenamefont {Geiser}\ and\ \citenamefont {Lindwasser}(2023)}]{Geiser:2022exp}%
  \BibitemOpen
  \bibfield  {author} {\bibinfo {author} {\bibfnamefont {N.}~\bibnamefont {Geiser}}\ and\ \bibinfo {author} {\bibfnamefont {L.~W.}\ \bibnamefont {Lindwasser}},\ }\href {\doibase 10.1007/JHEP04(2023)031} {\bibfield  {journal} {\bibinfo  {journal} {JHEP}\ }\textbf {\bibinfo {volume} {04}},\ \bibinfo {pages} {031} (\bibinfo {year} {2023})},\ \Eprint {http://arxiv.org/abs/2210.14920}{arXiv:2210.14920 [hep-th]}\BibitemShut {NoStop}%
\bibitem [{\citenamefont {Bargmann}\ and\ \citenamefont {Todorov}(1977)}]{Bargmann:1977gy}%
  \BibitemOpen
  \bibfield  {author} {\bibinfo {author} {\bibfnamefont {V.}~\bibnamefont {Bargmann}}\ and\ \bibinfo {author} {\bibfnamefont {I.~T.}\ \bibnamefont {Todorov}},\ }\href {\doibase 10.1063/1.523383} {\bibfield  {journal} {\bibinfo  {journal} {J. Math. Phys.}\ }\textbf {\bibinfo {volume} {18}},\ \bibinfo {pages} {1141} (\bibinfo {year} {1977})}\BibitemShut {NoStop}%
\bibitem [{\citenamefont {Francia}\ \emph {et~al.}(2007)\citenamefont {Francia}, \citenamefont {Mourad},\ and\ \citenamefont {Sagnotti}}]{Francia:2007qt}%
  \BibitemOpen
  \bibfield  {author} {\bibinfo {author} {\bibfnamefont {D.}~\bibnamefont {Francia}}, \bibinfo {author} {\bibfnamefont {J.}~\bibnamefont {Mourad}}, \ and\ \bibinfo {author} {\bibfnamefont {A.}~\bibnamefont {Sagnotti}},\ }\href {\doibase 10.1016/j.nuclphysb.2007.03.021} {\bibfield  {journal} {\bibinfo  {journal} {Nucl. Phys. B}\ }\textbf {\bibinfo {volume} {773}},\ \bibinfo {pages} {203} (\bibinfo {year} {2007})},\ \Eprint {http://arxiv.org/abs/hep-th/0701163}{arXiv:hep-th/0701163}\BibitemShut {NoStop}%
\bibitem [{\citenamefont {Fierz}(1939)}]{Fierz:1939zz}%
  \BibitemOpen
  \bibfield  {author} {\bibinfo {author} {\bibfnamefont {M.}~\bibnamefont {Fierz}},\ }\href@noop {} {\bibfield  {journal} {\bibinfo  {journal} {Helv. Phys. Acta}\ }\textbf {\bibinfo {volume} {12}},\ \bibinfo {pages} {3} (\bibinfo {year} {1939})}\BibitemShut {NoStop}%
\bibitem [{\citenamefont {Rarita}\ and\ \citenamefont {Schwinger}(1941)}]{PhysRev.60.61}%
  \BibitemOpen
  \bibfield  {author} {\bibinfo {author} {\bibfnamefont {W.}~\bibnamefont {Rarita}}\ and\ \bibinfo {author} {\bibfnamefont {J.}~\bibnamefont {Schwinger}},\ }\href {\doibase 10.1103/PhysRev.60.61} {\bibfield  {journal} {\bibinfo  {journal} {Phys. Rev.}\ }\textbf {\bibinfo {volume} {60}},\ \bibinfo {pages} {61} (\bibinfo {year} {1941})}\BibitemShut {NoStop}%
\bibitem [{\citenamefont {'t~Hooft}(1971{\natexlab{a}})}]{tHooft:1971akt}%
  \BibitemOpen
  \bibfield  {author} {\bibinfo {author} {\bibfnamefont {G.}~\bibnamefont {'t~Hooft}},\ }\href {\doibase 10.1016/0550-3213(71)90395-6} {\bibfield  {journal} {\bibinfo  {journal} {Nucl. Phys. B}\ }\textbf {\bibinfo {volume} {33}},\ \bibinfo {pages} {173} (\bibinfo {year} {1971}{\natexlab{a}})}\BibitemShut {NoStop}%
\bibitem [{\citenamefont {'t~Hooft}(1971{\natexlab{b}})}]{tHooft:1971qjg}%
  \BibitemOpen
  \bibfield  {author} {\bibinfo {author} {\bibfnamefont {G.}~\bibnamefont {'t~Hooft}},\ }\href {\doibase 10.1016/0550-3213(71)90139-8} {\bibfield  {journal} {\bibinfo  {journal} {Nucl. Phys. B}\ }\textbf {\bibinfo {volume} {35}},\ \bibinfo {pages} {167} (\bibinfo {year} {1971}{\natexlab{b}})}\BibitemShut {NoStop}%
\bibitem [{\citenamefont {Berends}\ \emph {et~al.}(1985)\citenamefont {Berends}, \citenamefont {Burgers},\ and\ \citenamefont {van Dam}}]{Berends:1984rq}%
  \BibitemOpen
  \bibfield  {author} {\bibinfo {author} {\bibfnamefont {F.~A.}\ \bibnamefont {Berends}}, \bibinfo {author} {\bibfnamefont {G.~J.~H.}\ \bibnamefont {Burgers}}, \ and\ \bibinfo {author} {\bibfnamefont {H.}~\bibnamefont {van Dam}},\ }\href {\doibase 10.1016/0550-3213(85)90074-4} {\bibfield  {journal} {\bibinfo  {journal} {Nucl. Phys. B}\ }\textbf {\bibinfo {volume} {260}},\ \bibinfo {pages} {295} (\bibinfo {year} {1985})}\BibitemShut {NoStop}%
\bibitem [{\citenamefont {Cortese}\ \emph {et~al.}(2014)\citenamefont {Cortese}, \citenamefont {Rahman},\ and\ \citenamefont {Sivakumar}}]{Cortese:2013lda}%
  \BibitemOpen
  \bibfield  {author} {\bibinfo {author} {\bibfnamefont {I.}~\bibnamefont {Cortese}}, \bibinfo {author} {\bibfnamefont {R.}~\bibnamefont {Rahman}}, \ and\ \bibinfo {author} {\bibfnamefont {M.}~\bibnamefont {Sivakumar}},\ }\href {\doibase 10.1016/j.nuclphysb.2013.12.005} {\bibfield  {journal} {\bibinfo  {journal} {Nucl. Phys. B}\ }\textbf {\bibinfo {volume} {879}},\ \bibinfo {pages} {143} (\bibinfo {year} {2014})},\ \Eprint {http://arxiv.org/abs/1307.7710}{arXiv:1307.7710 [hep-th]}\BibitemShut {NoStop}%
\bibitem [{\citenamefont {Labastida}(1989)}]{Labastida:1987kw}%
  \BibitemOpen
  \bibfield  {author} {\bibinfo {author} {\bibfnamefont {J.~M.~F.}\ \bibnamefont {Labastida}},\ }\href {\doibase 10.1016/0550-3213(89)90490-2} {\bibfield  {journal} {\bibinfo  {journal} {Nucl. Phys. B}\ }\textbf {\bibinfo {volume} {322}},\ \bibinfo {pages} {185} (\bibinfo {year} {1989})}\BibitemShut {NoStop}%
\end{thebibliography}%

\end{document}